\documentstyle[preprint,aps]{revtex}
\tighten
\draft
\widetext
\preprint{YITP-99-53, NSF-ITP-99-131}
\bigskip
\bigskip
\begin{document}
\title{Geometry of Orientifolds with NS-NS $B$-flux}
\medskip

\author{Zurab Kakushadze\footnote{E-mail: 
zurab@insti.physics.sunysb.edu}}

\bigskip
\address{C.N. Yang Institute for Theoretical Physics\\ 
State University of New York, Stony Brook, NY 11794\\
and\\
Institute for Theoretical Physics\\ 
University of California, Santa Barbara, CA 93106}

\date{January 31, 2000}
\bigskip
\medskip
\maketitle

\begin{abstract}
{}We discuss geometry underlying orientifolds with non-trivial NS-NS $B$-flux.
If D-branes wrap a torus
with $B$-flux the rank of the gauge group is reduced due to non-commuting 
Wilson lines whose presence is
implied by the $B$-flux. In the case of D-branes transverse
to a torus with $B$-flux the rank reduction is due to a smaller
number of D-branes required by tadpole cancellation conditions 
in the presence of $B$-flux as some of the orientifold planes
now have the 
opposite orientifold projection. We point out that T-duality
in the presence of $B$-flux is more subtle than in the case with
trivial $B$-flux, 
and it is precisely consistent with the qualitative difference 
between the aforementioned two setups. In the case where both types
of branes are present, the states in the mixed ({\em e.g.}, 59)
open string sectors come with
a non-trivial multiplicity, which we relate to a discrete gauge
symmetry due to non-zero $B$-flux, and construct vertex operators for the
the mixed sector states. Using these results we revisit 
K3 orientifolds with $B$-flux (where K3 is a $T^4/{\bf Z}_M$ orbifold) and
point out various subtleties arising in some of 
these models. For instance, 
in the ${\bf Z}_2$ case the conformal field theory
orbifold does not appear to be the consistent background for the corresponding
orientifolds with $B$-flux. This is related to the fact that non-zero $B$-flux 
requires the presence of both O5$^-$- as well as O5$^+$-planes at various 
${\bf Z}_2$ orbifold fixed points, which appears to be
inconsistent with the presence of the
{\em twisted} $B$-flux in the conformal field theory orbifold.
We also consider
four dimensional ${\cal N}=2$ and ${\cal N}=1$ supersymmetric orientifolds.
We construct consistent four dimensional models with $B$-flux
which do not suffer from difficulties encountered in the K3 cases.
 
\end{abstract}

\section{Introduction and Summary}

{}In recent years six and four dimensional orientifolds have been extensively
studied, and much progress has been made in understanding such string 
compactifications. Various ${\cal N}=1$ supersymmetric six dimensional 
orientifold vacua were constructed, for instance, in 
\cite{PS,GP,GJ,ABPSS,KST}. Generalizations of these constructions to 
${\cal N}=1$ supersymmetric four dimensional orientifold vacua have also been
discussed, for instance, in 
\cite{BL,ABPSS1,Ka,KS,Zw,AFIV,KST1,Ka1,ST,Ka2,LPT,BW,IRU,Ka3,cvetic}.

{}Most of the aforementioned discussions have been confined to orientifolds 
with vanishing NS-NS antisymmetric tensor backgrounds. However, 
generalizations to cases with non-trivial NS-NS $B$-flux have also been 
considered. Thus, in \cite{BPS}\footnote{For a more recent discussion
of toroidal Type I compactifications, see \cite{Bi}.} 
toroidal Type I compactifications with non-zero 
$B$-field were studied. In \cite{BPS} it was pointed out that even though the
NS-NS 2-form is projected out of the closed sting spectrum by the orientifold
projection $\Omega$, quantized expectation values of $B_{ij}$ are allowed
with $i,j$ corresponding to the toroidally compactified coordinates. In 
particular, since the components of $B_{ij}$ are defined up to unit
shifts $B_{ij}\rightarrow B_{ij}+1$, and $B_{ij}$ is odd under the world-sheet
parity reversal $\Omega$, the allowed background values for the components of 
$B_{ij}$ are 0 and 1/2. Furthermore, in \cite{BPS}
it was found that the rank of the gauge group
coming from D9-branes wrapped on tori with non-vanishing half-integer
$B$-flux is reduced
from 16 (that is, the rank of the original $SO(32)$ gauge group) down to
$16/2^{b/2}$, where $b$ is the rank (which is always even) of the matrix
$B_{ij}$.  
This rank reduction is evident once one considers the cylinder 
partition function for D9-branes wrapped on such tori. However, the partition
function alone does not provide a clear geometric interpretation of the rank
reduction phenomenon. In particular, one might conclude that 
in the cases with non-zero $B$-field we have only 
$32/2^{b/2}$ D9-branes instead of the usual 32 D9-branes. This, however, does
not appear to be the case. Thus, as was originally pointed out in \cite{Ka},
the $\Omega$ orientifold of Type IIB with non-zero $B$-flux can be viewed
as a toroidal Type I compactification with all 32 D9-branes
in the presence of {\em non-commuting} 
Wilson lines. This point was further elaborated in more detail in \cite{Ka4}. 
For the reasons that will become clear in a moment, in this paper we will 
review the approach of \cite{Ka,Ka4} in detail.

{}Further progress in understanding toroidal orientifolds with non-zero 
$B$-flux was made in \cite{Wi}, where D-branes transverse to tori with 
non-trivial $B_{ij}$ were considered. In particular, it was argued in \cite{Wi}
that if we, say, consider O7-planes transverse to a 2-torus $T^2$ with 
$B_{12}=1/2$ in the directions of $T^2$, then we have two types of
orientifold planes O7$^-$ and O7$^+$, where the O$7^-$-plane refers to the
usual orientifold plane with the $SO$ type of orientifold projection on the 
Chan-Paton charges, whereas the O$7^+$-plane refers to the orientifold 
plane with the $Sp$ type of projection. More concretely, out of the four
O7-planes (located at the 4 points on $T^2$ fixed under the reflection 
$R:X_{1,2}\rightarrow - X_{1,2}$, where $R$ is now a part of the orientifold 
projection $\Omega R (-1)^{F_L}$) three are of the O7$^-$ type, while one is of
the O7$^+$ type (to be contrasted with the case with trivial $B$-flux 
where all four O7-planes are of the O7$^-$ type). 
The R-R charges of the O7$^-$- and 
O7$^+$-planes are $-8$ and $+8$, 
respectively, so that the total R-R charge to be
canceled by D7-branes is $-16$. This implies that we must introduce 16
(instead of 32) D7-branes in the presence of non-zero $B_{12}$, hence ``rank
reduction''. However, as we will point out in the following, the mechanisms
for the rank reduction in the cases of D-branes wrapped on tori with non-zero 
$B_{ij}$ {\em vs.} D-branes transverse to such tori are {\em different}. In 
the former case we still have 32 D-branes, and the rank reduction occurs due
to non-commuting Wilson lines \cite{Ka,Ka4}. 
In the latter case the number of D-branes
is not 32 but $32/2^{b/2}$ due to the fact that some of the orientifold planes
are no longer of the $SO$ but $Sp$ type. Thus, in this case there is really
no ``rank reduction''. Moreover, the two cases are {\em not} T-dual to each 
in the usual sense as T-duality is more subtle in the presence of the $B$-flux
(some aspects of T-duality in this context were discussed in 
\cite{Wi}). We will make this more precise in the following using the approach
of \cite{Ka,Ka4}, which, in particular, makes it clear that 
one of the O7-planes must indeed be of the $Sp$ type.

{}So far we have mentioned toroidal orientifold compactifications which
preserve 16 supersymmetries. To obtain backgrounds with reduced
supersymmetry we need to consider compactifications with non-trivial
holonomy. A step in this direction was made in \cite{KST}, where K3 
orientifolds with non-zero $B$-field were discussed. More concretely,  
in \cite{KST} the following backgrounds were considered. Start with Type IIB
on ${\rm K3}=T^4/{\bf Z}_M$, where $M=2,3,4,6$ so that the orbifold action on
$T^4$ is crystallographic. Then consider the 
$\Omega$ orientifold, where $\Omega$
acts not as in the smooth case\footnote{Orientifolds of Type IIB on smooth K3
surfaces with non-zero $B$-flux were discussed in \cite{SS}.} but as in
\cite{GJ}, that is, $\Omega$ maps the $g^k$ twisted sector to the
$g^{M-k}$ twisted
sector \cite{Po}, where $g$ is the generator of the orbifold group ${\bf Z}_M$,
and $k=0,\dots,M-1$. Let us assume that there is a non-zero half-integer
$B$-flux of rank $b$ in the directions of K3. This is the setup of \cite{KST}.
{\em A priori} we expect that in the ${\bf Z}_3$ case we have no D5-branes 
(for the above choice of the orientifold projection), but some number of 
D9-branes. In the other three cases, namely, ${\bf Z}_2,{\bf Z}_4,{\bf Z}_6$,
we expect both D9- and D5-branes to be present. The aforementioned
question of how many D9-branes we have in these backgrounds, namely, whether
we have 32 or $32/2^{b/2}$ D9-branes, becomes extremely relevant, at least in
the cases where we have both D9- and D5-branes. The reason for this is the
following. As was originally pointed out in \cite{KST}, the 59 sector states
arise with multiplicity of $2^{b/2}$, which is 1 in the absence of the 
$B$-field, but becomes non-trivial whenever the $B$-field is non-zero. This
fact becomes evident, as was explained in \cite{KST}, if one examines the
boundary states for the D9- and D5-branes in the presence of the 
$B$-field\footnote{Equivalently, one can examine the annulus partition 
function, and see that the multiplicity of states in the 59 sector is indeed
$2^{b/2}$. This was done for the ${\bf Z}_2$ models in 
\cite{An}.}. In particular, this multiplicity of states is in accord with the
tadpole/anomaly cancellation requirements in, say, the ${\bf Z}_2$ model
\cite{KST}. 
However, just by looking at the boundary states (or, equivalently, the 
partition function), the geometric interpretation of this multiplicity of 
states in the 59 sector is obscure. In particular, in string theory we expect
that no two states should have identical vertex operators. Thus, one should be
able to distinguish the otherwise degenerate 59 states (in the presence of the
$B$-field) from each other by some quantum numbers. In this paper we will give
an explicit answer to this question. In particular, we will use the
approach of \cite{Ka,Ka4}, and point out that the number of D5-branes 
(if present) is $32/2^{b/2}$, whereas the number of D9-branes is always 32, 
albeit the rank of the 99 gauge group is $16/2^{b/2}$ (for the reasons 
mentioned above). The 59 sector degeneracy then is related to the fact that
in these sectors there are $2^{b/2}$ different vertex operators with otherwise
identical quantum numbers, and the former are distinguished by $2^{b/2}$
Chan-Paton degrees of freedom that 32 D9-branes have on top of those 
corresponding to the 
unbroken gauge group. In particular, as we will see in the following, these
degrees of freedom correspond to $2^{b/2}$ different charges carried by the 59
sector states under a discrete gauge symmetry, namely, 
$({\bf Z}_2)^{\otimes(b/2)}$, arising in the 99 sector.
In fact, one can turn this point around
and argue that the multiplicity of states in the 59 sector gives us a hint 
for the number of D9-branes being 32 and not $32/2^{b/2}$.

{}Understanding the geometry underlying 
toroidal orientifolds with NS-NS $B$-flux, which is
one of the aims of this paper, is 
a useful tool for shedding light on other non-trivial orientifold 
compactifications with $B$-flux, 
say, with reduced number of supersymmetries. In particular, in this paper
we will revisit the K3 orientifolds with $B$-flux originally discussed in
\cite{KST}. It turns out that there are various subtleties arising in these
models. For instance, as
we will 
argue in the following, the conformal field theory $T^4/{\bf Z}_2$ orbifold
does not appear to be a consistent background for the corresponding
orientifolds with $B$-flux. This is related to the fact that in the presence
of non-trivial $B$-flux we must have both O5$^-$- as well as 
O$5^+$-planes at various
${\bf Z}_2$ orbifold fixed points, and this is incompatible with the presence
of the {\em twisted} $B$-flux in the conformal field theory orbifold.
The ${\bf Z}_6$ as well as ${\bf Z}_4$ models with $B$-flux also
suffer from this problem. In fact, we will discuss other subtleties arising 
in the ${\bf Z}_6$ and ${\bf Z}_4$ models, which appear to be related, just
as it turns out to be the case for their ${\bf Z}_2$ 
counterparts, to the
fact that here we have 59 sectors whose presence appears to be incompatible
with the lack of vector structure dictated by non-zero $B$-flux. 
On the other hand, ${\bf Z}_3$ models with, say, D5-branes only (but
no mixed 59 sector) appear to be consistent\footnote{Here we should point out
that the other aforementioned models might be consistent in some other
sense, but we fail to find consistent constructions for them within the
(perhaps limited) orientifold framework.}.

{}The key points of the above discussion carry over to analogous 
compactifications on Calabi-Yau orbifolds. In particular, every time we have, 
say, a ${\bf Z}_3$ twist acting in the compact directions with non-zero 
$B$-flux, extra care is needed in making sure that tadpole cancellation
is compatible with geometric constraints. We illustrate these issues
by revisiting the four dimensional models of \cite{ST,Ka2}, in some of which 
we also observe various subtleties. 

{}The remainder of this paper is organized as follows. In section II we discuss
toroidal orientifolds with $B$-flux. Our discussion here is essentially a
generalization of non-commutative toroidal compactifications in the presence
of orientifold planes. This, generalization, however, is {\em not} completely
straightforward as including orientifold planes brings in additional subtle
issues. In section III we revisit K3 orientifolds with $B$-flux, 
and explain in detail the
aforementioned subtle inconsistencies arising in some of these backgrounds.
In section IV we discuss
four dimensional cases with $SU(3)$ as well $SU(2)$ holonomy. In particular, 
here we construct consistent orientifold models with $B$-flux which do not
suffer from difficulties encountered in the K3 cases. 
We comment on various 
issues in section V.   
 
\section{Toroidal Orientifolds with $B$-flux}

{}In this section we will review the effects of the $B$-field in toroidal
orientifolds. We will use the approach of \cite{Ka,Ka4}, where the
geometric interpretation of such backgrounds
is evident. In subsection A we will discuss
the cases where D-branes (and the corresponding orientifold planes) wrap
a torus with non-zero $B$-field. In subsection B will will consider the cases
where D-branes are transverse to such tori. In subsection C we will generalize
our discussion to the cases where both types of D-branes are present. There
we will also discuss in detail T-duality in orientifolds with $B$-flux, and,
in particular, argue that the results of this section are consistent with
T-duality considerations.

\subsection{D-branes and O-planes Wrapped on Tori with $B$-flux}

{}Consider Type IIB (it is straightforward to generalize our discussion to 
Type IIA) on ${\bf R}^{1,9-d}\otimes T^d$ in the presence of some number 
$n$ of D$p$-branes completely wrapping the $d$-torus ($p\geq d$). We would like
to study the effect of turning on a quantized (half-integer) $B$-field in the
directions of $T^d$. For the sake of clarity we will consider D9-branes 
wrapping $T^2$. Generalizations to other cases are completely straightforward.

{}Thus, let us start from Type IIB on ${\bf R}^{1,7}\otimes T^2$ in the 
presence of some number $n$ of D9-branes. Let us first assume the the 
$B$-field in the compactified directions is zero: $B_{12}=0$. Let $g_{ij}$ be
the metric on $T^2$. Then in the suitable normalization the left- and 
right-moving closed string momenta are given by
\begin{equation}
 P_{L,R}={1\over 2} {\widetilde e}^i m_i \pm e_i n^i~,
\end{equation}
where $e_i$ are two-component vielbeins satisfying $e_i\cdot e_j=g_{ij}$,
while ${\widetilde e}^i$ are their duals: ${\widetilde e}^i\cdot {\widetilde
e}^j={\widetilde g}^{ij}$, where ${\widetilde g}^{ij}$ is the inverse of 
$g_{ij}$. Also, the integers $m_i$ and $n^i$ are the momentum and winding
numbers, respectively. As usual, $T^2$ can be viewed as a quotient 
${\bf R}^2/\Lambda$, where the lattice $\Lambda\equiv \{e_i n^i\}$, and the
coordinates $X_i$ on $T^2$ are identified via $X_i\sim X_i+e_i$.

{}Next, consider a freely acting ${\bf Z}_2\otimes {\bf Z}_2$ orbifold of this
theory defined as follows. Let $S_i$ be a half-lattice shift in the $X_i$ 
direction, that is, $S_iX_i=X_i+e_i/2$. Note that the set 
$\{I,S_1,S_2,S_3\}$ forms a freely acting orbifold group isomorphic to 
${\bf Z}_2\otimes {\bf Z}_2$. Here $I$ is the identity element,
and $S_3\equiv S_1S_2$. Thus, in this orbifold
we have the untwisted sector labeled by $I$, and three twisted sectors labeled
by $S_1,S_2,S_3$. The left- and right-moving momenta can now be written as
\begin{equation}\label{LR}
 P_{L,R}(\alpha^1,\alpha^2)={1\over 2} {\widetilde e}^i m_i \pm e_i 
 (n^i+{1\over 2}\alpha^i)~,
\end{equation}
where $(\alpha^1,\alpha^2)=(0,0)$, $(1,0)$, $(0,1)$ and $(1,1)$ in the
untwisted, $S_1$, $S_2$ and $S_3$ twisted sectors, respectively. The
most general action of $S_i$ on the left- and right-moving momenta 
(compatible with modular invariance) is given by
\begin{equation}\label{projection}
 S_i|P_L,P_R\rangle_{(\alpha^1,\alpha^2)} =\epsilon_i(\alpha^1,\alpha^2)
 \exp(\pi i m_i)|P_L,P_R\rangle_{(\alpha^1,\alpha^2)}~,
\end{equation}
where $\epsilon_i(0,0)\equiv 1$, $\epsilon_1(1,0)=\epsilon_2(0,1)=1$,
$\epsilon_1(0,1)=\epsilon_2(1,0)=\epsilon_{1,2}(1,1)=\epsilon$. Here 
$\epsilon$ can take two values: $\pm 1$. If $\epsilon=+1$, then we have the
usual freely acting ${\bf Z}_2\otimes {\bf Z}_2$ orbifold. It is 
straightforward to show that the resulting theory corresponds to a 
compactification on $(T^2)^\prime$ with the metric $g^\prime_{ij}=g_{ij}/4$,
and zero $B$-field $B_{12}=0$. However, if $\epsilon=-1$, in which case we have
{\em discrete torsion} between the generators $S_1$ and $S_2$ of the two ${\bf
Z}_2$ subgroups of the orbifold group (that is, $S_1$ acts with an extra minus
sign in the $S_2$ twisted sector, and, consequently, $S_2$ acts with the extra
minus sign in the $S_1$ twisted sector; both $S_1$ and $S_2$ act with an extra
minus sign in the $S_3$ twisted sector), then it is straightforward to show 
that the resulting theory corresponds to a compactification on $(T^2)^\prime$
with the metric $g^\prime_{ij}=g_{ij}/4$, but now the $B$-field is 
{\em non-zero}: $B_{12}=1/2$. In deriving these results it is important to
note that (\ref{projection}) implies that only {\em even} momentum numbers
$m_i$ survive the orbifold projection in all four sectors 
if there is no discrete torsion (that is, $\epsilon=1$), while in the case with
discrete torsion (that is, $\epsilon=-1$) the momenta kept after the orbifold 
projection are given as follows. Both $m_1$ and $m_2$ are even in the untwisted
sector. In the $S_1$ twisted sector $m_1$ is even and $m_2$ is odd. In the
$S_2$ twisted sector $m_1$ is odd and $m_2$ is even. Finally, in the $S_3$ 
twisted sector both $m_1$ and $m_2$ are odd. Taking all of this into account,
we can see from (\ref{LR}) and (\ref{projection}) that the resulting left- and
right-moving momenta are given by
\begin{equation}\label{LR1}
 P_{L,R}={1\over 2} {\widetilde e}^{\prime i} 
(m^\prime_i - B_{ij} n^{\prime j}) \pm e^\prime_i 
 n^{\prime i}~,
\end{equation}
where the new momentum and winding numbers $m^\prime_i$ and $n^{\prime i}$
are now arbitrary integers, the new vielbeins $e^\prime_i$ and their duals 
${\widetilde e}^{\prime i}$ are related to the original ones via $e^\prime_i=
e_i/2$, ${\widetilde e}^{\prime i}=2{\widetilde e}^i$, and the $B$-field
is zero for $\epsilon=1$, while $B_{12}=1/2$ for $\epsilon=-1$. 

{}Thus, what we have learned from the above discussion is that we can
describe a compactification on $(T^2)^\prime$ 
with half-integer $B$-field as a freely
acting ${\bf Z}_2\otimes {\bf Z}_2$ orbifold that involves half-shifts
along the two cycles of $T^2$ (the metric on $(T^2)^\prime$ is four times
as small as that on $T^2$, that is, both of the cycles on $(T^2)^\prime$ are
half the size of the corresponding cycles on $T^2$) with discrete torsion 
between the generators $S_1$ and $S_2$ of the two ${\bf Z}_2$ subgroups of the
orbifold group ${\bf Z}_2\otimes {\bf Z}_2$. This approach proves to be
convenient in the context of D-branes (and O-planes)
wrapping tori with non-zero $B$-field as
we can recast the latter problem into the corresponding freely acting orbifold
of a setup where D-branes are wrapping a torus with {\em zero} $B$-field. In 
particular, as we will see in a moment, this provides a geometrization of 
the cases where D-branes wrap tori with non-zero $B$-field.

{}Let us now see what happens to open strings stretched between $n$ D9-branes
wrapped on $(T^2)^\prime$. As we have already mentioned, to study this 
system we can start from $n$ D9-branes wrapped on $T^2$ without the $B$-field, 
and then consider the aforementioned freely acting ${\bf Z}_2\otimes {\bf Z}_2$
orbifold. Thus, before orbifolding we have open string states
which can be obtained by the corresponding Kaluza-Klein (KK)
compactification of the
ten dimensional open string spectrum on $T^2$. In particular, among other
quantum numbers (corresponding to the open string oscillator modes) we have
the Kaluza-Klein momenta $m_i\in {\bf Z}$ with the contributions to the
masses of the corresponding open string states
given by $M_{\bf m}^2={1\over 2}P_{\bf m}^2$, where $P_{\bf m}
\equiv {\widetilde e}^i m_i$ (here ${\bf m}\equiv(m_1,m_2)$). 
The generators $S_1$ and $S_2$ of the ${\bf Z}_2\otimes {\bf Z}_2$ orbifold
group act on the momentum states as follows:
\begin{equation}
 S_i|P_{\bf m}\rangle = \exp(\pi i m_i) |P_{\bf m}\rangle~.
\end{equation}   
However, we must also specify the action of the orbifold group elements on the
Chan-Paton charges of D9-branes. It is described by $n\times n$ matrices
that form a representation of ${\bf Z}_2\otimes {\bf Z}_2$. Thus, we are free
to choose the Chan-Paton matrix $\gamma_I$ corresponding to the identity
element $I$ of the orbifold group as $\gamma_I=I_n$, where here and in the
following $I_m$ will denote the $m\times m$ identity matrix. As to the
twisted Chan-Paton matrices $\gamma_{S_1}$ and $\gamma_{S_2}$ (as well as
$\gamma_{S_3}$), they must satisfy certain constraints depending on the choice
of $\epsilon$, that is, depending upon whether we have discrete torsion or not.
In particular, if $\epsilon=1$, then $\gamma_{S_1}$ and $\gamma_{S_2}$ must 
commute. This follows from the fact that in this case we simply rescale the
two cycles on $T^2$ to obtain $(T^2)^\prime$ without the $B$-field, and if
the matrices $\gamma_{S_i}$ are non-trivial, then they act as (discrete)
Wilson lines corresponding to the two cycles on $T^2$. However, if 
$\epsilon=-1$, that is, if we have non-zero $B$-field ($B_{12}=1/2$), then
the string consistency (in particular, the closed to open string coupling
consistency) requires that $\gamma_{S_1}$ and $\gamma_{S_2}$ must {\em 
anticommute} \cite{Ka,Wi,Ka4}. In this case we can still view the action of the
orbifold group on the Chan-Paton Charges as having (discrete) Wilson lines, but
now they are {\em no} longer commuting \cite{Ka,Ka4}. Thus, to summarize,
the Chan-Paton matrices must satisfy
\begin{equation}
 \gamma_{S_1}\gamma_{S_2}=\epsilon\gamma_{S_2}\gamma_{S_1}~.
\end{equation}  
In the case without discrete torsion ($\epsilon=1$) we thus have gauge bundles
with vector structure. In the case with discrete torsion ($\epsilon=-1$)
we have gauge bundles {\em without} vector structure (and the corresponding
generalized second Stieffel-Whitney class is non-vanishing) \cite{Wi}.

{}Note that in the cases without discrete torsion the matrices $\gamma_{S_i}$
can essentially be arbitrary as long as they commute and form a (projective) 
representation of ${\bf Z}_2\otimes {\bf Z}_2$. In particular, their traces
are not fixed - there are no tadpoles associated with the twists $S_i$ as they
are freely acting. Thus, the action of the ${\bf Z}_2\otimes 
{\bf Z}_2$ orbifold on the Chan-Paton charges can be trivial, that is, 
$\gamma_{S_i}=I_n$, which corresponds to having trivial Wilson lines. If we 
do not include an orientifold plane, then the gauge symmetry is $U(n)$ with 
this choice of Wilson lines, while if Wilson lines are non-trivial, then
the gauge group $G$ is a subgroup of $U(n)$ with rank $r(G)=n$. In the case of
D9-branes the oriented open string theory suffers from massless tadpoles, which
can be canceled if we introduce the 
O9$^-$-plane (with the $SO$ type of orientifold projection on
the Chan-Paton charges), and choose the number $n$ of D9-branes to be $n=32$.
Then in the case of trivial Wilson lines we have the usual $SO(32)$ gauge 
group, while non-trivial Wilson lines break $SO(32)$ to its subgroup $G$
with rank $r(G)=16$.

{}However, in the case with discrete torsion the situation is qualitatively
different. In particular, all three twisted
Chan-Paton matrices $\gamma_{S_a}$, $a=1,2,3$, must be traceless. This follows
from the fact that for these matrices to form a (projective) representation of
${\bf Z}_2\otimes {\bf Z}_2$, they must satisfy
\begin{eqnarray}
 &&\gamma_{S_a}^2=\eta_{aa} I_n~,\\
 &&\gamma_{S_a}\gamma_{S_b}=\eta_{ab} \gamma_{S_c}~,~~~a\not=b\not=c~,
\end{eqnarray} 
where not all the nine structure constants $\eta_{ab}$ are independent but 
satisfy the following relations:
\begin{eqnarray}
 &&\eta_{33}=-\eta_{11}\eta_{22}~,\\
 &&\eta_{21}=-\eta_{12}~,\\
 &&\eta_{13}=-\eta_{31}=\eta_{11}\eta_{12}~,\\
 &&\eta_{23}=-\eta_{32}=-\eta_{22}\eta_{12}~.
\end{eqnarray}
Here $\eta_{11},\eta_{22},\eta_{12}$ {\em a priori} independently take values 
$\pm 1$. It then follows that ${\rm Tr}(\gamma_a)\equiv 0$. Moreover, even if
we consider the oriented open string theory, the number of D9-branes must be 
even: $n=2N$. In fact, we can now see the rank reduction phenomenon mentioned
in the previous section. Thus, the rank of the unbroken gauge group is no
longer $n$ but twice as small, that is, $N$. This follows from the fact that
the Wilson lines corresponding to the two cycles of $(T^2)^\prime$ do not 
commute, which can be seen 
from the fact that $\gamma_{S_1}$ and $\gamma_{S_2}$ 
{\em anticommute}. If we consider an unoriented open string theory, that is,
if we introduce an orientifold 9-plane, the rank of the gauge group is
then no longer $n/2$ (as is the case without discrete torsion), but rather
$N/2$ (in the presence of an orientifold plane $N$ must also be even). 
Intuitively it should be clear that to cancel all the tadpoles we must 
introduce the
O9$^-$-plane (and {\em not} the O9$^+$-plane), and choose the number of
D9-branes to be $n=32$. However, we would like to derive this result
rigorously as understanding this point will be important for the subsequent
discussions. 

{}To do this, let us start with the Klein bottle amplitude
${\cal K}$. It is obtained from
the torus amplitude by inserting the orientifold projection $\Omega$ into
the trace over the Hilbert space of closed string states. This implies that
the only states contributing into the Klein bottle amplitude are {\em 
left-right symmetric} closed string states. The oscillator modes are not going
to be important in the following as their contributions are the same with
or without the $B$-flux, so let us focus on the left- and right-momentum
contributions. Note, in particular, that the momentum numbers $m_i$ are
invariant under the action of $\Omega$, while the winding numbers $n^i$ change
sign under the action of $\Omega$. This implies that only the states with
zero winding (but arbitrary momentum) numbers contribute to the Klein bottle 
amplitude. Such states are the same regardless of the $B$-flux, which can be
readily seen from (\ref{LR1}). Thus, the Klein bottle amplitude is independent
of the $B$-field. We will write the Klein bottle amplitude in the language
of the ${\bf Z}_2\otimes {\bf Z}_2$ freely acting orbifold of $T^2$ with
discrete torsion (here we do not display the oscillator contributions):
\begin{equation}\label{Klein}
 {\cal K}=\left({1\over2}\right)\left({1\over 4}\right)
 \sum_{\bf m} q^{{1\over 2}P_{\bf m}^2}\Big[1+(-1)^{m_1}+(-1)^{m_2}+
 (-1)^{m_1+m_2}\big]~,
\end{equation}
where $P_{\bf m}={\widetilde e}^i m_i$, 
${\bf m}=(m_1,m_2)$, and $m_i\in {\bf Z}$. Also, the first numerical 
prefactor of 
$(1/2)$ is related to the orientifold projection, while the second numerical
prefactor of $(1/4)$ is related to the ${\bf Z}_2\otimes{\bf Z}_2$ orbifold 
projection. As we have already mentioned, the Klein bottle amplitude 
(\ref{Klein}) does not depend on whether we have discrete torsion or not, that
is, ${\cal K}$ in (\ref{Klein}) is the same as the Klein bottle amplitude
for the ${\bf Z}_2\otimes{\bf Z}_2$ freely acting orbifold {\em without}
discrete torsion.
Note that ${\cal K}$ is written in terms of the metric on $T^2$.
We can rewrite it in terms of the metric on $(T^2)^\prime$ as follows:
\begin{equation}\label{Klein1}
 {\cal K}=\left({1\over2}\right)
 \sum_{{\bf m}^\prime} q^{{1\over 2}P_{{\bf m}^\prime}^2}~,
\end{equation}
where where $P_{{\bf m}^\prime}={\widetilde e}^{\prime i} m^\prime_i$, 
${\bf m}^\prime=(m^\prime_1,m^\prime_2)$, and $m^\prime_i\in {\bf Z}$.
In arriving at (\ref{Klein1})
we have explicitly performed the ${\bf Z}_2\otimes{\bf Z}_2$ 
orbifold projections in (\ref{Klein}) which keep only the states with
$m_i\in 2{\bf Z}$. These states are then rewritten in terms of new momenta
$m^\prime_i\in {\bf Z}$ with the dual vielbeins ${\widetilde e}^{\prime i}=
2{\widetilde e}^i$ on $(T^2)^\prime$. 

{}Let us now discuss the annulus amplitude. We will write the latter in 
the language of D9-branes wrapped on $T^2$ with the subsequent 
${\bf Z}_2\otimes{\bf Z}_2$ freely acting orbifold action. Thus, the annulus 
amplitude is given by:
\begin{eqnarray}
 {\cal A}=\left({1\over2}\right)\left({1\over 4}\right)
 \sum_{\bf m} q^{{1\over 2}P_{\bf m}^2}\Big[ 
 ({\rm Tr}(\gamma_I))^2+&&(-1)^{m_1}({\rm Tr}(\gamma_{S_1}))^2 +
 \nonumber\\
 &&
  (-1)^{m_2}({\rm Tr}(\gamma_{S_2}))^2 +
 (-1)^{m_1+m_2}({\rm Tr}(\gamma_{S_3}))^2\Big]~.\label{annulus} 
\end{eqnarray}
Here we have chosen a specific orientation on the annulus so that the
traces over the Chan-Paton charges give
$({\rm Tr}(\gamma_{S_a}))^2$, while if we chose the opposite orientation, we
would instead have ${\rm Tr}(\gamma_{S_a}){\rm Tr}(\gamma^{-1}_{S_a})$.

{}Note that ${\rm Tr}(\gamma_I)=n$, and ${\rm Tr}(\gamma_{S_a})=0$,
$a=1,2,3$, and we can {\em formally} rewrite the annulus amplitude
via
\begin{eqnarray}\label{erroneous}
 {\cal A}_1=\left({1\over2}\right)
 \sum_{\bf m} q^{{1\over 2}P_{\bf m}^2} ({\rm Tr} (I_N))^2~, 
\end{eqnarray}
so that naively one could reinterpret this annulus amplitude as corresponding
to that of $N=n/2$ D9-branes (instead of $n$ D9-branes). This interpretation,
however, would be erroneous. Indeed, one should get a hint of this from
the sum over the momenta ${\bf m}$. These are the momenta written in terms of 
the metric on the original torus $T^2$. However, we would have to rewrite it
in terms of the metric on the torus $(T^2)^\prime$ - after all 
we are considering D9-branes wrapped $(T^2)^\prime$ (plus the $B$-field), 
and {\em not} on $T^2$, the latter merely being the starting point for the 
${\bf Z}_2\otimes {\bf Z}_2$ 
orbifold with discrete torsion which is a way of obtaining
$(T^2)^\prime$ with the $B$-field. However, if we chose to reinterpret the
annulus partition function via (\ref{erroneous}), where we essentially would 
be trying to ameliorate all the traces of the ${\bf Z}_2\otimes {\bf Z}_2$ 
orbifold, we would have to write the momenta in terms of the metric on 
$(T^2)^\prime$. Thus, we would ultimately arrive at the conclusion that 
the momenta would have to take {\em half}-integer (instead of usual integer)
values. Indeed, $P_{\bf m}$ in (\ref{erroneous}) is given by $P_{\bf m}=
{\widetilde e}^i m_i$, which can be rewritten in terms of the
dual vielbeins ${\widetilde e}^{\prime i}$ on $(T^2)^\prime$ as $P_{\hat 
{\bf m}}=
{\widetilde e}^{\prime i}{\hat m}_i$, where ${\hat m}_i\equiv m_i/2$ (recall 
that ${\widetilde e}^{\prime i}=2{\widetilde e}^i$), so that the new momenta
${\hat m}_i$ would be half-integer. This signals that such an interpretation
would indeed be erroneous\footnote{This erroneous interpretation has been
adopted in most of the literature on orientifolds with non-zero $B$-flux. 
Often it does not lead to inadequate description of the massless spectra of
such orientifolds. For instance, this interpretation, which was essentially
adopted in \cite{BPS}, gives the correct results in the case of toroidal 
orientifolds with $B$-flux due to their relative simplicity (these 
theories have 16 supercharges). In fact, this can be
understood from the fact that the annulus amplitude in (\ref{erroneous}) would
give the same massless spectrum as that in (\ref{annulus}). Moreover, it would
predict the same degeneracy of states
at the massive KK levels as (\ref{annulus}), but 
the vertex operators at the massive levels in the two interpretations would be
different. As we will see in the following, in more complicated cases such as
K3 or Calabi-Yau orientifolds with $B$-flux knowing the correct structure of
vertex operators becomes relevant already at the {\em massless} level, for 
instance, in the 59 open string sector.}. (In particular, note that in the 
Klein bottle amplitude (\ref{Klein1}) the sum is over integer momenta 
$m^\prime_i$ 
written in terms of the metric on $(T^2)^\prime$, while in the annulus 
amplitude (\ref{erroneous}) the sum is over {\em half}-integer momenta
${\hat m}_i$.)  
Instead, the correct physical interpretation is that
we have $n=2N$ (and {\em not} $N$) D9-branes. The effect of the $B$-field
then can be understood as in the interpretation provided by the annulus
amplitude (\ref{annulus}) via the ${\bf Z}_2\otimes {\bf Z}_2$ orbifold
construction. Thus, we see that merely examining the partition functions
is not sufficient to understand the geometric structure of orientifolds
with $B$-field as the former only provide us with the multiplicities of states
at various string levels, but may not carry the complete
information about the structure of 
the vertex operators.

{}At any rate, let us start with the annulus amplitude (\ref{annulus}), and
proceed further. Thus, we would like to discuss the M{\"o}bius strip amplitude
next. It is given by
\begin{eqnarray}
 {\cal M}=\lambda\left({1\over2}\right)\left({1\over 4}\right)
 \sum_{\bf m} q^{{1\over 2}P_{\bf m}^2}\Big[ 
 &&{\rm Tr}(\gamma_\Omega^{-1}\gamma_\Omega^T)+
 (-1)^{m_1}{\rm Tr}(\gamma_{\Omega S_1}^{-1}\gamma_{\Omega S_1}^T) +
 \nonumber\\
 &&
  (-1)^{m_2}{\rm Tr}(\gamma_{\Omega S_2}^{-1}\gamma_{\Omega S_2}^T) +
 (-1)^{m_1+m_2}{\rm Tr}(\gamma_{\Omega S_3}^{-1}\gamma_{\Omega S_3}^T)\Big]~,
 \label{Moebius} 
\end{eqnarray}
where $\lambda=\mp 1$ for the $SO/Sp$ orientifold projection (that is, 
$\lambda=\pm 1$ for the O9$^\pm$-plane). We can simplify this expression
by noting that
\begin{eqnarray}
 &&{\rm Tr}(\gamma_\Omega^{-1}\gamma_\Omega^T)={\rm Tr}(\gamma_I)~,\\
 &&{\rm Tr}(\gamma_{\Omega S_a}^{-1}\gamma_{\Omega S_a}^T)={\rm Tr}
 (\gamma_{(S_a)^2})=\eta_{aa}{\rm Tr}(\gamma_I)~.
\end{eqnarray}
In the second line we have used the fact that $\gamma_{(S_a)^2}=\gamma_{S_a}^2=
\eta_{aa} I_n$. Thus, the M{\"o}bius strip amplitude is given by
\begin{eqnarray}
 {\cal M}=\lambda\left({1\over2}\right)\left({1\over 4}\right)
 \sum_{\bf m} q^{{1\over 2}P_{\bf m}^2} 
 {\rm Tr}(\gamma_I)\Big[1+\eta_{11}(-1)^{m_1} +\eta_{22}(-1)^{m_2}+
 \eta_{33}(-1)^{m_1+m_2}\Big]~,
 \label{Moebius1} 
\end{eqnarray}
where $\eta_{33}=-\eta_{11}\eta_{22}$.

{}Note that the quantity in the square brackets in (\ref{Moebius1}) is always
$+2$ or $-2$. Thus, we can write
\begin{equation}
 1+\eta_{11}(-1)^{m_1} +\eta_{22}(-1)^{m_2}+
 \eta_{33}(-1)^{m_1+m_2}\equiv 2\rho_{\bf m}(\eta_{aa})~,
\end{equation}
where $\rho_{\bf m}(\eta_{aa})=\pm 1$. In fact, for a given choice of
$\eta_{aa}$, $\rho_{\bf m}$ by definition 
only depends on whether $m_1$ and $m_2$ are even or odd. So {\em formally}
we can 
rewrite the M{\"o}bius strip amplitude as follows:
\begin{eqnarray}
 {\cal M}_1=\lambda\left({1\over2}\right)
 \sum_{\bf m} q^{{1\over 2}P_{\bf m}^2} 
 {\rm Tr}(I_N)\rho_{\bf m}(\eta_{aa})~.
 \label{erroneous1} 
\end{eqnarray}
Thus, naively one could reinterpret this M{\"o}bius strip amplitude along the
lines of (\ref{erroneous}) as corresponding to that of $N=n/2$ D9-branes 
(instead of $n$ D9-branes), with the $SO/Sp$ orientifold 
projection at integer KK levels
${\hat m}_i$ (recall that ${\hat m}_i\equiv m_i/2$) if $\lambda\rho_{\bf m}
(\eta_{aa})=\mp 1$ for $m_1,m_2\in 2{\bf Z}$, while at the half-integer KK
levels ${\hat m}_i$ the type of the orientifold projection is determined
by the corresponding signs $\lambda\rho_{\bf m}(\eta_{aa})$ with either $m_1$
or $m_2$ or both odd. Such an interpretation, however, would be erroneous for 
the same reasons as in the annulus case discussed above.

{}Next, we turn to the tadpole cancellation conditions. To extract the 
tadpoles, we must rewrite the open string 
{\em loop-channel} Klein bottle ${\cal K}$, 
annulus ${\cal A}$ and M{\"o}bius strip ${\cal M}$ amplitudes in terms of
the corresponding closed string 
{\em tree-channel} exchange expressions. In doing so, as 
usual, one must be careful with the relative normalizations between the
proper times on these three surfaces. The modular transformations that map
the loop-channel expressions to the tree-channel expressions amount to
Poisson resummations of the momentum sums in ${\cal K}$, ${\cal A}$ and 
${\cal M}$ (they also act non-trivially on the characters corresponding to
the oscillator contributions). It is not difficult to see that after Poisson
resummations the terms in (\ref{Klein}), (\ref{annulus}) and (\ref{Moebius})
containing $(-1)^{m_1}$, $(-1)^{m_2}$ and $(-1)^{m_1+m_2}$ do {\em not}
contain terms corresponding to the
massless closed string exchange, and, therefore,
do not contribute to the tadpoles. This, actually, has been 
anticipated from the fact that the ${\bf Z}_2\otimes {\bf Z}_2$ orbifold is
freely acting. The remaining terms, which do contribute to the tadpoles, are
the same (up to a universal overall factor of $1/4)$ as those in the $\Omega$
orientifold of Type IIB on $T^2$ (with zero $B$-field) in the presence of $n$
D9-branes. This makes extracting the tadpoles in our case straightforward. 
Thus, the tadpoles are given by (here $c$ is a universal numerical constant):
\begin{eqnarray}
 &&{\rm Tad}({\cal K})=c~{\rm Vol}(T^2)~32^2~,\\
 &&{\rm Tad}({\cal A})=c~{\rm Vol}(T^2)~({\rm Tr}(\gamma_I))^2~,\\
 &&{\rm Tad}({\cal M})=c~{\rm Vol}(T^2)~64\lambda~{\rm Tr}(\gamma_I)~, 
\end{eqnarray}  
so that the total tadpole factorizes into a perfect square
\begin{equation}
 {\rm Tad}=c~{\rm Vol}(T^2)\big[32+\lambda~{\rm Tr}(\gamma_I)\big]^2~. 
\end{equation}
Since ${\rm Tr}(\gamma_I)=n$, we conclude that the orientifold projection
must be of the $SO$ type ($\lambda=-1$), that is, we must include the 
O9$^-$-plane (and {\em not} the O9$^+$-plane), as well as $n=32$ D9-branes.

{}Thus, we see that the number of D9-branes is indeed 32, and the orientifold
plane is of the O9$^-$ type, which induces the $SO$ type of projection on the
D9-branes. This, however, does {\em not} imply that we cannot obtain, say,
$Sp$ gauge symmetry, which is expected to arise in orientifolds with $B$-flux
\cite{BPS}. In order to understand this point we must study the possible
choices of the twisted Chan-Paton matrices $\gamma_{S_i}$, which is related to
the question of possible gauge bundles on $T^2$ in the presence of the 
$B$-field.

{}To begin with let us note that it suffices to consider a $2\times 2$ 
representation for $\gamma_{S_i}$ as the full $2N\times 2N$ matrices
can be obtained as the $N$-fold copy of the corresponding $2\times 2$ 
matrices. Thus, we can write
\begin{equation}
 \gamma_{S_a}=\gamma_a \otimes I_N~,
\end{equation}  
where the matrices $\gamma_a$, $a=1,2,3$, form the aforementioned $2\times 2$ 
representation. These matrices must satisfy $\gamma_a^2=\eta_{aa} I_2$ as 
well as ${\rm Tr}(\gamma_a)=0$ conditions, which, in particular, imply that
$\det(\gamma_a)=-\eta_{aa}$. Recall that $\eta_{33}=-\eta_{11}\eta_{22}$, so 
either all three $\gamma_a$ matrices have determinant $+1$ 
(if $\eta_{11}=\eta_{22}=-1$), or two have
determinant $-1$, while one has determinant $+1$ (for the other three choices
of $\eta_{11}$ and $\eta_{22}$). In the former case $\gamma_a$ are $SU(2)$ 
matrices, and form a 2-dimensional representation of the non-Abelian dihedral
$D_4$ subgroup of $SU(2)$. In the latter case $\gamma_a$ form a 2-dimensional
representation of the ``double cover'' of the $D_4$ subgroup of $SU(2)$, which
we will denote by $D_4^\prime$. Note that $D_4^\prime$ is not a subgroup of
$SU(2)$ but is a subgroup of $SO(3)$. Up to equivalent representations, we can
write $\gamma_a$ for the above two cases as follows:
\begin{eqnarray}\label{D4}
 &&D_4:~~~\gamma_1=i\sigma_3~,~~~\gamma_2=i\sigma_2~,~~~\gamma_3=i\eta_{12}
 \sigma_1~,\\
 \label{D4'}
 &&D_4^\prime:~~~\gamma_1=\sigma_3~,~~~\gamma_2=\sigma_1~,~~~\gamma_3=i
 \eta_{12}\sigma_2~.
\end{eqnarray}
Here $\sigma_1,\sigma_2,\sigma_3$ are the usual $2\times 2$ Pauli matrices.    
It is not difficult to show that the ${\bf Z}_2\otimes {\bf Z}_2$ orbifold
projection breaks the $SO(32)$ gauge group on 32 D9-branes at the O9$^-$-plane
down to $Sp(16)$ in the $D_4$ case\footnote{In our notations $Sp(2r)$ has rank
$r$.}, 
and $SO(16)$ in the $D_4^\prime$ case. 
More precisely, the unbroken gauge symmetries are $Sp(16)\otimes {\bf Z}_2$
and $SO(16)\otimes {\bf Z}_2$, respectively. 
The extra discrete ${\bf Z}_2$ gauge symmetry 
will be important in the subsequent discussions, so let us elaborate on its
appearance in more detail.

{}Let us first consider the $D_4^\prime$ case.
Note that the first ${\bf Z}_2$ twist $\gamma_{S_1}$ acting on the Chan-Paton
charges breaks $SO(32)$ down to $SO(16)\otimes SO(16)$. The second ${\bf Z}_2$
twist $\gamma_{S_2}$ breaks the latter down to its diagonal subgroup 
$SO(16)_{\rm diag}$ times the discrete ${\bf Z}_2$ gauge symmetry associated 
with the permutation of the two $SO(16)$ subgroups. Similarly, in the $D_4$
case $\gamma_{S_1}$ breaks $SO(32)$ down to $U(16)$, and $\gamma_{S_2}$
breaks the latter down to $Sp(16)$ times the discrete ${\bf Z}_2$ subgroup
of the $U(1)$ subgroup of $U(16)$ (under which, for instance, the fundamental 
and antifundamental representations of $SU(16)$ are charged).

{}Before we end this subsection, a few comments are in order. First, above
we have described how to obtain the $SO(16)$ and $Sp(16)$ gauge symmetries
via the ${\bf Z}_2\otimes {\bf Z}_2$ freely acting orbifold construction. 
Since these theories have 16 supercharges, it is clear that it should be 
possible to construct other points in the moduli space, whose generic points
have $U(1)^8$ gauge symmetry, but at other points one should be able to 
obtain enhanced unitary gauge subgroups with rank 8, 
in particular, $U(8)$. Moreover, one expects
that the moduli space of gauge bundles on $T^2$ without vector structure
should be connected, so that it should be possible to continuously interpolate
between the $SO(16)$ and $Sp(16)$ points. A detailed discussion of gauge
bundles on $T^2$ without vector structure can be found in \cite{Wi}. Here we 
will briefly review some of the basic relevant facts. Thus, the matrices 
$\gamma_{S_i}$ given above correspond to points in the moduli space which 
can be described via the ${\bf Z}_2\otimes {\bf Z}_2$ freely acting orbifold.
However, the matrices $\gamma_{S_i}$ can be more generally viewed as 
describing the gauge bundle on $T^2$ without vector structure if we think
about them as anticommuting Wilson lines. Thus, we can relax all the 
constraints we have imposed on $\gamma_{S_i}$ except for the non-commutative
property: $\gamma_{S_1}\gamma_{S_2}=-\gamma_{S_2}\gamma_{S_1}$. The most
general solution to this constraint can be (up to equivalent
representations) written as
\begin{equation}\label{gammaSi}
 \gamma_{S_i}=\gamma_i \otimes \Gamma_i~,
\end{equation}
where the unitary 
$N\times N$ matrices $\Gamma_i$ commute. In particular, note that
there is no longer a restriction on $\Gamma_i^2$. Thus, starting from the
solution corresponding to $D_4$ we can smoothly interpolate to the solution 
corresponding to $D_4^\prime$ with the intermediate points corresponding to
$U(N/2)$ or its subgroups of rank $N/2$.

{}Another point we would like to mention is generalization to higher tori
with the rank of $B_{ij}$ $b>2$. It is clear that, say, in the case of $T^4$
we can represent non-zero $B$-field essentially along the same lines as we
have done so far for $T^2$. Instead of being most general here, for 
illustrative purposes let us consider the case of $T^4=T^2\otimes T^2$ 
(generalizations should be clear). We can introduce two Wilson lines 
corresponding to the two cycles on the first $T^2$, call the corresponding 
Chan-Paton matrices $\gamma_{S_1},\gamma_{S_2}$, and two other Wilson lines
corresponding to the two cycles on the second $T^2$, call the corresponding
Chan-Paton matrices $\gamma_{T_1},\gamma_{T_2}$. If we take $\gamma_{S_1}$ and
$\gamma_{S_2}$ anticommuting, as well as $\gamma_{T_1}$ and $\gamma_{T_2}$
anticommuting, but $\gamma_{S_i}$ and $\gamma_{T_j}$ commuting, this 
corresponds to rank $b=4$ half-integer $B$-field on $T^4$. Such Chan-Paton 
matrices can be easily constructed. Thus, let $\gamma_a$ be a set of three
anticommuting $2\times 2$ matrices, and let $\beta_a$ be another set of
three anticommuting $2\times 2$ matrices, each set satisfying the constraints
discussed above in the $T^2$ case. Then we can choose the Chan-Paton matrices
corresponding to the four Wilson lines on $T^4$ as follows ($N^\prime\equiv
n/4$):
\begin{eqnarray}
 &&\gamma_{S_a}=\gamma_a\otimes I_2\otimes I_{N^\prime}~,\\  
 &&\gamma_{T_a}=I_2\otimes \beta_a\otimes I_{N^\prime}~.
\end{eqnarray}
Note that the type of the unbroken gauge group can be determined as follows.
If $\gamma_a$ and $\beta_a$ both are of the $D_4^\prime$ or $D_4$ type, 
then the unbroken gauge group is $SO(N^\prime)$. If one of them is of the
$D_4^\prime$ type while the other one is of the $D_4$ type, then the unbroken 
gauge group is $Sp(N^\prime)$. (More precisely, the unbroken gauge symmetry
includes the corresponding ${\bf Z}_2\otimes {\bf Z}_2$ subgroup.)
Other points in the moduli space interpolating
between these special configurations can be obtained in complete parallel with
the $T^2$ case. Also, generalizations to higher tori, in particular, $T^6$
should be evident from the above discussions.  

{}Finally, note that, say, in the case of $T^2$ having two non-commuting
Wilson lines implies that some of the components of the non-Abelian 
gauge field strength
$F_{12}$ (in the language of the original $SO(32)$ gauge symmetry)
in the compact directions are non-zero. (In the four dimensional language
this implies that some of the D-term components are non-zero.) This is
precisely the statement that the corresponding generalized second 
Stieffel-Whitney class of the gauge bundle in non-vanishing. Note that this is
perfectly consistent with the low energy supersymmetry as the 
moduli corresponding
to these directions are absent in the low energy spectrum of the $SO(16)$ or
$Sp(16)$ gauge theory with $16$ supercharges. In particular, the moduli 
space of gauge bundles of the $SO(32)$ theory on $T^2$ with vector structure
is $2\times 16$ dimensional, whereas that of gauge bundles of the 
$SO(16)/Sp(16)$ theory on $T^2$, which lacks vector structure, is $2\times 8$
dimensional. In fact, these two components of the moduli space of gauge bundles
on $T^2$ are disconnected. Under Type I-heterotic duality the component 
with vector structure maps to the corresponding part of the Narain moduli
space, while, as was originally pointed out in \cite{Ka}, the component without
vector structure maps to the corresponding part of the moduli space of CHL
strings \cite{CHL} (with rank 8 gauge symmetry) in 8 dimensions.

\subsection{D-branes and O-planes Transverse to Tori with $B$-flux}   

{}In this subsection we will discuss toroidal orientifolds with D-branes and
O-planes transverse to tori with non-zero $B$-field. From \cite{Wi} we expect
that in these cases we will have both O$^-$-planes and O$^+$-planes. We will 
see that this is indeed the case using the ${\bf Z}_2\otimes {\bf Z}_2$ freely
acting orbifold of \cite{Ka,Ka4} described in the previous subsection. However,
before we discuss orientifolds with non-zero $B$-flux, we will take a detour
into a discussion of possible types of O$^\pm$-planes as there are {\rm two
different} types of O$^+$-planes as well as O$^-$-planes\footnote{I would like
to thank Amihay Hanany for a valuable discussion on this point.}. 

{}To begin with, let us start with Type IIB in ten dimensions. {\em A priori}
we can orientifold Type IIB by two different orientifold actions, which we will
refer to as $\Omega_\pm$. One of these actions, namely, $\Omega_-$ (which we
have been denoting by $\Omega$ in the previous discussions) is the usual
orientifold projection which leads to the Type I theory in ten dimensions
with 16 supercharges and $SO(32)$ gauge group. Note that the action of 
$\Omega_-$ on the Chan-Paton charges of D9-branes is {\em antisymmetric} 
in both
Neveu-Schwarz and Ramond open string sectors. The corresponding orientifold 
plane has R-R charge $-32$, which requires introduction of 32 D9-branes to
cancel the R-R charge as each D9-brane has R-R charge $+1$. The R-R charge
cancellation implies that the corresponding R-R tadpoles are also canceled.
The fact that the NS-NS tadpoles also cancel then follows from supersymmetry.
However, in cases without supersymmetry NS-NS and R-R tadpole 
cancellations are {\em a priori} independent. To keep track of both
NS-NS and R-R tadpoles it is convenient to introduce the notion of the
``NS-NS charge'' for D-branes and O-planes as a {\em book-keeping} device 
(albeit there is no real conserved charge associated with NS-NS tadpoles).
Thus, let us assign NS-NS charge $+1$ to a single D9-brane. Then the NS-NS
charge of the orientifold plane corresponding to the action of $\Omega_-$
is $-32$. In this language the orientifold plane corresponding to the action
of $\Omega_+$ has R-R charge $+32$ and NS-NS charge $+32$. The action of
$\Omega_+$ on the Chan-Paton charges of D9-branes is now {\em symmetric} in
both NS and R open string sectors. Note that the gauge group on $M$ D9-branes
with the O-plane corresponding to the action of $\Omega_+$ is $Sp(M)$. In fact,
the corresponding gauge theory is supersymmetric. However, in ten dimensions
such a theory would be anomalous as we cannot cancel the R-R charge: both
the O-plane and D9-branes have positive R-R charges. Here we can ask if we
could cancel the R-R charges by introducing anti-D9-branes, which we will 
denote via D${\overline 9}$. Each D${\overline 9}$-brane has R-R charge
$-1$ and NS-NS charge $+1$. So if we introduce 32 D${\overline 9}$-branes
together with the aforementioned O-plane, we can cancel the R-R tadpoles, so
that the resulting theory would be anomaly free. However, this theory would be
non-supersymmetric, and NS-NS tadpoles would not be canceled. 

{}Let us systematize the above discussion. Let us denote the NS-NS and R-R
charges of a given object by $Q_{\rm NS}$ and $Q_{\rm R}$, respectively. Then
D9-branes and D${\overline 9}$-branes have the following 
$(Q_{\rm NS},Q_{\rm R})$
charges:
\begin{eqnarray}
 &&{\rm D9-brane}:~~~(+1,+1)~,\\
 &&{\rm D{\overline 9}-brane}:~~~(+1,-1)~.
\end{eqnarray}
We can introduce {\em four} different types of O9-planes according to their
NS-NS and R-R charges:
\begin{eqnarray}
 &&{\rm O}9^{--}:~~~(-32,-32)~~~~~~\Omega_-~,\\
 &&{\rm O}9^{++}:~~~(+32,+32)~~~~~~\Omega_+~,\\
 &&{\rm O}9^{+-}:~~~(+32,-32)~~~~~~\Omega_+(-1)^{F_L+F_R}~,\\
 &&{\rm O}9^{-+}:~~~(-32,+32)~~~~~~\Omega_-(-1)^{F_L+F_R}~,
\end{eqnarray}
where at the end of each line we have indicated the type of the
Type IIB orientifold which produces each of these objects ($F_L$ and $F_R$ are 
the usual left- and right-moving space-time fermion numbers). We will discuss
the above four orientifolds more explicitly in a moment. However, before we 
do this, let us discuss the action of the orientifold projections on D9-
and D${\overline 9}$-branes induced by the above orientifold planes. In the
following ``S'' refers to symmetrization on Chan-Paton charges, while ``A''
refers to antisymmetrization. The first entry corresponds to the NS (that is,
space-time bosonic) open string sector on the corresponding 
D9/D${\overline 9}$-branes, while the second entry corresponds to the R (that
is, space-time fermionic) sector. The orientifold projections of the above
O9-planes on the D9-branes are given by:
\begin{eqnarray}
 &&{\rm O}9^{--}:~~~(A,A)~,~~~~~~(-32+M,-32+M)~,\\
 &&{\rm O}9^{++}:~~~(S,S)~,~~~~~~(32+M,32+M)~,\\
 &&{\rm O}9^{+-}:~~~(S,A)~,~~~~~~(32+M,-32+M)~,\\
 &&{\rm O}9^{-+}:~~~(A,S)~,~~~~~~(-32+M,32+M)~,
\end{eqnarray}  
where at the end of each line we have indicated the total NS-NS and R-R charges
of the background with the corresponding O9-plane and $M$ D9-branes. Let us 
discuss these four cases in more detail.\\
$\bullet$ O9$^{--}$-plane plus $M$ D9-branes. The theory is supersymmetric, the
gauge group is $SO(M)$, and the fermions on D9-branes are in the antisymmetric
representation. The R-R tadpoles cancel for $M=32$, that is, we have an anomaly
free theory in this case. Note that NS-NS tadpoles also cancel. This is the
familiar Type I theory with $SO(32)$ gauge symmetry.\\
$\bullet$ O9$^{++}$-plane plus $M$ D9-branes. The theory is supersymmetric,
the gauge group is $Sp(M)$, and the fermions on D9-branes are in the symmetric
representation. The R-R (and NS-NS) tadpoles, however, cannot be canceled,
so that the theory is anomalous.\\
$\bullet$ O9$^{+-}$-plane plus $M$ D9-branes. The closed string sector is
supersymmetric at the tree level. The open string sector is non-supersymmetric.
The gauge group is $Sp(M)$, and the fermions on the D9-branes are in the
antisymmetric representation. The R-R tadpoles cancel for $M=32$, that is, we
have an anomaly free theory in this case. This is the non-supersymmetric 
$Sp(32)$ string theory recently proposed in \cite{Su}. Note, however, that
NS-NS tadpoles do {\em not} cancel in this theory, so that the flat Minkowski
metric with constant dilaton (more precisely, the corresponding
${\cal N}=1$ supergravity) does not appear to be the correct background for
this theory. At present it is unclear whether there is a consistent background
for this theory which could be reached via the Fischler-Susskind mechanism.\\
$\bullet$ O9$^{-+}$-plane plus $M$ D9-branes. The closed string spectrum
is supersymmetric at the tree level. The open string sector is 
non-supersymmetric. The gauge group is $SO(M)$, and the fermions on D9-branes
are in the symmetric representation. The NS-NS tadpoles cancel for $M=32$, but
the R-R tadpoles cannot be canceled, so that the theory is anomalous.

{}Next, let us discuss the orientifold projections induced by the above 
O9-planes on D${\overline 9}$-branes:
\begin{eqnarray}
 &&{\rm O}9^{--}:~~~(A,S)~,~~~~~~(-32+M,-32-M)~,\\
 &&{\rm O}9^{++}:~~~(S,A)~,~~~~~~(32+M,32-M)~,\\
 &&{\rm O}9^{+-}:~~~(S,S)~,~~~~~~(32+M,-32-M)~,\\
 &&{\rm O}9^{-+}:~~~(A,A)~,~~~~~~(-32+M,32-M)~,
\end{eqnarray}  
where, as before, at the end of each line 
we have indicated the corresponding total NS-NS and R-R
tadpoles for the system of the corresponding O9-plane and $M$ 
D${\overline 9}$-branes. It is not difficult to see that the system of the
O9$^{\alpha,\beta}$-plane 
plus $M$ D${\overline 9}$-branes gives the same theory as the system of the 
O9$^{\alpha,-\beta}$-plane plus $M$ D9-branes, where $\alpha,\beta=\pm$.

{}Here we would like to make one remark. In the above language we can
describe the Type O open plus closed string theory, which is the $\Omega_-$
orientifold of Type 0B string theory, as follows. Type 0B can be viewed as
the $(-1)^{F_L+F_R}$ orbifold of Type IIB. This implies that Type O can be
viewed as a Type IIB orientifold, where the orientifold group is ${\cal O}=\{1,
(-1)^{F_L+F_R},\Omega_-,\Omega_-(-1)^{F_L+F_R}\}$. Thus, we have two 
orientifold planes in this theory, namely, the 
O9$^{--}$-plane and the O9$^{-+}$-plane,
which together have the following NS-NS and R-R charges: $(Q_{\rm NS},Q_R)=
(-64,0)$. These can be canceled by introducing 32 D9-branes together with 32
D${\overline 9}$-branes\footnote{In Type 0B, unlike Type IIB, there are
two different types of D-branes and anti-D-branes, so other solutions to the
tadpole cancellation conditions also exist \cite{BeGa}.}. 
Note that in this language, for instance,
it is evident why
there are no fermionic states in the open string spectrum: the O9$^{--}$ 
orientifold projection on D9-brane fermions is antisymmetric, while the
O9$^{-+}$ orientifold projection on the same fermions is symmetric, so that
fermions are completely projected out. The same conclusion holds for the 
D${\overline 9}$-brane fermions. 
 
{}The reason why we discussed four different kinds of O-planes is that in 
our discussion of orientifolds with $B$-flux we will encounter O-planes
with the 
$Sp$ type of orientifold projection. However, as we have already pointed 
out, one must distinguish two possible O-planes of this type as well as two
O-planes with the $SO$ type of orientifold projection. In the following we will
only encounter O-planes of O$^{--}$ and O$^{++}$ type\footnote{In the following
we will discuss O$p$-planes with $p<9$ for which the above discussion of the
possible types of O-planes can be straightforwardly generalized. In particular,
we can start from the aforementioned four types of O9-planes and use the
standard T-duality arguments to arrive at the corresponding lower dimensional
O$p$-planes.}. 
We will refer to them
as O$^-$ and O$^+$, respectively. Also, just as in the previous sections,
we will refer to $\Omega_-$ as $\Omega$.
Note that the NS-NS and R-R charges for 
each of these O-planes are equal. The same holds for D-branes (we will not
need to introduce anti-D-branes in the following), so in the following 
it will suffice to just consider R-R charges. This is simply a manifestation 
of the fact that all the theories we will consider in the following are 
supersymmetric.

{}Next, we would like to return to orientifolds with $B$-flux, and consider
cases where O-planes and the corresponding D-branes are transverse to tori
with non-zero $B$-flux. To begin with, let us start with the simplest case.
Thus, consider the $\Omega R (-1)^{F_L}$ orientifold of
Type IIB on ${\bf R}^{1,7}\otimes T^2$, where $R$ acts as $RX_{1,2}=-X_{1,2}$
on the compact coordinates, and the $B$-flux on $T^2$ is zero. To consider
the case with non-zero $B$-flux, let us use the ${\bf Z}_2\otimes {\bf Z}_2$ 
freely acting orbifold construction discussed in the previous 
subsection. In fact, we will discuss both cases with and without discrete 
torsion, whose comparison will be helpful in understanding the types of 
orientifold planes arising in the former case.

{}Thus, let us consider the ${\bf Z}_2\otimes {\bf Z}_2$ orbifold of $T^2$
whose generators $S_1$ and $S_2$ act as shifts $S_i X_i =X_i+e_i/2$. Note that
there are four fixed points of $R$ on $T^2$ located at $(X_1,X_2)=(0,0),
(e_1/2,0),(0,e_2/2),(e_1/2,e_2/2)$. However, these four fixed points are
identified by the combined action of the $S_1$ and $S_2$ shifts, so that
we expect one orientifold 7-plane at $(X_1,X_2)=(0,0)$. On the other hand, we 
have additional fixed points due to the elements $RS_a$, $a=1,2,3$ (recall that
$S_3\equiv S_1S_2$). In particular, for a given 
element $RS_a$ we have four fixed points, which are identified by the
shifts $S_1$ and $S_2$. So we have one independent fixed point corresponding to
each element $RS_a$: $(X_1,X_2)=(e_1/4,0)$ for $RS_1$, $(X_1,X_2)=(0,e_2/4)$
for $RS_2$, and $(X_1,X_2)=(e_1/4,e_2/4)$ for $RS_3$. At each of these three
fixed points we expect one orientifold 7-plane.   

{}Next, we would like to understand what types of O7-planes we have in this
background. In the case without discrete torsion we have an O7$^-$-plane at 
each of the aforementioned four fixed points. In the case with discrete 
torsion, however, one of the O7-planes is of the O7$^+$ type, while the other
three are of the O7$^-$ type. To show this, let us consider the Klein bottle
in each case. Thus, we have four sectors of the ${\bf Z}_2\otimes {\bf Z}_2$
orientifold labeled by $(\alpha^1,\alpha^2)=(0,0),(1,0),(0,1),(1,1)$, and
the left- and right-moving closed string momenta are given in these sectors
by (\ref{LR}), where the windings $n^i$ are arbitrary integers, whereas the
momenta $m_i$ are integers
subject to the orbifold projections which can be read off
(\ref{projection}). Let us denote the
Klein bottle contribution coming from a given sector
$(\alpha^1,\alpha^2)$ via ${\cal K}(\alpha^1,\alpha^2)$. Let us denote the
contribution coming from the
tree-level closed string exchange between the O7-plane located at the fixed 
point labeled by  
$(\beta^1,\beta^2)$ and the O7-plane located at the fixed point labeled by
$(\gamma^1,\gamma^2)$ via ${\widetilde {\cal K}}
(\beta^1,\beta^2;\gamma^1,\gamma^2)$.
Here the fixed point labeled by $(\beta^1,\beta^2)$ is given by 
$(X_1,X_2)=(\beta^1
e_1/4,\beta^2e_2/4)$, $\beta^i=0,1$. Then we have the following relation:
\begin{equation}
 {\cal K}(\alpha^1,\alpha^2)=\sum_{\beta^1,\beta^2=0,1}
 {\widetilde {\cal K}}(\beta^1,\beta^2;\beta^1+\alpha^1~({\rm mod}~2),
 \beta^2+\alpha^2~({\rm mod}~2))~.  
\end{equation}
Note that for $\epsilon=1$ in the untwisted as well as twisted sectors the 
momenta $m_i$ are even integers. This implies that both the untwisted as well
as twisted sector contributions ${\cal K}(\alpha^1,\alpha^2)$ are 
non-vanishing - the $\Omega R (-1)^{F_L}$ orientifold projection keeps states
with zero momenta and arbitrary windings. This implies that all four 
orientifold planes must have the same R-R charge. This can be seen explicitly
by extracting the massless R-R tadpoles in each of these sectors, which are
actually identical. (To see this one can perform the Poisson resummation of
the corresponding winding sums.) However, for $\epsilon=-1$ the situation
is quite different. Thus, in the untwisted sector $(\alpha^1,\alpha^2)=(0,0)$
the momenta $m_i$ are even, so that the corresponding Klein bottle 
contribution ${\cal K}(0,0)$ is non-vanishing. On the other hand, in the
twisted sectors $(\alpha^1,\alpha^2)=(1,0),(0,1),(1,1)$ we have the 
momenta $(m_1\in 2{\bf Z},m_2\in 2{\bf Z}+1)$, 
$(m_1\in 2{\bf Z}+1,m_2\in 2{\bf Z})$ and $(m_1\in 2{\bf Z}+1,m_2\in 
2{\bf Z}+1)$, respectively. This implies that the corresponding Klein bottle
contributions ${\cal K}(\alpha^1,\alpha^2)$ vanish, and so do the
corresponding tadpoles. This is enough to deduce the R-R charges of the
orientifold planes located at the four fixed points. Let 
$Q_R(\beta^1,\beta^2)$ be the R-R charge of the O7-plane located at the
fixed point labeled by $(\beta^1,\beta^2)$. Then we have the following
conditions:
\begin{eqnarray}
 &&\sum_{\beta^1,\beta^2=0,1} Q^2_R(\beta^1,\beta^2)=4\times 8^2~,\\
 &&\sum_{\beta^1,\beta^2=0,1} Q_R(\beta^1,\beta^2)
 Q_R(\beta^1+\alpha^1~({\rm mod}~2),\beta^2+\alpha^2~({\rm mod}~2))=0~,
 ~~~(\alpha^1,\alpha^2)\not=(0,0)~.
\end{eqnarray} 
The first line follows from the fact that the untwisted sector contribution
${\cal K}(0,0)$ is the same as in the case without discrete torsion where
all four orientifold 7-planes are of the same type, and carry the R-R charge
$-8$ or $+8$ depending on the choice of the orientifold projection. It is 
not difficult to see that the solutions to the above conditions correspond to
having one O7-plane with R-R charge $\pm 8$ and the other three O7-planes
with R-R charge $\mp 8$ (where the signs are correlated). The solution that
we are interested in is the one where we have one O7$^+$-plane and three
O7$^-$-planes as in this case we can cancel the total R-R charge
of the O7-planes, which is $-16$, by introducing 16 D7-branes\footnote{For the
other solution where we have one O7$^-$-plane and three O7$^+$-plane
the total R-R charge is $+16$, so it cannot be canceled by introducing 
only D7-branes.}. This is precisely the result we wished to show.
 
{}Note that unlike the case of, say, D9-branes wrapped on a two-torus with
$B$-field, the ``rank reduction'' here is {\em not} due to non-commuting Wilson
lines but rather the fact that we have different types of O7-planes whose
total R-R charge adds up to $-16$ (rather than $-32$). In contrast, in the
case of D9-branes wrapped on a two-torus with $B$-flux the O9-plane is
of the O9$^-$ type, and the R-R charge cancellation requires 32 D9-branes.
The rank reduction in this case is due to the non-commuting Wilson lines.
The difference between these two cases is quite substantial. Thus, in the
case of D9-branes the gauge symmetry is $SO(16)\otimes {\bf Z}_2$ or
$Sp(16)\otimes {\bf Z}_2$. In the case of D7-branes the gauge symmetry is
$SO(16)$ if all D7-branes are placed at one of the O7$^-$-planes, and it is
$Sp(16)$ if all D7-planes are placed at the O7$^+$-plane\footnote{In these
configurations tadpoles are not canceled locally, so one expects a varying
dilaton background. Note that this is not the case in analogous 
configurations involving O$p$-planes and D$p$-branes with $p<7$ as they are
non-dilatonic.}. 
Thus, the gauge symmetry in the case of D7-branes does 
not contain a ${\bf Z}_2$ subgroup present in the case of D9-branes.

{}Before we end this subsection, let us mention the generalization to
higher tori. For instance, consider the case of $T^4$. For illustrative 
purposes let us actually concentrate on $T^4=T^2\otimes T^2$. We can obtain
the background with rank $b=4$ $B$-flux via separately orbifolding the two 
$T^2$'s by the respective ${\bf Z}_2\otimes {\bf Z}_2$ actions with discrete
torsion. Then, if we consider the $\Omega R$ orientifold of Type IIB in this
background, where $R$ inverts all four coordinates on $T^4$,  
we have 10 O5$^-$-planes and 6 O5$^+$-planes. Similarly, in the case of
the $\Omega R(-1)^{F_L}$ orientifold of Type IIB on $T^6$ with $b=6$, where
$R$ inverts all six coordinates on $T^6$, we have 36 O3$^-$-planes and
28 O3$^+$-planes. More generally, consider O$p$-planes transverse to
${\bf R}^{9-p-d}\otimes T^d$ with $B$-flux of rank $b$ on $T^d$ ($b\leq d$).
We have total of $n_f=2^d$ fixed points. Let the numbers of O$p^{\pm}$-planes
be $n_{f\pm}$. Then $n_f=n_{f+} +n_{f-}$, and $n_{f-}-n_{f+}=2^{d-b/2}$. Thus,
we have 
\begin{equation}
 n_{f\pm}={1\over 2}\left(2^d\mp 2^{d-b/2}\right)~.
\end{equation}
It is also straightforward to consider the cases where 
D-branes and O-planes are wrapped on tori with non-zero $B$-field, and, at the
same time, the tori transverse to these objects also have non-zero $B$-field
by combining the results of this and the previous subsections.

\subsection{Mixed Cases}

{}In this subsection we would like to discuss the cases where we have a
D$p$-D$p^\prime$ system (with $p-p^\prime=0~({\rm mod}~4)$ so that we preserve
some supersymmetries), where one set of branes is wrapped on a torus with
non-zero $B$-field, while the other set of branes is transverse to this torus.
The most general case is straightforward to treat, however, for 
illustrative purposes let us consider the case of the D5-D9 system, where
D9-branes are wrapped on $T^4$, while D5-branes are transverse to $T^4$.

{}Thus, let us assume that we have $n_5$ D5-branes and $n_9$ D9-branes. In 
fact, in this subsection we will not worry about tadpole/anomaly cancellation
as the point we would like to make here is purely geometric, and the former
issues are irrelevant for this discussion. So we will not introduce any 
orientifold planes. Moreover, here we will focus on the 59 sector states, so
that the conclusions we draw in this subsection are unchanged even after
introduction of orientifold planes (recall, for instance, that the 59 sector
states do not contribute to the M{\"o}bius strip amplitude).

{}To begin with, let us consider the above system with zero $B$-flux on $T^4$.
Then the gauge group is $U(n_5)\otimes U(n_9)$, and the 59 open string sector
hypermultiplets transform in the bifundamental representation $(n_5,n_9)$
(we will suppress the $U(1)$ charges which are straightforward to restore). Now
let us turn on a half-integer $B$-flux of rank $b=2$. For the sake of 
simplicity let us consider the case of $T^4=T^2\otimes T^2$ with non-zero
$B$-flux in the first $T^2$. Then, as in the previous subsections, we can 
view D9-branes wrapped on $T^4$ with $B$-flux in terms of turning on 
non-commuting Wilson lines corresponding to the two cycles of the first $T^2$.
The 55 gauge group is unchanged as the Wilson lines are turned on in the 
directions transverse to D5-branes. The 99 gauge group, however, is broken down
to $U(N)\otimes {\bf Z}_2$, 
where $N\equiv n_9/2$. (Recall that the first Wilson line brakes
$U(n_9)$ to $U(N)\otimes U(N)$, while the second Wilson line breaks the latter
to its diagonal subgroup $U(N)\otimes {\bf Z}_2$.) Even though the 99 gauge 
group is broken by the Wilson lines, the 59 sector states are not affected
(we will explain this in a moment). In
particular, the number of the 59 hypermultiplets is still $n_5 n_9$. More 
concretely, we have two hypermultiplets in $(n_5,N)$ of $U(n_5)\otimes U(N)$. 
Actually, the gauge group is $U(n_5)\otimes U(N)\otimes {\bf Z}_2$, and the
59 hypermultiplets are given by $(n_5,N)_{+1}$ and $(n_5,N)_{-1}$, where 
the subscript indicates the corresponding ${\bf Z}_2$ charge. This is precisely
the phenomenon first observed in \cite{KST} - the 59 open string states
come with a non-trivial multiplicity, which depends on the rank of the 
$B$-field. This multiplicity is given by 
\begin{equation}
 \xi_{59}=2^{b/2}~.
\end{equation}
Here we can understand this multiplicity completely geometrically - it 
corresponds to the ${\bf Z}_2^{\otimes (b/2)}$ 
discrete gauge symmetry, that is, the $2^{b/2}$ states, which have otherwise
identical quantum numbers, carry all different quantum numbers under the
${\bf Z}_2^{\otimes (b/2)}$ discrete gauge symmetry. This answers the question
of the vertex operators in the 59 sector. For instance, in the $b=2$ case
we can write down the vertex operators for the 59 states as follows. The
first Wilson line breaks $U(n_9)$ to $U(N)\otimes U(N)$. The original 
hypermultiplet in the bifundamental $(n_5,n_9)$ of $U(n_5)\otimes U(n_9)$
thus gives rise to the following hypermultiplets charged under $U(n_5)\otimes
U(N)\otimes U(N)$: $(n_5,N,1)$ and $(n_5,1,N)$. Next, the second Wilson line 
breaks $U(N)\otimes U(N)$ down to its diagonal subgroup $U(N)\otimes 
{\bf Z}_2$. The aforementioned hypermultiplets can now be combined into two
linear combinations
\begin{equation}
 |(n_5,N)_{\pm 1}\rangle = {1\over \sqrt{2}}\left(|(n_5,N,1)\rangle \pm 
 |(n_5,1,N)\rangle\right)~,
\end{equation} 
which carry definite $U(n_5)\otimes U(N)\otimes {\bf Z}_2$ gauge quantum 
numbers\footnote{Actually, the appearance of the ${\bf Z}_2^{\otimes(b/2)}$
discrete gauge symmetry in orientifolds with $B$-flux was originally pointed 
out in \cite{Ka2}.}.

{}Now let us explain why the 59 sector states are not affected by turning on
Wilson lines even though the 99 sector states are. The point is that the 59
sector states have no momentum (or winding) excitations - the 99 strings
have only momenta on $T^4$, while the 55 strings have only windings on $T^4$
(this is precisely why 55 sector states are unaffected by the Wilson lines
which act only on the momenta). Thus, the Wilson lines do {\em not} act on the
59 sector states.

{}Before we end this section we would like to discuss one other point. 
As we have seen 
from the previous discussions, the 99 and 55 sectors feel the presence of the
$B$-flux in qualitatively different ways. Thus, the orientifold 9-planes
are unaffected by the presence of the $B$-field, while the D9-brane gauge 
symmetry suffers rank reduction. On the other hand, the structure of
orientifold 5-planes is modified in the presence of the $B$-flux,
while D5-branes are unaffected. More generally, the above conclusions hold for
O-planes and D-branes wrapped on tori with $B$-flux {\em vs.} O-planes and
D-branes transverse to such tori. At first this might seem puzzling in the 
light of T-duality, which one might expect to map the aforementioned two setups
into each other. However, as we will see in a moment, such an expectation would
be erroneous, and there is no puzzle here.

{}To understand this, let us consider the simplest case of D9-branes wrapped 
on $T^2$ with the $B$-flux $B_{12}=1/2$. Let the metric on $T^2$ be $g_{ij}$.
In the following it will be convenient to work with $G_{ij}\equiv 2g_{ij}$,
and introduce the following matrix (here we are closely following the 
discussion in \cite{GR}):
\begin{equation}\label{E}
 E_{ij}\equiv G_{ij}+B_{ij}= \left(\begin{array}{cc}
               G_{11} & G_{12}+B_{12}\\
               G_{12}-B_{12} & G_{22}
              \end{array}\right)~.
\end{equation}
The T-duality group in the case of $T^2$ is $SO(2,2,{\bf Z})$ whose
elements can be described in terms of $4\times 4$ matrices
\begin{equation}\label{T-duality}
 \left(\begin{array}{cc}
               \alpha & \beta\\
               \gamma &  \delta
              \end{array}\right)~,
\end{equation} 
where $\alpha,\beta,\gamma,\delta$ are $2\times 2$ matrices with integer
entries, and satisfy the following constraints:
\begin{eqnarray}
 &&\gamma^T\alpha+\alpha^T\gamma=0~,\nonumber\\
 &&\delta^T\beta+\beta^T\delta=0~,\nonumber\\
 &&\gamma^T\beta+\alpha^T\delta=I~.\nonumber
\end{eqnarray}
Here the superscript $T$ stands for transposition, and $I$ denotes 
the $2\times 2$ identity matrix. The above T-duality element acts on 
$E_{ij}$ as follows:
\begin{equation}
 E\rightarrow E^\prime=(\alpha E+\beta)(\gamma E+\delta)^{-1}~.
\end{equation}
In these notations the familiar $S$- and $T$-transformations are described
as follows. The $S$-transformation 
corresponds to taking $\alpha=0$, $\beta=I$, $\gamma=I$ and $\delta=0$, and
amounts to mapping a 2-torus with metric $G$ and zero $B$-field to another
2-torus (with zero $B$-field) whose metric is given by the inverse of $G$. 
This is just the usual ``$R\rightarrow 1/R$'' type of T-duality
transformation. On the other hand, the $T$-transformation corresponds to 
taking $\alpha=I$, $\beta=\Sigma$, $\gamma=0$ and $\delta=I$, and
amounts to unit shifts of the $B$-field (but does not affect the metric on
$T^2$), where $\Sigma$ is the $2\times 2$ antisymmetric matrix with 
$\Sigma_{12}=1$.

{}Here we can ask what happens to D9-branes wrapped on $T^2$ with $B_{12}=1/2$
under the $S$-transformation. Note that for the $S$-transformation
the new matrix $E^\prime$ is simply
the inverse of $E$, so that we have 
\begin{equation}
 E^\prime=G^\prime +B^\prime ={1\over \det(G)+B_{12}^2}
 \left(\begin{array}{cc}
               G_{22} & -G_{12}-B_{12}\\
               -G_{12}+B_{12} & G_{11}
              \end{array}\right)~.
\end{equation}
Thus, the new $B$-field has the non-zero component given by
\begin{equation}
 B^\prime_{12}= -{B_{12}\over {\det(G)+B_{12}^2}}~.
\end{equation}
Since $B_{12}=1/2$, and we must require that $B_{12}^\prime$ must also be 
half-integer (or else the orientifold in the T-dual picture would not be well
defined), it follows that the $S$-transformation can only be performed 
for\footnote{Actually, there are additional solutions to this constraint,
but all of them are equivalent to the one we discuss here by T-duality
transformations. Thus, for instance,
we can start from the point $\det(G)=1/12$, 
$B_{12}=1/2$, which is mapped by the $S$-transformation 
to the point $\det(G^\prime)=3/4$, $B_{12}^\prime=-3/2$. However, the point 
$\det(G)=1/12$, $B_{12}=1/2$ is equivalent to the point $\det(G)=3/4$,
$B_{12}=1/2$ via the T-duality transformation $STS$. I would like to thank
Ofer Aharony for bringing this point to my attention.}
$\det(G)=3/4$. The point in the moduli space of K{\"a}hler structures
of $T^2$ where $B_{12}=1/2$ and $\det(G)=3/4$ is one of the self-dual points.
That is, the momentum and winding states at this point are indistinguishable,
and, therefore, D9-branes wrapped on such a 2-torus are indistinguishable from
D7-branes transverse to such a torus\footnote{As we will explain in a moment,
however, because of half-integer $B$-field the precise map between these 
objects is {\em different} from that at the self-dual point without the 
$B$-field.}. 
Thus, the $S$-transformation, which
maps D9-branes to D7-branes, can only be performed at the self-dual point where
the two types of branes are identical. This avoids the aforementioned puzzle 
with T-duality. Indeed, at a generic point in the K{\"a}hler structure moduli 
space D9-branes wrapped on $T^2$ with $B$-flux are {\em not} T-dual to
D7-branes transverse to (another) $T^2$ with $B$-flux. As to the self-dual
point in the K{\"a}hler structure moduli space, there are two {\em a priori}
consistent setups, one of which can be continuously deformed into D9-branes
wrapped on a generic $T^2$ with half-integer 
$B$-flux, and the other one can be continuously
deformed into D7-branes transverse to a generic 
$T^2$ with half-integer $B$-flux. The fact that these
two setups are indeed different is evident - at generic points in the 
respective moduli spaces the number of D9-branes
is {\em twice} the number of D7-branes for a given rank of the gauge group.
  
{}Even though the difference between the aforementioned two setups is evident,
we would like to understand the origin of these two choices\footnote{Parts of 
our discussion here have appeared in a footnote in \cite{NC}.}. 
To do this, let us consider D9-branes wrapping $T^2$ with $B_{12}=1/2$. 
Following \cite{BPS} we can assume that the closed strings that couple to
D-branes satisfy the ``no momentum flow'' condition in the directions of
$T^2$, that is, these states have the left- and right-moving momenta 
(note that $p^{L,R}_i\equiv 2e_i\cdot P_{L,R}$)
\begin{eqnarray}
 &&p^L_i=m_i+E_{ji}n^j~,\\
 &&p^R_i=m_i-E_{ij} n^j~,
\end{eqnarray}
which satisfy
\begin{equation}\label{NMF}
 p^L_i=-p^R_i~.
\end{equation}
This constraint implies that the closed strings coupled to D9-branes
have the momenta and windings such that
\begin{equation}
 m_i-B_{ij}n^j=0~,
\end{equation}
from which it follows that the windings $n^i$ must be 
{\em even}\footnote{This is in accord with (\ref{erroneous}). More precisely,
(\ref{erroneous}) gives the loop-channel annulus amplitude. Upon the modular
transformation $t\rightarrow 1/t$, where $t$ is the proper time on the 
cylinder, which involves the appropriate Poisson resummation of the momentum
sum in (\ref{erroneous}), we arrive at the tree-channel annulus amplitude
corresponding to the closed string exchanges between D-branes. The latter 
amplitude is in agreement with the aforementioned conclusion that the closed
string states that couple to D-branes have even windings. Note that, as we 
explained in detail in subsection A, strictly speaking the interpretation
corresponding to (\ref{erroneous}), which arises in the approach of
\cite{BPS}, is somewhat imprecise. However, it suffices for our purposes here.
The above analysis can be repeated in the language of the freely acting 
${\bf Z}_2\otimes
{\bf Z}_2$ orbifold discussed in the previous subsections, which
gives the precise description of D-branes wrapped on $T^2$ with half-integer 
$B$-field. Here we have chosen the approach of \cite{BPS} for illustrative 
purposes, as it suffices to explain the point we are trying to make here.}. 
This, in
particular, leads to the rank reduction for the 99 Chan-Paton gauge group
\cite{BPS},
which we have understood in terms of non-commuting Wilson lines in the
previous subsections. Note that in this case we also have a consistent
coupling between the D9-branes and the O9-plane. Indeed, the loop-channel Klein
bottle amplitude receives contributions from the closed string states with
the left- and right-moving momenta satisfying $p^L_i=p^R_i$. Thus, in the 
loop channel we have arbitrary momenta $m_i$ and zero windings $n^i$. After
the modular transformation $t\rightarrow 1/t$, which maps the loop-channel 
Klein
bottle amplitude to the tree-channel Klein bottle amplitude, we obtain a sum
over arbitrary integer windings with the metric $g_{ij}$. That is, the closed
string states that couple to the O9-plane have the left- and right-moving 
momenta $p^L_i=-p^R_i=G_{ij}n^j$, so that $P_L=-P_R=e_in^i$,
and $P_L^2=P_R^2=g_{ij}n^i n^j={1\over 2}G_{ij}n^i n^j$. (Note that these
states are the same as in the case without the $B$-field.) Thus, the closed
string states that couple to D9-branes wrapped on $T^2$ with half-integer 
$B$-field are a subset of the closed string states that couple to the O9-plane,
so that D9-branes consistently couple to the O9-planes, in particular,
the M{\"o}bius strip amplitude is consistent\footnote{Once again, as
we explained in subsection A, a more precise description of this coupling is
given in terms of the freely acting ${\bf Z}_2\otimes {\bf Z}_2$ orbifold, but
the above description is adequate for our purposes here.}.  

{}However, {\em a priori} there is another choice we can make for the
constraint on the left- and right-moving momenta of the closed string states
that couple to D-branes. This second choice has been recently discussed in
\cite{HKMS}, and is given by\footnote{There is a misprint in \cite{HKMS}, 
which amounts to a missing minus sign in (\ref{NMF1}).}
\begin{equation}\label{NMF1}
 p^L_i=-{{\cal R}_i}^j p^R_j~,
\end{equation} 
where ${\cal R} \equiv E^T E^{-1}$. Note that (\ref{NMF1}) reduces to
(\ref{NMF}) for $B_{ij}=0$, but is different otherwise. It is not difficult to
show that the solution of (\ref{NMF1}) is as follows. The closed string states
coupled to D-branes have zero momenta $m_i$ and {\em arbitrary} windings $n^i$
(and in this case we do {\em not} expect rank reduction). This
implies that these states have the following left- and right-moving momenta: 
\begin{equation}
 p^L_i= E_{ji}n^j~,~~~p^R_i=-E_{ij}n^j~.
\end{equation}
Note that these states have the expected property 
\begin{equation}\label{9P}
 P_L^2=P_R^2={1\over 2} {\cal G}_{ij}n^in^j~,
\end{equation}
where
\begin{equation}
 {\cal G}\equiv G - BG^{-1} B~.
\end{equation}
Note, however, that at generic points (for half-integer $B$-flux) these
states are quite different from the states that couple to the O9-plane for 
which from the above discussion we have $P_L^2=P_R^2={1\over 2} G_{ij}n^in^j$.
In fact, these two sets of closed string states coincide only for zero 
$B$-flux. Thus, D9-branes defined via (\ref{NMF1}) cannot be consistently 
coupled to the O9-plane for half-integer $B$-field.

{}There is, however, a setup where we can consistently couple such D9-branes
to orientifold planes, except that these are not O9- but O7-planes. Thus,
consider D7-branes transverse to $T^2$ with half-integer $B$-flux. Recall that
in the case of D9-branes wrapped on such a $T^2$ we have imposed the ``no
momentum flow'' condition (\ref{NMF}). In the case of D7-branes transverse to
such a $T^2$ the analogous condition is that of ``no winding flow'' \cite{KST}:
\begin{equation}\label{NWF}
 p^L_i=p^R_i~.
\end{equation}  
This constraint implies that $n^i$ are zero and $m_i$ are arbitrary. Thus,
the left- and right-moving momenta in this case are given by $p^L_i=p^R_i=m_i$,
and
\begin{equation}\label{7P}
 P_L^2=P_R^2={1\over 2} G^{ij} m_im_j~,
\end{equation}
where $G^{ij}$ is the inverse of $G_{ij}$. Let us compare (\ref{9P}) and
(\ref{7P}). They are {\em identical} at the self-dual point in the moduli space
of K{\"a}hler structures where $\det(G)=3/4$ and $B_{12}=1/2$. Indeed,
note that at a generic point in this moduli space we have 
\begin{equation}
 {\cal G}_{ij}=\left(1+{B_{12}^2\over \det(G)}\right) G_{ij}~.
\end{equation} 
At the self-dual point we have 
\begin{equation}\label{calGSD}
 {\cal G}_{ij}={4\over 3}G_{ij}={4\over 3} \left(\begin{array}{cc}
               G_{11} & G_{12}\\
               G_{12} & G_{22}
              \end{array}\right). 
\end{equation}
On the other
hand, for the inverse metric $G^{ij}$ at the self-dual point we have
\begin{equation}\label{invGSD}
 G^{ij}={1\over \det(G)} \left(\begin{array}{cc}
               G_{22} & -G_{12}\\
               -G_{12} & G_{11}
              \end{array}\right)=
 {4\over 3} \left(\begin{array}{cc}
               G_{22} & -G_{12}\\
               -G_{12} & G_{11}
              \end{array}\right)~.
\end{equation}
Even though (\ref{calGSD}) and (\ref{invGSD}) generically are different, the
corresponding squared momenta $P^2_{L,R}$ in (\ref{9P}) and (\ref{7P}) are
the same. Moreover, the corresponding momentum states are the same once we 
make the appropriate identification
\begin{equation}
 n^i=\epsilon^{ij}m_j~,
\end{equation}
where $\epsilon^{ij}$ is a unit antisymmetric $2\times 2$ matrix (that is,
$\epsilon^{12}=-\epsilon^{21}=+1$ or $-1$). Thus, at the self-dual point 
(with half-integer $B$-field)
D7-branes satisfying the ``no winding flow'' condition have the same spectrum
as D9-branes satisfying the condition (\ref{NMF1}), and {\em not} the usual
``no momentum flow'' condition (\ref{NMF}). At first this might seem a bit
strange, as we can ask what is the analogous statement for D9-branes that  
satisfy the usual ``no momentum flow condition'' (\ref{NMF}). The answer to 
this question, actually, is very simple. It is not difficult to show that
D9-branes that satisfy the usual ``no momentum flow'' condition have the same
spectrum as D7-branes satisfying the ``no winding flow'' condition at the
self-dual point in the moduli space of K{\"a}hler structures corresponding
to $\det(G)=1$ and $B_{12}=0$. 

{}Now everything falls in place, and we arrive at
the following consistent picture. D-branes 
transverse to a torus are the same whether the $B$-field on the 
torus is zero or non-zero. However, D-branes wrapped on a torus are sensitive
to whether the $B$-field is zero or non-zero. In the former case they are
T-dual to the corresponding D-branes transverse to the torus
even at generic points in the moduli space of K{\"a}hler structures, and at the
self-dual point without the $B$-field the two types of D-branes are 
indistinguishable provided that they satisfy the usual ``no momentum flow'' 
and ``no winding flow'' conditions, respectively. In the case with non-zero
$B$-field there is no T-duality between the two types of D-branes (in
the presence of orientifold planes) at generic
points of the K{\"a}hler structure moduli space as the T-duality 
$S$-transformation would map a torus
with half-integer $B$-field into a torus with non-zero $B$-field which does not
take half-integer values. At the self-dual point with half-integer $B$-field
the $S$-transformation is consistent with the orientifold action, but precisely
at this point D-branes transverse to the torus and D-branes wrapped on the
torus are indistinguishable, except that the former still satisfy the usual 
``no winding flow'' condition, while the latter satisfy the modified constraint
(\ref{NMF1}). Here it is crucial that the latter type of
D-branes {\em cannot} be 
consistently coupled to the orientifold planes of the same spatial 
dimensionality (but, at the self-dual point, couple consistently to the
corresponding O-planes transverse to the torus as they are indistinguishable 
from the corresponding D-branes transverse to the torus).
Note that above we arrived at these conclusions in the case of
$T^2$, where we have two inequivalent self-dual points in the K{\"a}hler 
structure moduli space. The generalization to higher tori should be evident
once we observe that even though the number of inequivalent self-dual points
grows, the latter always have {\em half}-integer (or zero) $B$-field, 
so that different self-dual points correspond to different half-integer 
$B$-field configurations. The above discussion should make it evident that
the geometric picture of orientifolds with $B$-flux we described in this 
section is completely consistent with T-duality considerations. In particular,
the fact that D-branes transverse to a torus with $B$-flux behave quite
differently from D-branes wrapped on such a torus is no longer
mysterious.

{}Finally, let us ask to what extent T-duality can be useful in the usual 
sense. In particular, suppose we have D-branes wrapped on a small volume
(in the
string units) torus with non-zero $B$-field, 
and we would like to perform a T-duality transformation
that maps this torus to a large volume torus. Is there such a T-duality
transformation? This question has already been answered in \cite{Wi,SeWi,NC},
but we would like to reiterate this point here for the sake of 
completeness. Here we will be a bit more general, and closely follow the 
discussion in \cite{NC}. Thus, consider D9-branes wrapped on $T^2$ with
the metric $G_{ij}$ and $B$-field $B_{ij}=B_{12}\Sigma$, where $B_{12}=1/k$,
$k\in {\bf N}-\{1\}$. Now consider the T-duality transformation 
(\ref{T-duality}) with 
\begin{equation}
 \alpha=I~,~~~\beta=0~,~~~\gamma=k\Sigma~,~~~\delta=I~. 
\end{equation}
We will denote this T-duality transformation by $P$. The corresponding matrix
$E^\prime=G^\prime+B^\prime$ is given by
\begin{equation}
 E^\prime=-BG^{-1}B-B={B_{12}^2\over \det(G)} ~G-B~.
\end{equation}
Thus, the T-duality transformation $P$ amounts to 
\begin{equation}
 G\rightarrow G^\prime={B_{12}^2\over \det(G)}~ G~,~~~
 B\rightarrow B^\prime=-B~.
\end{equation}
In particular, a small volume torus with $B$-flux is mapped to a large volume
torus with opposite $B$-flux. (Note that $\det(G^\prime)=B_{12}^4/\det(G)$.)

{}Next, let us see what happens to D9-branes under the T-duality 
transformation $P$. It is not difficult to see that the transformation $P$ can
be written as
\begin{equation}
 P=ST^kS~.
\end{equation}
Under the first $S$-transformation D9-branes are mapped to D7-branes,
the subsequent $T$-transformations do not affect the dimensionality of the 
branes, and the last $S$-transformation maps D7-branes back to D9-branes.
Thus, the T-duality transformation $P$ maps D$p$-branes wrapped on a small 
volume $T^2$ with $B$-flux ($B_{12}=1/k$) to D$p$-branes wrapped on
a large volume $T^2$ with opposite 
$B$-flux\footnote{In fact, this was used in \cite{NC} in the discussion of
unification via Kaluza-Klein thresholds in gauge theories compactified 
on non-commutative tori, that is, tori with non-zero $B$-flux.}. 
(Generalizations to higher tori should be clear.)  

\section{K3 Orientifolds with $B$-flux}

{}Having understood toroidal orientifolds with $B$-flux, we would like to
discuss K3 orientifolds with $B$-flux next. The setup of this section is
mostly\footnote{In the ${\bf Z}_3$ cases we will also consider $\Omega R$
orientifolds, where $R$ reverses the sign of all coordinates on K3.
In such orientifolds we have D5-branes but no D9-branes.}
going to be the $\Omega$ orientifold\footnote{In the ${\bf Z}_3,{\bf Z}_4,
{\bf Z}_6$ cases $\Omega$ will actually be accompanied by an additional 
action as in \cite{GJ} - see subsections B,C,D for details.} 
of Type IIB on ${\bf R}^{1,5}\otimes
{\rm K3}$, 
where ${\rm K3}=T^4/{\bf Z}_M$. Here the orbifold group generator $g$
acts as $gz_1=\omega z_1$, $gz_2=\omega^{-1}z_2$, where $z_{1,2}$ are the 
complex coordinates parametrizing $T^4$, and $\omega\equiv\exp(2\pi i/M)$. 
Note that in order for the orbifold action to be crystallographic, $M$ must
be 2,3,4 or 6. The resulting background has ${\cal N}=1$ supersymmetry in six
dimensions, and contains D9-branes for $M=3$, and both D9- and D5-branes
for $M=2,4,6$. K3 orientifolds without $B$-flux have been studied in detail
in \cite{GJ}. The cases with rank $b=2,4$ $B$-flux in the directions of K3
were discussed in \cite{KST}. Here we would like to revisit K3 orientifolds
with $B$-flux using our improved understanding of their underlying geometric
structure. In subsections A, B, C, D we will discuss the ${\bf Z}_2,{\bf Z}_3,
{\bf Z}_6,{\bf Z}_4$ models, respectively. In subsection E we briefly summarize
the results of this section.

\subsection{The ${\bf Z}_2$ Models}

{}Let us first consider the $\Omega$ orientifold of Type IIB on ${\bf R}^{1,5}
\otimes (T^4/{\bf Z}_2)$, where the generator $R$ of ${\bf Z}_2$ acts as 
$Rz_{1,2}=-z_{1,2}$ on the complex coordinates parametrizing $T^4$. In fact,
for our purposes here it will suffice to consider $T^4=T^2\otimes T^2$, where
the first and the second 2-tori are parametrized by $z_1$ and $z_2$, 
respectively.

{}The orientifold group is ${\cal O}=\{1,R,\Omega,\Omega R\}$. Here and in the
following $\Omega=\Omega_-$, that is, $\Omega$ induces the $SO$ type of 
orientifold projection on D9-branes. Note that the presence of $\Omega$ among
the orientifold group elements implies that we have the O9$^-$-plane. Suppose
the $B$-field in the compact directions is trivial. Then the presence of the
$\Omega R$ orientifold group element implies that we have 16 O5$^-$-planes
as well. This has a non-trivial implication on the action of the ${\bf Z}_2$ 
orbifold group element $R$ on the D9- and D5-brane Chan-Paton charges.
Let $\gamma_{R,9}$ and $\gamma_{R,5}$ be the corresponding Chan-Paton 
matrices. (Here for definiteness we will assume that all D5-branes are placed
at the same O5$^-$-plane located at the origin $z_1=z_2=0$.) Then we can show
that (up to equivalent representations) \cite{GP} 
\begin{equation}\label{R95}
 \gamma_{R,9}=\gamma_{R,5}=i\sigma_3\otimes I_{16}~.
\end{equation} 
To show this, let us first note that the presence of the O9$^-$-plane implies
that we must introduce 32 D9-branes to cancel the 10-form 
R-R charge. Similarly,
the presence of 16 O5$^-$-planes implies that we must introduce 32 D5-branes
to cancel the 6-form R-R charge. Thus, all Chan-Paton matrices will be 
$32\times 32$ dimensional. Next, the matrix $\gamma_{\Omega,9}$ is 
symmetric \cite{GP}:
\begin{equation}\label{Omega9T}
 \gamma_{\Omega,9}^T=+\gamma_{\Omega,9}~.
\end{equation}
This follows from the fact that the orientifold projection on D9-branes is
of the $SO$ type. (Note that in this case we always include the appropriate
minus sign in the definition of the M{\"o}bius strip amplitude as in 
(\ref{Moebius}).) On the other hand, $\gamma_{\Omega,5}$ is 
antisymmetric \cite{GP}:
\begin{equation}\label{Omega5}
 \gamma_{\Omega,5}^T=-\gamma_{\Omega,5}~.
\end{equation}
This follows from the fact that the $\Omega$ projection on D5-branes must be
of the $Sp$ type \cite{GP} (which is consistent with \cite{Wi1}). 
One way to see this is to consider the action of $\Omega^2$ in the 59 sector,
where $\Omega^2=-1$ (note that in the 99 and 55 sectors $\Omega^2=+1$) 
\cite{GP}. On the
other hand, the orientifold 5-planes are of the O5$^-$ type. This implies
that $\gamma_{\Omega R,5}$ must be symmetric:
\begin{equation}\label{OmegaR5}
\gamma_{\Omega R,5}^T=+\gamma_{\Omega R,5}~.
\end{equation}
{}{}{}From (\ref{Omega5}) 
and (\ref{OmegaR5}) together with the fact that we must
have (the choice of the sign is immaterial)
\begin{equation}\label{OR5}
 \gamma_{\Omega R,5}=\pm\gamma_{\Omega,5}\gamma_{R,5}~,
\end{equation}
we obtain
\begin{equation}\label{R5}
 \gamma_{R,5}^T=-\gamma_{\Omega,5}\gamma_{R,5}\gamma_{\Omega,5}~.
\end{equation}
Now consider the basis where $\gamma_{R,5}$ is diagonal. Then 
$\gamma_{R,5}^T=+\gamma_{R,5}$, which together with 
(\ref{R5}) and the fact that we must have
\begin{equation}\label{Omega5^2}
 \gamma_{\Omega,5}^2=\gamma_{\Omega R,5}^2=1~, 
\end{equation}
implies that 
\begin{equation}
 \gamma_{R,5}^2=-1~.
\end{equation}
This implies that the eigenvalues of the matrix $\gamma_{R,5}$ are $\pm i$.
For the ${\bf Z}_2$ orbifold projection in the 59 open string sector to be
consistent (that is, so that we get only ${\bf Z}_2$ valued phases
from the action of the orbifold group on the Chan-Paton charges in the
59 sector), we must then require that the eigenvalues of the $\gamma_{R,9}$
matrix must also be $\pm i$, so that 
\begin{equation}\label{gammaR9^2}
 \gamma_{R,9}^2=-1~.
\end{equation}
Note that this is consistent with the analog of (\ref{Omega5}) for the
D9-branes, namely,
\begin{equation}
 \gamma_{\Omega R,9}^T=-\gamma_{\Omega R,9}~.
\end{equation}
Finally, the twisted tadpole cancellation condition implies that (in the
aforementioned setup) the matrices $\gamma_{R,9}$ and $\gamma_{R,5}$ are
traceless \cite{PS,GP}. This then implies (\ref{R95}).

{}Note that in the above discussion we have assumed that the O5-planes are of
the O5$^-$ type. The physical reason for this is clear - had we assumed
that the O5-planes were of the O5$^+$ type, we would not have been able
to cancel all tadpoles by introducing D5-branes. Another important point is
that the twisted Chan-Paton matrices $\gamma_{R,9}$ and $\gamma_{R,5}$ both
must have eigenvalues $\pm i$ or $\pm 1$, which, as we have explained
above, follows from the requirement that the ${\bf Z}_2$ orbifold
projection be consistent in the 59 sector. In the above model (with trivial
$B$-field) we must choose these matrices to have eigenvalues $\pm i$ - had
we chosen $\gamma_{R,5}$ with eigenvalues $\pm 1$, we would have found that
the O5-planes are of the O5$^+$ type. In fact, the constraint
(\ref{gammaR9^2}) can be alternatively derived by considering 
the M{\"o}bius strip amplitude in this model. Thus, we can
organize the latter into four terms according to the Chan-Paton factors that
multiply them:
\begin{eqnarray}
 {\cal M}=-\left({1\over 2}\right)^2\Big[ &&{\rm Tr}(\gamma_{\Omega,9}^{-1}
 \gamma_{\Omega,9}^T){\cal Z}(\Omega,9) +
 {\rm Tr}(\gamma_{\Omega R,9}^{-1}
 \gamma_{\Omega R,9}^T){\cal Z}(\Omega R,9)+\nonumber\\
 &&{\rm Tr}(\gamma_{\Omega R,5}^{-1}
 \gamma_{\Omega R,5}^T){\cal Z}(\Omega R,5)+
 {\rm Tr}(\gamma_{\Omega,5}^{-1}
 \gamma_{\Omega,5}^T){\cal Z}(\Omega,5)\Big]~,\label{MoebiusZ_2}
\end{eqnarray} 
where the overall factor of $(1/2)^2$ arises due to the orientifold and 
orbifold projections. Let us discuss each term in the above M{\"o}bius
strip amplitude. The first term containing ${\cal Z}(\Omega,9)$ when
rewritten in the closed string tree-channel corresponds to the closed
string exchange between D9-branes and the 
O9$^-$-plane. In fact, (\ref{Omega9T})
is precisely the statement that the O9-plane is of the O9$^-$ type (note
the overall minus sign in the definition of the M{\"o}bius strip amplitude
in (\ref{MoebiusZ_2})). Similarly, the term containing 
${\cal Z}(\Omega R,5)$ when
rewritten in the closed string tree-channel corresponds to the closed
string exchange between D5-branes and the O5-planes. Moreover, the 
characters ${\cal Z}(\Omega,9)$ and ${\cal Z}(\Omega R,5)$ are actually
identical except for the corresponding momentum respectively winding sums
(that is, the string oscillator contributions to these characters are
identical). Note that (\ref{OmegaR5}) is the statement that the O5-planes
are of the O5$^-$ type. Next, consider the term
containing ${\cal Z}(\Omega R,9)$.
The latter corresponds to the 99 sector M{\"o}bius contribution with the
${\bf Z}_2$ orbifold projection inserted on the boundary. Similarly, the
term containing ${\cal Z}(\Omega,5)$ corresponds to the 55 M{\"o}bius
contribution with the ${\bf Z}_2$ orbifold projection inserted on the
boundary. In fact, the characters ${\cal Z}(\Omega R,9)$ and ${\cal Z}
(\Omega,5)$ are actually identical\footnote{These characters actually
vanish. More precisely, all the characters vanish by
supersymmetry. However, the bosonic and fermionic pieces in 
the aforementioned characters vanish separately. Thus, in the open string
loop-channel the NS and R contributions to these characters vanish
separately. Moreover, in the closed string 
tree-channel the NS-NS and R-R contributions
to these characters also vanish separately. This, in particular, implies
that the corresponding terms in the M{\"o}bius strip amplitude do not
contribute to massless tadpoles. In fact, these terms are often dropped
when discussing this model (as in, {\em e.g.}, \cite{GP}). However, these
terms are important to keep when discussing the spectrum and vertex
operators in this model. In particular, one has to make sure that the ${\bf
Z}_2$ orbifold projection is consistent, which is precisely
the constraint we are
going to discuss in a moment.}. Putting all of the above together, we can
now derive a non-trivial constraint on $\gamma_{R,9}$. Thus, note that the
traces ${\rm Tr}(\gamma_{\Omega,9}^{-1}\gamma_{\Omega,9}^T)={\rm
Tr}(\gamma_I)=32$ and ${\rm Tr}(\gamma_{\Omega R,5}^{-1}\gamma_{\Omega
R,5}^T)= {\rm Tr}(\gamma_I)=32$ in front of the characters ${\cal
Z}(\Omega,9)$ respectively ${\cal Z}(\Omega R,5)$ are identical. This
implies that the traces ${\rm Tr}(\gamma_{\Omega R,9}^{-1}
\gamma_{\Omega R,9}^T)={\rm Tr}(\gamma_{R,9}^2)$ and 
${\rm Tr}(\gamma_{\Omega,5}^{-1}\gamma_{\Omega,5}^T)=-{\rm
Tr}(\gamma_I)=-32$ in front of the characters ${\cal
Z}(\Omega R,9)$ respectively ${\cal Z}(\Omega,5)$ must also be identical
for the ${\bf Z}_2$ orbifold projection in the 99 sector to be consistent
with that in the 55 sector. This then implies (\ref{gammaR9^2}). As we will
see in the following, requiring consistency of the orbifold projection in
the M{\"o}bius strip amplitude will result in additional non-trivial
constraints in the models with non-zero $B$-flux.

{}Before we consider the cases with non-zero $B$-flux, for later convenience
let us review the ${\bf Z}_2$ model of \cite{PS,GP} without the 
$B$-field. The closed string sector contains the six dimensional 
${\cal N}=1$ supergravity multiplet, one 
untwisted (self-dual) tensor supermultiplet, 4 untwisted hypermultiplets, and
16 twisted hypermultiplets. Note that the 16 fixed points of the ${\bf Z}_2$ 
orbifold give rise to hypermultiplets but no (anti-self-dual) 
tensor multiplets as all 16 fixed points are even under the orientifold action.
This follows from the fact that all 16 orientifold 5-planes located at the
fixed points are of the O5$^-$ type. Next, let us discuss the open string
spectrum. The gauge group is $U(16)_{99}\otimes U(16)_{55}$, and the
massless matter
consists of the following hypermultiplets:
\begin{eqnarray}
 2\times &&({\bf 120};{\bf 1})_{99}~,\\
 2\times &&({\bf 1};{\bf 120})_{55}~,\\
         &&({\bf 16};{\bf 16})_{95}~.
\end{eqnarray} 
Here semi-colon separates the 99 and 55 gauge quantum numbers. Note that 
$U(1)$'s are actually anomalous, and are broken via the generalized 
Green-Schwarz mechanism \cite{BLPSSW,Serone}, so that the gauge group is 
actually $SU(16)_{99}\otimes SU(16)_{55}$. 

{}Here we would like to stress one important point. In particular, the
$\Omega$ projection on D5-branes is of the $Sp$ type, while the O5-planes
located at the 16 ${\bf Z}_2$ fixed points are of the O5$^-$ type, that
is, they induce the $SO$ type of orientifold projection on the Chan-Paton
charges of D5-branes. The above discussion relates this to the fact that
the latter projection is determined by $\gamma_{\Omega R,5}$, and {\em not}
by $\gamma_{\Omega,5}$. Nonetheless, at first it might appear a bit strange
that the projection on D5-branes is of the $SO$ type - from the arguments
of \cite{Wi1} we would expect (subgroups of) symplectic gauge groups
coming form D5-branes. In particular, appearance of antisymmetric 
representations (namely, ${\bf 120}$ of $SU(16)$) in the 55 sector might
seem puzzling in the context of \cite{Wi1}. However, there is no puzzle
here as there is a geometric explanation of this point. To arrive at this
explanation, however, we will first need to review the difference between the
O5$^-$- and O5$^+$-planes. Here we will be closely following the 
discussion in \cite{Ha}.

{}Thus, consider Type IIB on ${\bf R}^{1,9}$ in the presence of an O5-plane.
The O5-plane is located at the fixed point at the origin of
${\bf R}^4/{\bf Z}_2$,
where the ${\bf Z}_2$ action simultaneously reflects all four coordinates of
${\bf R}^4$ transverse to the O5-plane. The orientifold replaces a 3-sphere 
${\bf S}^3$ around the origin of ${\bf R}^4$ by 
${\bf RP}^3={\bf S}^3/{\bf Z}_2$. (Recall that the real projective $n$-space
${\bf RP}^n$ is defined as the quotient ${\bf S}^n/{\bf Z}_2$
of the $n$-sphere ${\bf S}^n$ defined via $\sum_{i=1}^{n+1} 
x_i^2=\rho^2$, where the 
action of ${\bf Z}_2$ on the coordinates $x_i$, $i=1,\dots,n+1$, is given by
$x_i\rightarrow -x_i$, and $\rho$ is the radius of the $n$-sphere 
${\bf S}^n$.) Now consider unorientable closed world-sheets $\Sigma=
{\bf RP}^2$. Such a world-sheet is embeddable in ${\bf RP}^3$. Thus, we can 
define a ${\bf Z}_2$ charge for the O5-plane as follows. Consider a constant
NS-NS 
$B$-flux. Then the world-sheets $\Sigma={\bf RP}^2$ contribute to the path 
integral with an extra phase
\begin{equation}
 \exp\left(i\int_\Sigma B\right)~,
\end{equation}      
which is $+1$ for the trivial $B$-flux, and $-1$ for the half-integer 
$B$-flux (recall that the $B$-flux is quantized in the presence of an 
orientifold plane).
The aforementioned ${\bf Z}_2$ charge assignments are then as follows. In the
former case we assign charge 0, while in the latter case we assign charge
1, and the ${\bf Z}_2$ charge is defined modulo 2. 
One can then show that the O5-plane with the ${\bf Z}_2$ charge 0 is
of the O5$^-$ type, while the one with the charge 1 is of the O5$^+$ type
\cite{Ha}. This, for instance, can be seen by considering the following
BPS configuration with eight supercharges. Let the O5-plane fill the 
coordinates $x_0,x_1,x_2,x_3,x_4,x_5$ of ${\bf R}^{1,9}$, and a 
${1\over 2}$NS5-brane\footnote{In our conventions a 
${1\over 2}$NS5-brane is the 
S-dual of a D5-brane,
whose R-R charge is $+1$, while O5$^\pm$-planes have the R-R charges $\pm 2$,
respectively. Note that an NS5-brane is S-dual of a pair of 
D5-branes, which combine
into a dynamical 5-brane - in the presence of an O5-plane D5-branes always move
in pairs.}
fill the coordinates $x_0,x_1,x_2,x_3,x_4,x_6$. That is, the O5-plane and
the ${1\over 2}$NS5-plane 
intersect at 90 degrees in the 56-plane. Now consider the origin
of the 789 space transverse to both the O5-plane and the 
${1\over 2}$NS5-brane. 
The orientifold replaces a 2-sphere around the origin of this
space by ${\bf RP}^2$. The NS-NS $B$-flux couples magnetically to NS5-branes.
Thus, with the appropriate normalization 
$\int_{{\bf RP}^2} B$ counts the number of ${1\over 2}$NS5-branes 
modulo 2. This
implies that the aforementioned ${\bf Z}_2$ charge of the O5-plane, 
which with this
normalization can be identified with $\int_{{\bf RP}^2} B$, changes by 1
every time
the O5-plane crosses the 
${1\over 2}$NS5-brane\footnote{Here we note that for an O$p$-plane
with $p\leq 5$ one can define another
${\bf Z}_2$ charge (in the appropriate normalization) 
via $\int_{{\bf RP}^{5-p}} C^{(5-p)}$ \cite{Ha}, 
where $C^{(5-p)}$
is a Ramond-Ramond form. The relevant brane configuration here is that of
an O$p$-plane intersecting with a D$(p+2)$-brane such that we have 
8 unbroken supercharges. The aforementioned
${\bf Z}_2$ charge 
then changes by 1 every time the O$p$-plane crosses the
D$(p+2)$-brane. This additional ${\bf Z}_2$ charge, however, will not be
important in the subsequent discussions.}. 
To see that the O5$^-$-plane
has the ${\bf Z}_2$ charge 0, consider the T-dual version 
of the above setup, where we have an O3-plane. Then from Montonen-Olive 
self-duality of the $SO(2k)$ gauge theories we conclude that the O3$^-$-plane
must have zero ${\bf Z}_2$ charge or else it would not be 
invariant under the $SL(2,{\bf Z})$ symmetry of Type IIB.

{}Now we can explain why the O5-planes are of the O5$^-$ type in the
aforementioned model of \cite{PS,GP} without the $B$-field. Thus, naively
we expect that the O5-planes are of the O5$^+$ type as the orientifold 
projection is of the $SO$ type on D9-branes implying that the orientifold 
projection is of the $Sp$ type on the D5-branes. 
This would require that the corresponding ${\bf Z}_2$ charge related to the
$B$-field for the orientifold planes is 1, that is, we have an odd-half-integer
$B$-flux in ${\bf RP}^2$. However, the O5-planes are
located at the ${\bf Z}_2$ orbifold fixed points, and each ${\bf Z}_2$ orbifold
fixed point corresponds to a collapsed 2-sphere ${\bf P}^1$. As was pointed 
out in \cite{Asp}, in the {\em conformal field theory} ${\bf Z}_2$ orbifold
there is an odd-half-integer {\em twisted} 
$B$-field\footnote{The twisted $B$-field plays an important role in the context
of orientifolds in a number of setups - see, {\em e.g.}, \cite{Sen} and 
\cite{KST1}.} 
stuck inside of each 
collapsed ${\bf P}^1$. The orientifold replaces these ${\bf P}^1$'s by 
${\bf RP}^2$'s, so that we have an odd-half-integer twisted $B$-flux stuck 
inside of each collapsed ${\bf RP}^2$. This additional twisted $B$-flux
converts the would-be O5$^+$-planes into O5$^-$-planes, which is the result we 
wished to explain. 

{}Next, let us consider the $\Omega$ orientifold of Type IIB on $T^4/{\bf Z}_2$
in the presence of the $B$-field. Here we will mainly 
focus on the case of $T^4=
T^2\otimes T^2$ with $B$-field of rank $b=2$ turned on in the directions of the
first $T^2$ (while the $B$-field in the directions of the second $T^2$ is
trivial). Generalizations to generic $T^4$'s with $B$-field of rank $b=2$ as
well as $b=4$ should be evident.

{}In the case of $B$-field of rank $b$ we have $n_{f-}=
8(1+1/2^{b/2})$ O5$^-$-planes and $n_{f+}=8(1-1/2^{b/2})$ O5$^+$-planes. 
In the closed string ${\bf Z}_2$ twisted sector the
orientifold projection at the ${\bf Z}_2$ orbifold fixed points where the 
O5$^-$-planes are located gives rise to hypermultiplets, while at the fixed
points where the O5$^+$-planes are located it gives rise to (anti-self-dual)
tensor multiplets. On the
other hand, the untwisted closed string
sectors of the models with $b=2,4$ are the same as that of the $b=0$ model.
Thus, the closed string sector of the model with $B$-field of rank $b$ contains
the six dimensional ${\cal N}=1$ supergravity multiplet, one untwisted 
(self-dual) tensor supermultiplet, 4 untwisted hypermultiplets, $n_{f-}$ 
twisted hypermultiplets and $n_{f+}$ twisted (anti-self-dual) tensor 
multiplets.

{}Next, let us discuss the open string sector. As we have already mentioned, we
will specialize to the case of $b=2$. First, let us understand the geometric
structure of the orientifold planes. Let the vielbeins on the first $T^2$
(where we have non-zero $B$-flux) be $e_i$, $i=1,2$, 
while the vielbeins on the second
$T^2$ (where the $B$-flux is trivial) be $d_i$, $i=1,2$. Let us define $e_3
\equiv -e_1-e_2$, and $d_3\equiv -d_1-d_2$. The 16 ${\bf Z}_2$ fixed points are
located at $(0,0),(e_a/2,0),(0,d_a/2),(e_a/2,d_b/2)$, $a,b=1,2,3$. Without
loss of generality we can 
choose the following distribution for the O5-planes. At
$(0,0),(0,d_a/2),(e_i/2,0),(e_i/2,d_a/2)$ we have 12 O5$^-$-planes, while at
$(e_3/2,0),(e_3/2,d_a/2)$ we have 4 O5$^+$-planes. 

{}At first it might seem that the above setup is completely consistent, in
particular, that there is no difficulty with having O5$^-$- and O5$^+$-planes
at the same time. Thus, this is certainly consistent in the toroidal case, so 
it might seem that in the $T^4/{\bf Z}_2$ orbifold
case this should also be consistent.
However, there is a subtlety here. In particular, note that the orientifold
projection is of the $SO$ type on D9-branes, and, therefore, it is of the
$Sp$ type on D5-branes. The odd-half-integer 
twisted $B$-flux is present at all 16 fixed points of the $T^4/{\bf Z}_2$
orbifold. Thus, repeating the above argument we would conclude that all 16
O5-planes must be of the O5$^-$ type. So there seems to be a puzzle here.
Let us, therefore, try to understand this point better.

{}To begin with, let us note that there are two separate issues
here. First,
by studying the orientifold action in the closed string sector we
unambiguously arrive at the conclusion that we have 12 O5$^-$- and 4 
O5$^+$-planes as in the toroidal case. However, this does not guarantee
that, once we introduce {\em both} 32 D9-branes and 16 D5-branes, the
orientifold action in the open string sector is indeed consistent. This 
statement might seem surprising at first as the orientifold action in the
99 and 55 sectors certainly seems to be consistent. This follows from our
analyses of the corresponding toroidal orientifolds with $B$-flux (and we
do not expect any obstruction in consistently orbifolding the 99 and 55
sectors). However, the subtlety here can (and, as we will point out in a
moment, does) arise in the 59 sector. In particular, it is not always
sufficient to consider, say, the Klein bottle amplitude (and the
corresponding tadpoles) to conclude that we have some numbers of O$p^-$-
and O$p^+$-planes. Rather, we must also make sure that in any given setup
these objects indeed induce the $SO$ and $Sp$ type of orientifold
projections on D$p$-branes (that is, we must make sure that the couplings
between the O$p$-planes and D$p$-branes are consistent).

{}Since the issue we are discussing here appears to be subtle, we would
like to proceed step-by-step. Thus, first let us make sure that the
orientifold projection in the closed string sector is indeed such that we
have $n_{f-}$ twisted hypermultiplets and $n_{f+}$ twisted tensor
multiplets. Thus, we must show that the orientifold action on $n_{f+}$
fixed points has an extra minus sign compared with that on the other
$n_{f-}$ fixed points. One way to see this explicitly was already discussed
in \cite{KST}. Thus, let us consider a special point in the moduli space of
$T^4$'s corresponding to the $SO(8)$ symmetry. At this point the $B$-field
has rank $b=2$ \cite{KST}, and in the conformal field theory we have
the $[SO(8)]_L \otimes [SO(8)]_R$ current algebra at level 1 on the closed
string world-sheet. The ${\bf Z}_2$ orbifold action reduces this current
algebra to $[SU(2)^4]_L\otimes [SU(2)^4]_R$ at level 1. The vertex
operators for the 16 fixed points are especially simple at this point. In
particular, they carry the following quantum numbers under 
$[SU(2)^4]_L\otimes [SU(2)^4]_R$: $({\bf 2},{\bf 1},{\bf 1},{\bf 1}||{\bf
2}, {\bf 1},{\bf 1},{\bf 1})$, plus states obtained by simultaneous
permutations of the $SU(2)$ factors for both left- and right-movers. The
orientifold action reduces the $[SU(2)^4]_L\otimes [SU(2)^4]_R$ current algebra
to $[SU(2)^4]_{\rm diag}$, and the fixed points now carry the following
quantum numbers: $({\bf 3}_s\oplus{\bf 1}_a,{\bf 1},{\bf 1},{\bf 1})$, plus
states obtained by permutations of the $SU(2)$ factors. Note that the
states
transforming in ${\bf 3}_s$'s are symmetric under $\Omega$, while the
states transforming in ${\bf 1}_a$'s are antisymmetric. Thus, we indeed
have 12 twisted hypermultiplets and 4 twisted tensor multiplets in this 
case\footnote{It is not difficult to see that in the $b=4$ case we then
have 10 twisted hypermultiplets and 6 twisted tensor multiplets
\cite{KST}.}. In particular, this is consistent with the fact that we have
$n_{f-}$ O5$^-$-planes located at the fixed points giving rise to twisted
hypermultiplets, and $n_{f+}$ O5$^+$-planes located at the fixed points
giving rise to twisted tensor multiplets (as can be explicitly seen by
considering the Klein bottle amplitude in this model).

{}Next, let us discuss the open string sector. In the 99 sector we have
Chan-Paton matrices $\gamma_{S_a,9}$ corresponding to non-commuting Wilson
lines\footnote{As in the previous section, here we are going to view
half-integer $B$-flux on the first $T^2$ with vielbeins $e_i$ in terms of
the ${\bf Z}_2\otimes {\bf Z}_2$ freely acting orbifold (with discrete
torsion) of the torus ${\widetilde T}^2$ 
with vielbeins $E_i=2e_i$ and zero $B$-flux. The 
non-commuting Wilson lines then correspond to half-lattice 
shifts $S_i X_i=X_i+E_i/2$ on ${\widetilde T}^2$ acting on the Chan-Paton
degrees of freedom.}, 
and also the twisted Chan-Paton matrix $\gamma_{R,9}$. We can write the
corresponding Chan-Paton matrices as follows:  
\begin{eqnarray}\label{gammaSa90}
 &&\gamma_{S_a,9}=\gamma_a\otimes I_2\otimes I_8~,\\
 &&\gamma_{R,9}=I_2\otimes \nu\sigma_3\otimes I_8~,\label{gammaR90}
\end{eqnarray}
where $\gamma_a$ are the $2\times 2$ matrices corresponding to the
$D_4$ or $D_4^\prime$ type of Wilson lines given in 
(\ref{D4}) respectively (\ref{D4'}). On the other hand, for the twisted
Chan-Paton matrix $\gamma_{R,9}$ {\em a priori} we have two inequivalent
choices: $\nu^2=+1$ and $\nu^2=-1$. However, as we will show
in a moment, these 
choices are correlated with the choices of the non-commuting Wilson lines
as follows:
\begin{eqnarray}
 &&D_4:~~~\nu^2=+1~,\\ 
 &&D_4^\prime:~~~\nu^2=-1~.
\end{eqnarray} 
To show this, let us consider the M{\"o}bius strip amplitude:
\begin{eqnarray}
 {\cal M}=-\left({1\over 2}\right)^2\Big[ 
 {\cal Y}(\Omega, 9)+{\cal Y}(\Omega R,9) +{\cal Y}(\Omega R,5) +
 {\cal Y}(\Omega, 5)\Big]~.
\end{eqnarray}
Here the characters ${\cal Y}$, which contain the corresponding Chan-Paton 
factors, are defined as follows. The character ${\cal Y}(\Omega, 9)$
corresponds to the 99 sector contribution with the identity element of the
${\bf Z}_2$ orbifold group inserted on the boundary of the M{\"o}bius
strip; the character ${\cal Y}(\Omega R,9)$ corresponds to the 99 sector
contribution with element $R$ of the ${\bf Z}_2$ orbifold group inserted on
the boundary. Similarly, the character ${\cal Y}(\Omega R,5)$ corresponds
to the 55 sector contribution with the identity element inserted on the
boundary; the character ${\cal Y}(\Omega,5)$ corresponds to the 55 sector
contribution with the element $R$ inserted on the boundary. Starting from
(\ref{Moebius}), it is not difficult to see that the character ${\cal
Y}(\Omega R,9)$ is given by:
\begin{eqnarray}
 {\cal Y}(\Omega R,9)&&={1\over 4}{\cal Z}(\Omega R,9)\Big[
 {\rm Tr}(\gamma_{\Omega R,9}^{-1}\gamma_{\Omega R,9}^T)+
 \sum_{a=1}^3{\rm Tr}(\gamma_{\Omega RS_a,9}^{-1}\gamma_{\Omega RS_a,9}^T)
 \Big]
 \nonumber\\
 &&={1\over 4}
 {\cal Z}(\Omega R,9){\rm Tr}(\gamma_{R,9}^2) \Big[1+\sum_{a=1}^3
 \eta_{aa}\Big]~,
\end{eqnarray}  
where the character ${\cal Z}(\Omega R,9)$ here is the same as in
(\ref{MoebiusZ_2}). 
Next, let us consider the character ${\cal
Y}(\Omega,5)$. Since there are no Wilson lines in the 55 sector, we have 
\begin{equation}
 {\cal Y}(\Omega,5)={\rm Tr}(\gamma_{\Omega,5}^{-1}\gamma_{\Omega,5}^T)
 {\cal Z}(\Omega,5)~,
\end{equation}  
where the character ${\cal Z}(\Omega,5)$ is the same as in (\ref{MoebiusZ_2}).
Now repeating the argument after (\ref{MoebiusZ_2}) for the above model, we 
conclude that in
the M{\"o}bius strip amplitude 
for the ${\bf Z}_2$ orbifold projection in the 99 sector 
to be consistent with that in the 55 sector, the following constraint must
be satisfied:
\begin{equation}
 {1\over 4} {\rm Tr}(\gamma_{R,9}^2)\Big[1+\sum_{a=1}^3\eta_{aa}]=
 {\rm Tr}(\gamma_{\Omega,5}^{-1}\gamma_{\Omega,5}^T)~.
\end{equation}  
We can simplify this constraint as follows. First, note that 
${\rm Tr}(\gamma_{\Omega,5}^{-1}\gamma_{\Omega,5}^T)=-16$. Also, 
${\rm Tr}(\gamma_{R,9}^2)=32 \nu^2$. Finally, $\sum_{a=1}^3 \eta_{aa}=+1$
for the $D_4^\prime$ type of Wilson lines (\ref{D4'}), 
and $\sum_{a=1}^3 \eta_{aa}=-3$
for the $D_4$ type of Wilson lines (\ref{D4}). Putting all of this
together, we conclude that for the $D_4$ type of Wilson lines we must have
$\nu^2=+1$, whereas for the $D_4^\prime$ type of Wilson lines we must have
$\nu^2=-1$.

{}The constraint we have just derived is the analogue of the corresponding
constraint in the case without $B$-flux. In the latter case, as we have
already discussed above, relaxing this
constraint would result in a model where tadpoles/anomalies cannot be 
canceled completely. In the case with non-zero $B$-field relaxing the
corresponding constraint would also result in inconsistent models. Thus,
for instance, consider taking the $D_4$ type of Wilson lines with
$\nu^2=-1$. The resulting model has the following massless
spectrum. We have already
discussed the closed string sector (which is independent of the choice of
the Chan-Paton matrices) - it contains 16 hypermultiplets and 5 tensor
multiplets (for $b=2$). In the open string sector the gauge group is
$[U(8)\otimes {\bf Z}_2]_{99}\otimes U(8)_{55}$, and the massless matter is
given by the following hypermultiplets:
\begin{eqnarray}
 2\times &&({\bf 36};{\bf 1})_{99}~,\\
 2\times &&({\bf 1};{\bf 36})_{55}~,\\
         &&({\bf 8}_+;{\bf 8})_{95}~,\\
         &&({\bf 8}_-;{\bf 8})_{59}~,
\end{eqnarray}  
where the subscript $\pm$ in the 95 sector refers to the 99 ${\bf Z}_2$
charge. Note that this spectrum is anomalous. In particular, the
irreducible $R^4$ gravitational anomaly does not cancel. A similar 
conclusion applies to the case where the Wilson lines are of the 
$D_4^\prime$ type, while the ${\bf Z}_2$ twisted Chan-Paton matrix is
chosen such that $\nu^2=+1$.

{}Next, we would like to study the models arising for the aforementioned
two inequivalent choices satisfying the above consistency condition.
To begin with, let
us consider the following choice:
\begin{eqnarray}\label{gammaSa9}
 &&\gamma_{S_a,9}=\gamma_a\otimes I_2\otimes I_8~,\\
 &&\gamma_{R,9}=I_2\otimes i\sigma_3\otimes I_8~,\label{gammaR9}
\end{eqnarray} 
where the $\gamma_a$ matrices correspond to the
$D_4^\prime$ type of Wilson lines given in (\ref{D4'}).
In the 55 sector we must also specify the choice of $\gamma_{R,5}$, which
is a $16\times 16$ matrix (and not a $32\times 32$ matrix as we have only 16
D5-branes). The choice of this matrix must be consistent with the
orientifold action. In particular, let us consider an O5$^-$-plane, say,
that located at the fixed point at the origin $(0,0)$. Then, if we place 16
D5-branes on top of this orientifold plane, according to the above
arguments we must choose $\gamma_{R,5}$ such that its eigenvalues are $\pm
i$. On the other hand, suppose we consider an O5$^+$-plane, say, that
located at the fixed point $(e_3/2,0)$. Then, if we placed D5-branes on top
of this O5$^+$-plane,
we would encounter an inconsistency. To see this let us repeat the argument
given in the beginning of this subsection for the case where the O5-plane
is of the O5$^+$ type. Thus, in this case we have
\begin{equation}
 \gamma_{\Omega R,5}^T=-\gamma_{\Omega R,5}~.
\end{equation}
Together with (\ref{Omega5}) and (\ref{OR5}) this implies that
\begin{equation}\label{trans}
 \gamma_{R,5}^T=+\gamma_{\Omega,5}\gamma_{R,5}\gamma_{\Omega,5}~.
\end{equation}
Now consider the basis where $\gamma_{R,5}$ is diagonal. Then
$\gamma_{R,5}^T=+\gamma_{R,5}$, which together with (\ref{trans}) and
(\ref{Omega5^2}) implies that
\begin{equation}\label{gammaR5^2=1}
 \gamma_{R,5}^2=+1~.
\end{equation} 
This implies that the eigenvalues of the matrix $\gamma_{R,5}$ are $\pm
1$. Note, however, that the eigenvalues of the matrix $\gamma_{R,9}$ are
$\pm i$. That is, in the 59 sector the ${\bf Z}_2$ orbifold action is
inconsistent as the 59 states would have phases $\pm i$ (instead of $\pm
1$). At first it might seem that this difficulty can be simply avoided by
concluding that in this background 
it is inconsistent to place D5-branes at the O5$^+$-planes,
albeit it is consistent to place them at the O5$^-$-planes. However, this
naive resolution
fails due to the following simple argument\footnote{This effective 
field theory argument was pointed out to me by Alex Buchel.} - 
there is no obstruction to
moving D5-branes from an O5$^-$-plane to an O5$^+$-plane in this model. 

{}Thus, let us consider the $T^4/{\bf Z}_2$ model with $B$-field of rank
$b=2$. Let us place all 16 D5-branes at the O5$^-$-plane located at the
origin $(0,0)$. Then the consistent choice of $\gamma_{R,5}$ is 
(up to equivalent representations) given by
\begin{equation}\label{gammaR5i}
 \gamma_{R,5}=i\sigma_3\otimes I_8~.
\end{equation}
The gauge group of this model is $[U(8)\otimes {\bf Z}_2]_{99}\otimes 
U(8)_{55}$. The open string massless 
matter consists of the following hypermultiplets 
\begin{eqnarray}
 2\times &&({\bf 28};{\bf 1})_{99}~,\\
 2\times &&({\bf 1};{\bf 28})_{55}~,\\
         &&({\bf 8}_+;{\bf 8})_{95}~,\\
         &&({\bf 8}_-;{\bf 8})_{95}~.  
\end{eqnarray}
Here the subscript $\pm$ in the 95 sector refers to the 99 ${\bf Z}_2$ 
discrete gauge charge
of the corresponding state. Note that $U(1)$'s are actually anomalous, and
are broken by the generalized Green-Schwarz mechanism \cite{BST}. In
particular, the states that participate in the generalized Green-Schwarz
mechanism are (certain linear combinations of) the R-R scalars in the 
twisted hypermultiplets. Thus, the gauge group is actually $[SU(8)\otimes
{\bf Z}_2]_{99}\otimes SU(8)_{55}$. Moreover, the corresponding {\em two} 
twisted
hypermultiplets are eaten in the process of Higgsing $U(1)$'s. Note that
the above spectrum is free of the irreducible $R^4$ and $F^4$ anomalies.

{}Note that we can Higgs the 55 gauge group by giving VEVs to the
hypermultiplets $2\times ({\bf 1};{\bf 28})_{55}$. Thus, for instance, we
can break the 55 gauge group $SU(8)_{55}$ down to $Sp(8)_{55}$ by giving
appropriate VEVs to the aforementioned hypermultiplets - under
the breaking $SU(8)\rightarrow Sp(8)$, the antisymmetric representation
${\bf 28}$ of $SU(8)$ decomposes as ${\bf 28}={\bf 1}+{\bf 27}$ in terms of
the $Sp(8)$ representations. Actually, we can start from the $U(8)_{55}$
gauge group with the anomalous $U(1)_{55}$ factor, and break $U(8)_{55}$
down to $Sp(8)_{55}$. In this process 28 out of the 56 hypermultiplets $2\times
({\bf 1};{\bf 28})_{55}$ are eaten in the Higgs mechanism\footnote{Note
that a singlet of $Sp(8)_{55}$ 
is eaten in Higgsing the $U(1)_{55}$ factor. This
implies that only one (instead of two) of the closed twisted
hypermultiplets is eaten, namely, in the process of Higgsing the anomalous 
$U(1)_{99}$ factor.}. The leftover 28 hypermultiplets transform in ${\bf
1}\oplus{\bf 27}$ of $Sp(8)_{55}$. 
Note that this Higgsing corresponds to nothing
but moving together all 16 D5-branes off the O5$^-$-plane into the
bulk. The VEV of the leftover singlet  
hypermultiplet then
corresponds to the location of D5-branes in the bulk (note that there are
4 real scalars in a hypermultiplet, and those in the singlet hypermultiplet
parametrize the location of D5-branes in four real dimensions of K3). 
On the other hand, we can further break the $Sp(8)$ gauge group by
giving a VEV to the leftover hypermultiplet which is in ${\bf
27}$ of $Sp(8)$. This Higgsing corresponds to pulling D5-branes apart from
each other in groups of 4 (or multiples thereof) - each dynamical 5-brane,
which consists of 4 D5-branes\footnote{Here two pairings take place - one
due to the orientifold projection, and the other one due to the orbifold
projection.}, separately gives rise to an $Sp(2)$ gauge
group. On the other hand, $k$ coincident dynamical 5-branes (corresponding
to $4k$ coincident D5-branes) give rise to
$Sp(2k)$ gauge group, with all four dynamical 5-branes coincident 
giving rise to the $Sp(8)$ gauge symmetry. 

{}Even though the general case is
straightforward to discuss, from now on it will suffice for our purposes
here to consider moving together all 16 D5-branes off the O5$^-$-plane, so
that in the bulk they give rise to the $Sp(8)$ gauge symmetry.       
In this subspace of the moduli space the gauge group is $[SU(8)\otimes 
{\bf Z}_2]_{99}\otimes Sp(8)_{55}$, and the open string massless matter
consists of the following hypermultiplets:
\begin{eqnarray}
 2\times &&({\bf 28};{\bf 1})_{99}~,\\
         &&({\bf 1};{\bf 1}\oplus {\bf 27})_{55}~,\\
         &&({\bf 8}_+;{\bf 8})_{95}~,\\
         &&({\bf 8}_-;{\bf 8})_{95}~. 
\end{eqnarray}
In fact, this spectrum can be derived directly in the orientifold
language if we consider placing 16 D5-branes in the bulk (that is, away
from the ${\bf Z}_2$ orbifold fixed points). In particular, let us place 8
D5-branes at a generic point $(z_1,z_2)$ 
in the bulk. Then we must place the other 8
D5-branes at the point $(-z_1,-z_2)$, so that the entire background is
invariant under the action of the $\Omega R$ element of the orientifold
group. If we did not have to perform the further ${\bf Z}_2$ orbifold
projection with respect to the element $R$, the 55 gauge group at such a
generic point would be $U(8)_{55}$. However, the ${\bf Z}_2$ orbifold 
projection reduces $U(8)_{55}$ to its subgroup $Sp(8)_{55}$. Note that the
reason why the rank of the unbroken gauge group $Sp(8)_{55}$ 
is halved compared
with that of $U(8)_{55}$ 
is that $\gamma_{\Omega R,5}$ and $\gamma_{R,5}$ do not
commute, in fact, they anticommute as can be seen from the relation 
$\gamma_{R,5}=-\gamma_{\Omega,5}\gamma_{R,5}\gamma_{\Omega,5}$, which holds
in the basis where $\gamma_{R,5}$ is diagonal. Furthermore, the 55 gauge
bosons are in the symmetric representation (so that the gauge group is
symplectic), while the 55 hypermultiplets are in the antisymmetric
representation (which is reducible for symplectic gauge groups). 
The latter fact is due to the extra minus sign that the ${\bf
Z}_2$ twist $R$ has when acting on the
hypermultiplets compared with when it acts on the gauge bosons. 

{}Next, we would like to ask what would happen if we bring D5-branes to one
of the O5$^+$-planes. Here we expect that the 55 gauge symmetry should be 
enhanced, and, moreover, the 55 gauge group must be a subgroup of $Sp(16)$.
The latter is the gauge group we would obtain in the toroidal case, and in
the orbifold case the 55 gauge group would have to be determined by the
choice of $\gamma_{R,5}$. Thus, if we choose $\gamma_{R,5}$ to be
\begin{equation}\label{gammaR5}
 \gamma_{R,5}=\sigma_3\otimes I_8~,
\end{equation}
then the gauge group coming from 16 D5-branes on top of an O5$^+$-plane
is $Sp(8)\otimes Sp(8)$. Thus, the gauge group of the model at this point 
in the moduli space is $[SU(8)\otimes {\bf Z}_2]_{99}\otimes [Sp(8)\otimes
Sp(8)]_{55}$, and the open string massless matter is given by:
\begin{eqnarray}
 2\times &&({\bf 28};{\bf 1},{\bf 1})_{99}~,\\
         &&({\bf 1};{\bf 8},{\bf 8})_{55}~,\\
      {1\over 2}&&({\bf 8}_+;{\bf 8},{\bf 1})_{95}~,\\
      {1\over 2}&&({\bf 8}_+;{\bf 1},{\bf 8})_{95}~,\\  
      {1\over 2}&&({\bf 8}_-;{\bf 8},{\bf 1})_{95}~,\\
      {1\over 2}&&({\bf 8}_-;{\bf 1},{\bf 8})_{95}~.
\end{eqnarray} 
Note that the 99 spectrum is the same as before, and the 55 spectrum can be
deduced by keeping the states in the toroidal compactification invariant
under the ${\bf Z}_2$ orbifold action. The states in the 95 sector,
however, cannot be deduced in this way as the ${\bf Z}_2$ action in this
sector is inconsistent - recall that the eigenvalues of $\gamma_{R,9}$ are
$\pm i$, while the eigenvalues of $\gamma_{R,5}$ are $\pm 1$. The above 95
spectrum has been written down as follows. First, the number of 95
hypermultiplets must be the same whether D5-branes are on top of the
O5$^+$-plane or in the bulk. Second, the two $Sp(8)$ subgroups in the 55
sector are on the equal footing, so that the spectrum should possess a
symmetry under the permutation of these two subgroups. This then fixes the
spectrum as above. This spectrum, however, is {\em anomalous}. Indeed, it
contains half-hypermultiplets in complex representations. Note that had we
not distinguished the 99 ${\bf Z}_2$ discrete gauge quantum numbers, naively we
might have thought that we have one hypermultiplet in $({\bf 8};{\bf
8},{\bf 1})$, and one hypermultiplet in $({\bf 8};{\bf 1},{\bf 8})$ of
$SU(8)_{99} \otimes [Sp(8)\otimes Sp(8)]_{55}$. This is, however, not the
case, and, as we see, the 99 ${\bf Z}_2$ discrete gauge symmetry does
indeed play an important role. 

{}Thus, we have arrived at an inconsistency - we started from a seemingly
consistent setup where we had all 16 D5-branes on top of an O5$^-$-plane,
and then moved them to one of the O5$^+$-planes. Even though in the bulk
the 55 gauge theory appears to be consistent, at the O5$^+$-plane it is
anomalous. We have seen this in the effective field theory language, and
in the orientifold language this inconsistency is translated to that in the
${\bf Z}_2$ orbifold action in the 59 sector due to the fact that
$\gamma_{R,5}$ is given by (\ref{gammaR5}), while $\gamma_{R,9}$ is given
by (\ref{gammaR9}). Here we can ask if we could possibly have chosen
$\gamma_{R,5}$ as in (\ref{gammaR5i}) even if D5-branes are on top of an
O5$^+$-plane. It is, however, not difficult to see that this choice is 
inconsistent. Thus, we know from our discussion which led to 
(\ref{gammaR5^2=1}) 
that the choice
consistent with having D5-branes on top of an O5$^+$-plane is that given in
(\ref{gammaR5}), and not in (\ref{gammaR5i}). However, we can readily see
what goes wrong if we make this inconsistent choice 
in the language of the effective field theory as well. Thus, let us
assume for a moment that $\gamma_{R,5}$ is given by (\ref{gammaR5i}) even
though D5-branes are placed on top of an O5$^+$-plane. Then the gauge group
of the model would be $[U(8)\otimes {\bf Z}_2]_{99}\otimes U(8)_{55}$,
and the massless open string spectrum would read (here we have kept
anomalous $U(1)$'s for the convenience reasons that will become clear in a
moment):
\begin{eqnarray}
 2\times &&({\bf 28};{\bf 1})_{99}~,\\
 2\times &&({\bf 1};{\bf 36})_{55}~,\\
         &&({\bf 8}_+;{\bf 8})_{95}~,\\
         &&({\bf 8}_-;{\bf 8})_{95}~.
\end{eqnarray} 
Note that the 95 part of this spectrum now looks consistent. However, this
spectrum is actually anomalous once we take into account the massless
content of the closed string sector, which consists of the six dimensional
${\cal N}=1$ supergravity multiplet, 5 tensor
supermultiplets, and 16 hypermultiplets. Thus, for instance, the $R^4$ 
gravitational anomaly does not cancel in this model. This is due to the
fact that we have 55 hypermultiplets in ${\bf 36}$ (symmetric)
representation of $SU(8)$ instead of ${\bf 28}$ (antisymmetric), the reason
being that D5-branes now are at an O5$^+$-plane instead of an O5$^-$-plane.
Let us mention that we have kept the anomalous $U(1)$ factors merely for the
counting convenience - eventually they are broken, so we can drop them, but
then instead of 16 closed string hypermultiplets we would have only 14 as
two twisted hypermultiplets (or, more precisely, certain linear
combinations thereof) are eaten in the corresponding Higgs mechanism. 

{}Thus, the assumption that we can have both O5$^-$- and O5$^+$-planes 
at various fixed points of the conformal field theory orbifold $T^4/{\bf Z}_2$ 
does not seem to be self-consistent. This in accord with our discussion of
the role of the twisted $B$-flux in the collapsed ${\bf P}^1$'s at the
orbifold fixed points. In
particular, according to this discussion in the case of the  
conformal field theory orbifold $T^4/{\bf Z}_2$ we expect that only
O5$^-$-planes can be placed at the orbifold fixed points. The obstruction
for placing O5$^+$-planes at the fixed points comes precisely from the fact
that we have twisted $B$-flux. This, in turn, gives us a hint of how to
possibly remedy the situation in the case of the orientifold of 
Type IIB on $T^2/{\bf Z}_2$
with $B$-flux. More precisely, we can attempt to guess what the correct
background for such an orientifold might be (the orientifold of the
conformal field theory orbifold $T^4/{\bf Z}_2$ does not appear to be the
correct background). 

{}The key observation here is that to have an O5$^+$-plane at the orbifold
fixed point we must have {\em trivial} twisted $B$-flux in the
corresponding collapsed ${\bf P}^1$. However, if we simply turn off the
twisted $B$-flux, the corresponding background could no longer be described
within the world-sheet (that is, the conformal field theory)
approach. Indeed, in this case we have a true geometric ${\bf A}_1$
singularity at each of the fixed points with the twisted $B$-field turned
off. To avoid this, we would have to {\em blow up} the orbifold singularity
by giving a VEV to the corresponding twisted hypermultiplet. Note, however,
that if we first consider the $T^4/{\bf Z}_2$ orbifold with collapsed ${\bf
P}^1$'s and then orientifold, the fixed points where we have O5$^+$-planes
would give rise to twisted tensor multiplets only, so such a blow-up would
not be possible. However, we can proceed as follows. Consider Type IIB on
K3, where K3 is defined as follows. Consider the $T^4/{\bf Z}_2$ orbifold with
$B$-flux of rank $b=2$. At the 12 fixed points where we expect
O5$^-$-planes we have collapsed ${\bf P}^1$'s with half-integer twisted
$B$-flux, and these points locally can be described in the conformal field
theory. At the other 4 fixed points where we expect O5$^+$-planes we have
${\bf P}^1$'s of non-zero size but with trivial twisted $B$-flux. Moreover,
let us assume that the size of K3 is large compared with the size of these
blow-ups. Then we can indeed locally describe the other 12 fixed points in
the exactly solvable conformal field theory language of the ${\bf C}^2/{\bf
Z}_2$ orbifold. 
On the other hand, the 4 blown-up
fixed points without the twisted $B$-flux can no longer be described in
terms of an exactly solvable orbifold conformal field theory (albeit there
should exist some conformal field theory description of such a K3
which is not exactly
solvable but corresponds to some non-trivial sigma-model). Next, start from
Type IIB on the K3 surface we have just described, and consider its
$\Omega$ orientifold. At the 12 fixed points with the twisted $B$-flux we
have O5$^-$-planes, and these fixed points give rise to twisted
hypermultiplets, while at the 4 fixed points without the twisted $B$-flux
we have O5$^+$-planes, and these fixed points give rise to twisted tensor
multiplets. Since there are no twisted hypermultiplets at these 4 fixed
points, we cannot blow down the corresponding ${\bf P}^1$'s. In this sense,
the orientifolding procedure does not commute with the blowing-up
procedure. This, in turn, might signal that there could be a caveat in the
above discussion. In particular, it is not completely evident that the
boundary states at the 4 blown-up fixed points, if such are at all present,
carry the correct R-R charges to be interpreted as O5$^+$-planes. In fact,
technical issues in conformal field theories with boundaries might, at
least partially, be responsible for difficulties in making this point 
quantitatively more precise. Nonetheless, we can attempt to proceed further
in understanding the underlying qualitative picture 
assuming that the boundary states at the blown-up fixed points indeed
correspond to the O5$^+$-planes. 

{}What can we say about this orientifold? With the aforementioned
assumption, it is reasonable to assume that
if we place 16 D5-branes at one of the O5$^-$-planes, then we get the
spectrum described above, which (modulo the missing 99 ${\bf Z}_2$ discrete
symmetry) is the same as that given in \cite{KST}. This spectrum is
consistent, and with the aforementioned interpretation of the K3 background
might adequately describe physics at the corresponding point in the moduli
space. Once we move D5-branes off the O5$^-$-plane, the 55 gauge symmetry
in the bulk is $Sp(8)$ (or an appropriate subgroup thereof). The key
question, however, is what happens when we approach an O5$^+$-plane - after
all this was where we have encountered trouble in the above discussion in
the context of the conformal field theory orbifold. Note, however, that in
the case of the K3 surface under consideration the moduli space
corresponding to the motion of D5-branes is no longer flat as it was in the
case of the conformal field theory orbifold (more precisely, in the latter case
it is flat everywhere except for the fixed points). Because of the non-zero
size of the corresponding ${\bf P}^1$'s, D5-branes actually might not be
able to come on
top of the O5$^+$-planes (whose locations in K3 are given by points inside
of the blown-up ${\bf P}^1$'s). If so, this would avoid 
the contradiction we have
encountered in the conformal field theory orbifold case. That is, the 55 gauge 
symmetry would never be 
enhanced to $Sp(8)\otimes Sp(8)$ - in the bulk it is at most
$Sp(8)$, at an O5$^-$-plane it is $SU(8)$, while the points 
corresponding to D5-branes being on top of O5$^+$-planes might possibly be
thought of as being at infinite distance in the moduli space. If so, the
aforementioned K3 surface might indeed be the correct consistent background
for the corresponding orientifold with $B$-flux. Once again, however, it is
not completely clear how to make the above discussion quantitatively more
precise due to the fact that the conformal field theory corresponding to
such a K3 would not be exactly solvable\footnote{In the next section we
will discuss four dimensional orientifolds with ${\cal N}=2$ and ${\cal
N}=1$ supersymmetry where the aforementioned difficulties are avoided within
exactly solvable conformal field theory compactifications.}.

{}Before we summarize the findings of this subsection, we would like to give
two further pieces of evidence that in considering 
the $\Omega$ orientifold of Type IIB on 
the conformal field theory $T^4/{\bf Z}_2$ orbifold with $B$-flux one indeed
runs into various subtle inconsistencies. The first piece of such evidence
comes from studying the 99 sector moduli space in the above model. The
second piece of evidence comes from considering the second seemingly
consistent setup, namely, that with the $D_4$ type of Wilson lines
accompanied with the twisted Chan-Paton matrix $\gamma_{R,9}$ with
$\nu^2=+1$. 

{}Thus, so far we have focused on the 55 sector moduli space. Similar
considerations apply to the 99 sector. Thus, suppose we start from the
point in the moduli space where all 16 D5-branes are placed at an
O5$^-$-plane. At this point we have $[SU(8)\otimes {\bf
Z}_2]_{99}\otimes SU(8)_{55}$ gauge symmetry. We can Higgs the $SU(8)_{99}$
gauge group down to $Sp(8)_{99}$ or its appropriate subgroups, and this
Higgsing now corresponds to turning on Wilson lines. We can describe these
Wilson lines as in (\ref{gammaSi}), except that now we also have the
twisted Chan-Paton matrix $\gamma_{R,9}$. Thus, consider the following
choice of $\gamma_{S_a,9}$ and $\gamma_{R,9}$:
\begin{eqnarray}
 &&\gamma_{S_1,9}=\sigma_3\otimes\rho(a)\otimes I_8~,\\
 &&\gamma_{S_2,9}=\sigma_1\otimes\rho(a^{-1})\otimes I_8~,\\
 &&\gamma_{S_3,9}=i\sigma_2\otimes I_2\otimes I_8~,\\
 &&\gamma_{R,9}=I_2\otimes i\sigma_1 \otimes I_8~.
\end{eqnarray}   
Here $\rho(a)\equiv{\rm diag}(a,a^{-1})$, where $a$ is a complex
phase. For definiteness we have chosen
$\gamma_{S_3,9}=+\gamma_{S_1,9}\gamma_{S_2,9}$. Note that $\gamma_{R,9}$
commutes with $\gamma_{S_3,9}$. However, for $\gamma_{S_i,9}$, $i=1,2$,
we have:
\begin{equation}\label{conjugation}
 \gamma_{R,9} \gamma_{S_i,9}=\gamma_{S_i,9}^{-1}\gamma_{R,9}~.
\end{equation}
Note that for $a^2=1$ 
$\gamma_{R,9}$ actually commutes with $\gamma_{S_i,9}$.  

{}Before we proceed further, the following remarks are in order. First, 
for $a^2=1$ (as well as $a^2=-1$) the aforementioned Wilson lines can be
thought of in terms of the freely acting ${\bf Z}_2\otimes {\bf Z}_2$
orbifold (with discrete torsion). However, at generic points they
correspond to more general Wilson lines that do not possess a simple
(freely acting) orbifold description. However, at rational values of the
phase, say, $a=\exp(2\pi i/k)$, $k\in {\bf Z}$, such an orbifold
description does exist. Thus, for instance, let $k\in 2{\bf N}+1$. Then 
$\gamma_{S_1,9}^2$ generates a ${\bf Z}_k$ discrete group. In the language
of a freely acting orbifold this can be understood as having ${\bf Z}_k$
valued shifts of the torus lattice. Note that such a shift is mapped to its
inverse by the reflection of the torus coordinates described by the twist
$R$. This is precisely the reason why we have chosen the Chan-Paton
matrices $\gamma_{S_a,9}$ and $\gamma_{R,9}$ so that they satisfy the
conjugation relation (\ref{conjugation}). In fact, this conjugation
relation should also hold for generic (that is, irrational) phases as
well. To see this, consider a non-trivial Wilson line (which we
schematically write as a diagonal $N\times N$ matrix)
\begin{equation}
 W=\exp\left(i \int_{C} A\right)~,
\end{equation} 
where $A={\rm diag}(\theta_1,\dots,\theta_N)$ 
is a constant gauge field, and $C$ is a non-trivial 1-cycle. 
Now consider a ${\bf Z}_2$ orbifold action such that it reverses the
coordinate parametrizing $C$. Then the action of the ${\bf Z}_2$ orbifold 
generator $R$ on the Wilson line is given by (note that $R^2=1$)
\begin{equation}
 R W R = W^{-1}~,
\end{equation}
which follows from the fact that $A$ reverses its sign under the action of
$R$. This then implies that the conjugation relation (\ref{conjugation}) also
holds for generic values of the phase $a$.  

{}Next, note that at, say, $a=1$ the above choice of $\gamma_{S_a,9}$ and 
$\gamma_{R,9}$ is equivalent to that in (\ref{gammaSa9}) and
(\ref{gammaR9}). In particular, the 99 gauge group is $[SU(8)\otimes {\bf
Z}_2]_{99}$, while the 55 gauge group is $SU(8)_{55}$, and all the
D5-branes are placed at an O5$^-$-plane. Now let us continuously deform the
above Wilson lines from the point $a=1$ to, say, the point $a=i$. At this
point the Wilson lines $\gamma_{S_a,9}$ are of the $D_4$ type (while the
Wilson lines are of the $D_4^\prime$ type at $a=1$) - the eigenvalues of
all three matrices $\gamma_{S_a,9}$ are $\pm i$. So at this point the 99
gauge group would be $Sp(16)\otimes {\bf Z}_2$ 
if we did not have a further ${\bf Z}_2$
orbifold action. Let us see what the gauge group is once we perform the
${\bf Z}_2$ orbifold projection with respect to $\gamma_{R,9}$. Actually,
$\gamma_{R,9}$ does not commute with $\gamma_{S_i,9}$, $i=1,2$, at $a=i$,
in fact, they anticommute. Thus, we should find a basis where the twisted
Chan-Paton matrix commutes with the Wilson lines. Such a basis is given by
$\gamma_{S_a,9}$ together with $\gamma_{{\widetilde R},9}$, where
${\widetilde R}\equiv RS_3$, and we can choose 
\begin{equation}
 \gamma_{{\widetilde R},9}=\gamma_{R,9}\gamma_{S_3,9}=i\sigma_2\otimes
 i\sigma_1\otimes I_8~.  
\end{equation}
Note that the eigenvalues of $\gamma_{{\widetilde R},9}$ are no longer $\pm
i$ but $\pm 1$, which is consistent with the fact that for the $D_4$ type
of Wilson lines we must have $\nu^2=+1$. 
This then implies that the 99 gauge group is actually
$[Sp(8)\otimes Sp(8)\otimes {\bf Z}_2]_{99}$. 
On the other hand, the 55 gauge group is unchanged - 
it is still $SU(8)_{55}$. It is then not 
difficult to see that here we are running into a problem similar to that in
the case where we attempted to place D5-branes on top of an O5$^+$-plane -
the resulting spectrum in the 59 sector is anomalous. The reason for this
is that the ${\bf Z}_2$ projection in the 99 sector corresponds to the
twisted Chan-Paton matrix $\gamma_{{\widetilde R},9}$ with eigenvalues $\pm
1$, while that in the 55 sector corresponds to $\gamma_{R,5}$, whose
eigenvalues are $\pm i$. That is, here we have the same type of
inconsistency in the 59 sector as that encountered in the case of D5-branes
sitting on top of an O5$^+$-plane. Note that if we attempt to interpret
$\gamma_{R,9}$ (instead of $\gamma_{{\widetilde R},9}$) as the twisted
Chan-Paton matrix with respect to which we must perform the ${\bf Z}_2$
orbifold projection, then the latter might naively 
appear to be consistent in the 59 sector. However, it is not difficult to
see that having performed the ${\bf Z}_2$ orbifold projection in the 59 
sector with respect to $\gamma_{R,9}$ (together with $\gamma_{R,5}$), the
resulting 59 sector states would {\em not} transform in representations of the
$[Sp(8)\otimes Sp(8)\otimes {\bf Z}_2]_{55}$ gauge group, which is due to
the fact that $\gamma_{R,9}$ and $\gamma_{S_i,9}$ do not commute at this
point in the moduli space\footnote{We will encounter similar situations in
other models in the following subsections, where we will discuss this point
in more detail.}.    

{}Note that the aforementioned inconsistency arises if we choose 
the background to be the $\Omega$ orientifold of Type IIB on 
the conformal field theory
orbifold $T^4/{\bf Z}_2$ as in this case there is no obstruction to
continuously deforming the above Wilson lines from the point $a=1$ to the
point $a=i$. However, if we consider the partially blown-up K3 surface 
described above as the consistent
orientifold background, such a continuous deformation of
Wilson lines might no longer be 
possible as the corresponding moduli space is no
longer flat. As in the discussion of the 55 moduli space, however, it is
not clear how to make this point quantitatively precise. At any rate, if
the aforementioned K3 surface is indeed a consistent background for the
above orientifold, there would have to exist an obstruction to continuously
deforming the Wilson lines from the points with $a^2=+1$ to those with
$a^2=-1$, as well as an obstruction to placing D5-branes on top of an
O5$^+$-plane. If this is indeed the case, then the above $b=2$ model (where
the Wilson lines are of the $D_4^\prime$ type, $\nu^2=-1$, and all 16
D5-branes are placed at an O5$^-$-plane) would be consistent. In fact, then
we could also consider its $b=4$ counterpart as follows. Consider the
$\Omega$ orientifold of Type IIB
on K3, where K3 is a $T^4/{\bf Z}_2$ orbifold with $b=4$ $B$-flux, and 10
of the ${\bf Z}_2$ orbifold fixed points locally can be described in the
language of the conformal field theory orbifold ${\bf C}^2/{\bf Z}_2$,
while the other 6 fixed points are blown-up, and the corresponding twisted
$B$-flux is trivial. The former 10 fixed points are those at which we have
O5$^-$-planes, while the latter 6 fixed points are those at which we have
O5$^+$-planes. The $b=4$ $B$-flux can be described as follows. For the sake
of simplicity consider $T^4=T^2\otimes T^2$. On the first $T^2$ we have
non-commuting Wilson lines $\gamma_{S_i,9}$, while on the second $T^2$ we
have non-commuting Wilson lines $\gamma_{T_i,9}$ (these two sets of Wilson
lines, however, commute). Now consider the case where both sets of Wilson
lines are of the $D_4^\prime$ or $D_4$ type. Then the consistent choice for
the twisted Chan-Paton matrix $\gamma_{R,9}$ is that with $\nu^2=-1$. The
twisted Chan-Paton matrix $\gamma_{R,5}$ (which is an $8\times 8$ matrix as
we have 8 D5-branes in this case) 
is then also fixed. If we place all D5-branes at an
O5$^-$-plane, then the gauge group is $[U(4)\otimes {\bf
Z}_2\otimes {\bf Z}_2]_{99}\otimes U(4)_{55}$, and the open string sector
massless hypermultiplets are given by:
\begin{eqnarray}  
 2\times &&({\bf 6};{\bf 1})_{99}~,\\
 2\times &&({\bf 1};{\bf 6})_{55}~,\\
         &&({\bf 4}_{++};{\bf 4})_{95}~,\\
         &&({\bf 4}_{+-};{\bf 4})_{95}~,\\
         &&({\bf 4}_{-+};{\bf 4})_{95}~,\\
         &&({\bf 4}_{--};{\bf 4})_{95}~,
\end{eqnarray}
where the subscript $\pm\pm$ in the 95 sector refers to the 99 ${\bf
Z}_2\otimes {\bf Z}_2$ discrete gauge 
charges. This spectrum (together with that
from the closed string sector) is free of the irreducible $R^4$ and $F^4$
anomalies. The $U(1)$ factors are anomalous as usual, and are broken via
the generalized Green-Schwarz mechanism. Note that (modulo the missing 99
${\bf Z}_2\otimes {\bf Z}_2$ discrete gauge symmetry) this is the spectrum
of the $b=4$ ${\bf Z}_2$ model in \cite{KST}. 

{}Next, in the $b=2$ case, 
we would like to discuss the second {\em a priori} consistent
choice for the Wilson lines and the twisted Chan-Paton matrix, namely,
where the Wilson lines are of the $D_4$ type, and the twisted Chan-Paton
matrix is chosen such that $\nu^2=+1$. Note that in this case we must have
$\gamma_{R,5}^2=+1$, from which it follows that D5-branes must be placed at
an O5$^+$-plane. Note that just as in the previous case the conformal field
theory $T^4/{\bf Z}_2$ orbifold does not appear to be the correct
background for such an orientifold. In particular, in the conformal field
theory $T^4/{\bf Z}_2$ orbifold case we would have all the troubles we have
encountered for the previous choice of the Wilson lines and the twisted
Chan-Paton matrix. Thus, for instance, there would be no obstruction to
moving D5-branes from an O5$^+$-plane to an O5$^-$-plane, and at the latter
point in the moduli space the spectrum of the model would be
anomalous. However, in the present case (that is, where 
the Wilson lines are of the $D_4$ type, and the twisted Chan-Paton
matrix is chosen such that $\nu^2=+1$) there is an additional puzzling
issue. In particular, from our discussion of the role of the twisted
$B$-flux it actually follows that we could not have an O5$^+$-plane at a
fixed point (locally) corresponding to the conformal field theory orbifold
${\bf C}^2/{\bf Z}_2$. In fact, if we can at all have an O5$^+$-plane at a
${\bf Z}_2$ fixed point, we expect that the latter would have to be
blown-up, and the twisted $B$-field would have to be trivial. Moreover, in
this case it would be impossible to place D5-branes on top of such an
O5$^+$-plane. 

{}In the light of the above discussion, we would like to see whether we can
find any inconsistency in the model where the Wilson lines are of the $D_4$ 
type, the twisted Chan-Paton
matrix is chosen such that $\nu^2=+1$, and all D5-branes are placed at an
O5$^+$-plane. It is not difficult to see that the gauge group of this model
is $[Sp(8)\otimes Sp(8)\otimes {\bf Z}_2]_{99}\otimes [Sp(8)\otimes
Sp(8)]_{55}$, and the massless open string hypermultiplets are given by:
\begin{eqnarray}
 &&({\bf 8},{\bf 8};{\bf 1},{\bf 1})_{99}~,\\
 &&({\bf 1},{\bf 1};{\bf 8},{\bf 8})_{55}~,\\
 {1\over 2} &&({\bf 8}_+,{\bf 1};{\bf 8},{\bf 1})_{95}~,\\
 {1\over 2} &&({\bf 8}_-,{\bf 1};{\bf 8},{\bf 1})_{95}~,\\
 {1\over 2} &&({\bf 1},{\bf 8}_+;{\bf 1},{\bf 8})_{95}~,\\
 {1\over 2} &&({\bf 1},{\bf 8}_-;{\bf 1},{\bf 8})_{95}~, 
\end{eqnarray}   
where the subscript $\pm$ in the 95 sector refers to the 99 ${\bf Z}_2$
discrete gauge charge of the corresponding state. Note that the $R^4$
gravitational anomaly cancels in this model. However, the above spectrum is
not completely anomaly free - it contains half-hypermultiplets in real
representations\footnote{Note that the fundamental 
representation of a symplectic
gauge group is pseudoreal. However, here we have bifundamental
representations of a product symplectic group, and the former are real.}.
Note that had we not distinguished the 99 ${\bf Z}_2$ discrete gauge
quantum numbers, naively we might have thought that in the 95 sector 
we have one hypermultiplet in $({\bf 8},{\bf 1};{\bf 8},{\bf 1})$, and  
one hypermultiplet in $({\bf 1},{\bf 8};{\bf 1},{\bf 8})$, which would
give a consistent spectrum. This is, however, not the case here, and, once
again, the 99 ${\bf Z}_2$ discrete gauge symmetry indeed plays an important
role. Thus, the above spectrum with the $[Sp(8)\otimes Sp(8)\otimes 
{\bf Z}_2]_{99}\otimes [Sp(8)\otimes Sp(8)]_{55}$ gauge symmetry does not
appear to be completely consistent\footnote{The $Sp(8)^4$ model with the
above spectrum but with missing 99 ${\bf Z}_2$ discrete gauge symmetry was
originally constructed in \cite{PS} via the ``rational construction''
equivalent to the conformal field theory orbifold $T^4/{\bf Z}_2$ at the
special point in the moduli space of $T^4$'s corresponding to the $SO(8)$
symmetry. This model was discussed in the context of general $T^4/{\bf
Z}_2$ compactifications with $b=2$ $B$-flux in the second reference in
\cite{Ka3}, and more recently in \cite{An}. However, all of these
references missed the importance of the 99 
${\bf Z}_2$ discrete gauge symmetry, whose presence, as we have just
pointed out, leads to a subtle
inconsistency in the model.}, and, as we have discussed
above, this gives additional
evidence for the conclusion that we cannot have an O5$^+$-plane at the
conformal field theory ${\bf C}^2/{\bf Z}_2$ orbifold fixed
point\footnote{Note that the above conclusions also apply to the analogous 
$b=4$ model, which is constructed as follows. Consider the $(T^2\otimes
T^2)/{\bf Z}_2$ orbifold. The $b=4$ $B$-field can be described in terms of
the two sets of Wilson lines $\gamma_{S_i}$ and $\gamma_{T_i}$. Choose one of
these sets to be of the $D_4$ type, while the other set to be of the
$D_4^\prime$ type. Then it is not difficult to show that the consistent
choice for the twisted Chan-Paton matrix is such that
$\nu^2=+1$. Consequently, we must place D5-branes at an O5$^+$-plane in
this model. The gauge group of this model is $[Sp(4)\otimes Sp(4)\otimes
{\bf Z}_2\otimes {\bf Z}_2]_{99}\otimes [Sp(4)\otimes Sp(4)]_{55}$. In the
95 sector we again have half-hypermultiplets in real representations, which
are charged non-trivially under the 99 ${\bf Z}_2\otimes {\bf Z}_2$
discrete gauge symmetry.}.

{}Let us summarize the results of this subsection. We have considered the 
$\Omega$ orientifold of Type IIB on $T^4/{\bf Z}_2$ with non-zero
$B$-flux. The latter requires that $n_{f+}$ of the 16 O5-planes be of the
O5$^+$ type. However, the latter cannot be placed at conformal field
theory orbifold fixed points due to the non-zero twisted $B$-flux inside of
the corresponding collapsed ${\bf P}^1$'s (while it is precisely an
O5$^-$-plane that can be consistently placed at such a fixed point). 
A possible way around this would
be to blow up the orbifold fixed point and turn off the twisted
$B$-flux. The price one would have to pay for this, however, is that the
corresponding K3 no longer corresponds to an exactly solvable conformal
field theory. Because of this, it is not entirely clear whether the
corresponding models are completely consistent, albeit their massless
spectra appear to be.

{}We have presented various pieces of evidence that attempts to interpret
the aforementioned orientifolds in the context of the conformal field
theory $T^4/{\bf Z}_2$ orbifold with $B$-flux run into various
inconsistencies visible already at the massless level, in particular, in
the 59 sector. In fact, here we would like to suggest a simple geometric
interpretation of these inconsistencies. Note that a peculiar feature of 
the aforementioned orientifolds is that we have D9-branes wrapping a torus
(or, more precisely, an orbifold thereof) with $B$-flux together with
D5-branes transverse to this torus. On the one hand, the gauge bundles of
branes wrapped on such tori lack vector structure as the corresponding
Stieffel-Whitney class is non-vanishing. On the other hand, the presence of
D9- and D5-branes implies that we have the 59 sector, where the states
transform in the bifundamental representations of the gauge group. The
latter require non-trivial vector structure which is in conflict with the
lack of vector structure for the 99 gauge bundles. However, in the next 
section we will be able to avoid this difficulty in non-trivial {\em four} 
dimensional ${\cal N}=2$ and ${\cal N}=1$ supersymmetric
orientifolds with $B$-flux.

\subsection{The ${\bf Z}_3$ Models}

{}In this subsection we will consider orientifolds of Type IIB on 
${\bf R}^{1,5}\otimes (T^4/{\bf Z}_3)$ with non-zero $B$-flux. We will denote 
the generator of the ${\bf Z}_3$ orbifold group via $\theta$, whose action on 
the complex coordinates $z_1,z_2$ parametrizing $T^4$ is given by $\theta z_1= 
\omega z_1$, $\theta z_2=\omega^{-1} z_2$, where $\omega\equiv \exp(2\pi i/3)$.
In fact, for our purposes here it will suffice to consider $T^4=T^2\otimes 
T^2$,
where the first and the second 2-tori are parametrized by $z_1$ and $z_2$,
respectively.

{}The orientifolds we will discuss here are defined as follows. The orientifold
action is given by $\Omega J$, where $\Omega$ interchanges the left- and 
right-movers (and its action is the same as in the smooth K3 
case), while $J=J^\prime$, or $J=RJ^\prime$. Here $J^\prime$ acts as follows.
Its action on the untwisted sector fields is trivial, however, it 
interchanges the $\theta$ twisted sector with its inverse $\theta^{-1}$ twisted
sector. Thus, as explained in \cite{Po}, the $\Omega J^\prime$ orientifold is
precisely
that discussed in \cite{GJ}. The geometric meaning of the $J^\prime$ action was
discussed in detail in \cite{KST1}. Next, $R$ is the simultaneous reflection of
the coordinates on $T^4$: $Rz_{1,2}=-z_{1,2}$. Note that in the $\Omega 
J^\prime$ orientifold we expect 32 D9-branes but no D5-branes, while in the
$\Omega RJ^\prime$ orientifold we expect $32/2^{b/2}$ D5-branes but no 
D9-branes, where, as before, $b$ is the rank of the untwisted NS-NS $B$-field.

{}First, let us discuss the $\Omega J^\prime$ orientifold with $b=2$ (for 
definiteness we will assume that the $B$-flux is turned on in the direction of
the first $T^2$). In this
case we have 32 D9-branes but no D5-branes. The $B$-flux
can be described in terms of the freely acting ${\bf Z}_2\otimes 
{\bf Z}_2$ orbifold with discrete torsion acting on the first $T^2$, 
whose action on the D9-brane Chan-Paton charges is given by the matrices 
$\gamma_{S_a}$. In fact, we are now going to show that these matrices must be
of the $D_4$ type, that is, it would be inconsistent to choose them of the 
$D_4^\prime$ type. This can be seen as follows. First, the first $T^2$ (as
well as the second $T^2$) must have ${\bf Z}_3$ symmetry. This implies that
the vielbeins $e_i$ of the first $T^2$
are rotated by the action of $\theta$ as follows:
\begin{equation}
 \theta e_1=e_2~,~~~\theta e_2=e_3~,~~~\theta e_3=e_1~,
\end{equation}
where $e_3\equiv -e_1-e_2$. Next, note that $S_a$ are half-lattice shifts in 
the directions of $e_a$. This implies that we must have the following relations
between the ${\bf Z}_2\otimes {\bf Z}_2$ orbifold groups elements $S_a$ and the
${\bf Z}_3$ orbifold group element $\theta$:
\begin{equation}
 \theta S_1\theta^{-1}=S_2~,~~~\theta S_2\theta^{-1}=S_3~,~~~
 \theta S_3\theta^{-1}=S_1~.
\end{equation}
That is, the orbifold group elements $\theta$ and $S_a$ do {\em not} commute,
and, in fact, they generate a non-Abelian {\em tetrahedral} subgroup of 
$SU(2)$ (or a double cover thereof)\footnote{Note that the discrete torsion 
between the ${\bf Z}_2\otimes {\bf Z}_2$ elements is compatible with the 
${\bf Z}_3$ orbifold action.}. 
This then implies that the corresponding 
Chan-Paton matrices must satisfy the following relations:
\begin{equation}
 \gamma_\theta \gamma_{S_1}\gamma_\theta^{-1}=\gamma_{S_2}~,~~~
 \gamma_\theta \gamma_{S_2}\gamma_\theta^{-1}=\gamma_{S_3}~,~~~
 \gamma_\theta \gamma_{S_3}\gamma_\theta^{-1}=\gamma_{S_1}~. 
\end{equation}
This, in particular, implies that
\begin{equation}
 \gamma_{S_1}^2=\gamma_{S_2}^2=\gamma_{S_3}^2~.
\end{equation}
It then follows that the Wilson lines must be of the $D_4$ (and {\em not} 
$D_4^\prime$) type\footnote{Recall that $\gamma_{S_a}^2=\eta_{aa} I_{32}$,
where all three $\eta_{aa}$ equal $-1$ for the $D_4$ type of Wilson
lines, while two of them equal $+1$ and the third one equals $-1$ for the
$D_4^\prime$ type of Wilson lines.}.

{}Next, let us discuss solutions to the above conditions. Up to equivalent 
representations we can write them as follows (here we are using the fact that
$\gamma_{S_3}=\eta_{12}\gamma_{S_1}\gamma_{S_2}$):
\begin{eqnarray}
 &&\gamma_{S_1}=i\sigma_3\otimes I_{16}~,\\
 &&\gamma_{S_2}=i\sigma_2\otimes I_{16}~,\\
 &&\gamma_{S_3}=i\eta_{12}\sigma_1\otimes I_{16}~,\\
 &&\gamma_\theta =\xi_\theta\otimes \Gamma_\theta~,
\end{eqnarray} 
where 
\begin{equation}\label{xitheta}
 \xi_\theta\equiv \left(-{1\over 2}\right)\left[I_2+
 i\sigma_1+i\eta_{12}\sigma_2+i\eta_{12}\sigma_3\right]~,
\end{equation}
and the $16\times 16$ 
matrix $\Gamma_\theta$ is a diagonal matrix with non-zero entries
taking values $1,\omega,\omega^{-1}$. Note that the $2\times 2$ matrix 
$\xi_\theta$ has eigenvalues $\omega$ and $\omega^{-1}$, so that the matrix
$\gamma_\theta$ has eigenvalues taking values $1,\omega,\omega^{-1}$.

{}Note that in the above solution we still have to fix the form of the matrix
$\Gamma_\theta$. It is uniquely determined (up to equivalent representations) 
once we impose twisted tadpole cancellation conditions. The latter can be
deduced as follows. Note that the orientifold with $b=2$ $B$-flux is the
${\bf Z}_2\otimes {\bf Z}_2$ orbifold of the orientifold without $B$-flux. The 
trace of the twisted Chan-Paton matrix $\gamma_\theta$ is then fixed as in the
${\bf Z}_3$ model of \cite{GJ}:
\begin{equation}
 {\rm Tr}(\gamma_\theta)=8~.
\end{equation}
Traces of all the other twisted matrices such as $\gamma_{\theta S_a}$ are then
fixed unambiguously by the above relations. In particular:
\begin{equation}
 {\rm Tr}(\gamma_{\theta S_a})={\rm Tr}(\gamma_\theta \gamma_{S_a})=-
 \eta_{12} {\rm Tr}(\gamma_\theta)=\eta_{12} {\rm Tr}(\Gamma_\theta)~.
\end{equation}
Note that ${\rm Tr}(\Gamma_\theta)=-{\rm Tr}(\gamma_\theta)=-8$. This implies
that up to equivalent representations we have
\begin{equation}\label{Gammatheta}
 \Gamma_\theta={\rm diag}(\omega,\omega^{-1})\otimes I_8~.
\end{equation}
Note that in the diagonal basis $\gamma_\theta$ can be written as
$\gamma_\theta={\rm diag}(1,1,\omega,\omega^{-1})\otimes I_8$.

{}Next, let us determine the massless
spectrum of this model. First, let us discuss the 
closed string spectrum. It contains the six dimensional ${\cal N}=1$ 
supergravity multiplet, one untwisted tensor multiplet, 2 untwisted 
hypermultiplets, 9 twisted tensor multiplets and 9 twisted hypermultiplets. The
open string spectrum can be determined as follows. First, note that the 
$2\times 2$ matrices
\begin{eqnarray}
 \gamma_1\equiv i\sigma_3~,~~~\gamma_2\equiv i\sigma_2~,~~~\gamma_3\equiv
 i\eta_{12}\sigma_1~,~~~\gamma^{(k)}_\theta\equiv
 \omega^k \xi_\theta
\end{eqnarray}  
define three irreducible two dimensional representations of the tetrahedral 
subgroup ${\cal T}$ of $SU(2)$ labeled by the integer $k=0,1,2$. The 
aforementioned set of matrices $\gamma_{S_a},\gamma_\theta$ with 
$\Gamma_\theta$ given by (\ref{Gammatheta}) corresponds to taking 8 copies
of the two dimensional representation labeled by $k=1$ together with 8
copies of the two dimensional representation labeled by $k=2$. The ${\bf Z}_2
\otimes {\bf Z}_2$ orbifold action (that is, the action of the 
Chan-Paton matrices 
$\gamma_{S_a}$) breaks the original $SO(32)$ gauge group down to its $Sp(16)
\otimes {\bf Z}_2$ subgroup. The further ${\bf Z}_3$ orbifold action (that is,
the action of the twisted Chan-Paton matrix $\gamma_\theta$) breaks this gauge
group down to its $U(8)$ subgroup\footnote{Here we note that the ${\bf Z}_3$
twist breaks the ${\bf Z}_2$ discrete subgroup in the product $Sp(16)\otimes
{\bf Z}_2$. This point will become important in the next subsection, and we 
will discuss it there in more detail. However, the fate of the ${\bf Z}_2$ 
discrete gauge symmetry will not be important for our purposes here as no
massless states carry non-trivial ${\bf Z}_2$ charges.}. In fact, the open 
string massless spectrum is given by the ${\cal N}=1$ gauge 
supermultiplet in the
adjoint of $U(8)$ plus one hypermultiplet in ${\bf 36}$ of $U(8)$. (Note that
the adjoint of $Sp(16)$ decomposes in terms of the $U(8)$ representations as
follows: ${\bf 136}={\bf 64}(0)\oplus {\bf 36}(+2)\oplus{\overline{\bf 36}}
(-2)$, where we have given the $U(1)$ charges in parentheses, and the latter 
are normalized so that the fundamental of $SU(8)$ has the $U(1)$ charge $+1$.)
One can directly 
derive\footnote{We will give the details of this derivation in the next
subsection.} 
this spectrum by keeping the states invariant under the action 
of the non-Abelian tetrahedral group ${\cal T}$. In particular, note that the
action of the $\gamma_\theta$ matrix on the $Sp(16)$ part of the gauge group, 
which is given by the $16\times 16$ matrix $\Gamma_\theta$,
breaks $Sp(16)$ down to $U(8)$ as $\Gamma_\theta={\rm diag}(\omega,\omega^{-1})
\otimes I_8$.
Alternatively, we can use the
following trick. The open string partition function ${\cal Z}[{\cal T}]$ of 
the full ${\cal T}$ orbifold model can be expressed in terms of the partition
functions ${\cal Z}[{\bf Z}_2\otimes {\bf Z}_2]$, ${\cal Z}[{\bf Z}_3]$ and
${\cal Z}[1]$ as follows:
\begin{equation}
 {\cal Z}[{\cal T}]={\cal Z}[{\bf Z}_3]+{1\over 3}{\cal Z}[{\bf Z}_2\otimes
 {\bf Z}_2]-{1\over 3}{\cal Z}[1]~,
\end{equation}  
where ${\cal Z}[1]$ is the partition function of the model corresponding to the
toroidal compactification {\em without} the $B$-flux (this model has 
${\cal N}=2$ supersymmetry and $SO(32)$ gauge group), ${\cal Z}[{\bf Z}_2
\otimes {\bf Z}_2]$ is the partition function of the model corresponding to the
toroidal compactification with $b=2$ $B$-flux, that is, the ${\bf Z}_2\otimes 
{\bf Z}_2$ freely acting orbifold model (this model has ${\cal N}=2$ 
supersymmetry and $Sp(16)\otimes {\bf Z}_2$ gauge group), and 
${\cal Z}[{\bf Z}_3]$ is the partition function of the model corresponding to 
the $T^4/{\bf Z}_3$ compactification {\em without} the $B$-flux (this is the
${\bf Z}_3$ model of \cite{GJ} with ${\cal N}=1$ supersymmetry, $SO(16)\otimes
U(8)$ gauge group and massless hypermultiplets in $({\bf 16},{\bf 8})\oplus
({\bf 1},{\bf 28})$ of $SO(16)\otimes U(8)$). If we now write the spectra of 
the ${\cal N}=2$ models in the ${\cal N}=1$ language, we can then read off the 
numbers of gauge bosons and massless hypermultiplets in the full ${\cal T}$ 
orbifold model from ${\cal Z}[{\cal T}]$ defined as above. 

{}Here we note that the above closed plus open string spectrum, which is the 
same as that given in \cite{KST}, is free of
the irreducible $R^4$ and $F^4$ anomalies. The $U(1)$ factor is anomalous as 
usual, and is broken via the generalized Green-Schwarz mechanism in a way 
similar to the ${\bf Z}_3$ model of \cite{GJ} without the $B$-flux.

{}Next, let us discuss the $\Omega J^\prime$ orientifold model with $b=4$ 
$B$-flux. The latter can be described in terms of two ${\bf Z}_2\otimes 
{\bf Z}_2$ freely acting orbifolds acting on the first respectively second 
$T^2$. We can choose the corresponding Chan-Paton matrices as follows:
\begin{eqnarray}
 &&\gamma_{S_1}=i\sigma_3\otimes I_2\otimes I_8~,\\
 &&\gamma_{S_2}=i\sigma_2\otimes I_2\otimes I_8~,\\
 &&\gamma_{S_3}=i\eta_{12}\sigma_1\otimes I_2\otimes I_8~,\\
 &&\gamma_{T_1}=I_2\otimes i\sigma_3\otimes I_8~,\\
 &&\gamma_{T_2}=I_2\otimes i\sigma_2\otimes I_8~,\\
 &&\gamma_{T_3}=I_2\otimes i\eta^\prime_{12}\sigma_1\otimes I_8~,\\ 
 &&\gamma_\theta=\xi_\theta\otimes
 \xi^\prime_\theta \otimes {\widetilde \Gamma}_\theta~, 
\end{eqnarray}  
where $\xi^\prime_\theta$ is given by the the same expression 
(\ref{xitheta}) with $\eta^\prime_{12}$ instead of $\eta_{12}$. Note
the $4\times 4$ matrix $\xi_\theta\otimes \xi^\prime_\theta$ has
eigenvalues $1,1,\omega,\omega^{-1}$. This then implies that to satisfy the 
twisted tadpole cancellation condition ${\rm Tr}(\gamma_\theta)=8$, we must 
take the $8\times 8$ matrix ${\widetilde \Gamma}_\theta$ to be the identity
matrix $I_8$. It is then not difficult to see that the gauge group of this 
model is $SO(8)$, and we have no massless hypermultiplets in the open string 
spectrum\footnote{Note that in the diagonal basis the $16\times 16$ matrix
$\xi_\theta^\prime\otimes {\widetilde \Gamma}_\theta$ 
can be written as ${\rm diag}(\omega,
\omega^{-1})\otimes I_8$, which is consistent with $\Gamma_\theta$ given in
(\ref{Gammatheta}).}. 
In particular, note that the Wilson lines $\gamma_{S_a}$ and 
$\gamma_{T_a}$ break the original $SO(32)$ gauge group down to $SO(8)
\otimes ({\bf Z}_2)^2$. The action of the ${\bf Z}_3$ orbifold group on the
$SO(8)$ part of the gauge group is trivial\footnote{As in the $b=2$ case, 
however, it breaks the $({\bf Z}_2)^2$ discrete gauge symmetry.} 
as it is given by ${\widetilde
\Gamma}_\theta=I_8$. 
The closed string spectrum is the same as in the $b=2$ model. Note 
that the spectrum of the $b=4$ model, which is the same as that given in
\cite{KST}, is free of the irreducible $R^4$ and $F^4$ anomalies.

{}Now we would like to discuss the $\Omega R J^\prime$ orientifolds with 
$B$-flux. Let us start with the $b=2$ case. As before, let us assume that the
$B$-field is turned on in the direction of the first $T^2$. Let $e_i$ be
the vielbeins of the first $T^2$, and $d_i$ be the vielbeins of the second 
$T^2$. As before, we will use the notation $e_3\equiv -e_1-e_2$, 
$d_3\equiv -d_1-d_2$. In this model we have 16 O5-planes located at the 16
points fixed under the action of the reflection $R$. These fixed points are
located at $(0,0),(0,d_a/2),(e_a/2,0),(e_a/2,d_a/2)$. Note that 12 out of these
O5-planes must be of the O5$^-$ type, while 4 must be of the O5$^+$ type. 
Together with the requirement that the entire background be ${\bf Z}_3$ 
symmetric (so that the further ${\bf Z}_3$ orbifold is consistent), this
uniquely fixes the allowed distribution of O5-planes. Thus, the O5-planes
located at the fixed points $(0,0),(0,d_a/2)$ must be of the O5$^+$ type, while
the rest of the O5-planes are of the O5$^-$ type.

{}In this orientifold we have 16 D5-branes. The latter must be distributed in
a ${\bf Z}_3$ 
symmetric fashion. First, let us consider placing all 16 D5-branes
on top of the O5$^+$-plane at the origin $(0,0)$ of $T^4/{\bf Z}_3$. Before the
${\bf Z}_3$ orbifold projection the gauge group is $Sp(16)$. The action of the 
${\bf Z}_3$ orbifold on the Chan-Paton charges is given by the twisted 
$16\times 16$ Chan-Paton matrix $\gamma_\theta$. The only constraint on this
matrix comes from the twisted tadpole cancellation condition, which in this 
case reads
\begin{equation}
 {\rm Tr}(\gamma_\theta)=-8~.
\end{equation} 
Note that this is the same twisted tadpole cancellation condition as in the 
case of D9-branes wrapped on $T^4/{\bf Z}_3$ without $B$-field except for
the extra minus sign\footnote{Here we note that generally twisted tadpole
cancellation conditions for D5-branes transverse to the ${\bf C}^2/{\bf Z}_M$
orbifold are different from those for D9-branes with the ${\bf C}^2/{\bf Z}_M$
orbifold in their world-volumes. However, for a particular case of the 
${\bf Z}_3$ orbifold group they actually coincide with the overall sign 
depending on the type of the corresponding orientifold plane.}. 
This minus sign is, in fact, due to the O5-plane here
being of the O5$^+$ type (while the twisted tadpole cancellation condition 
in the case of D9-branes was derived in \cite{GJ} for the configuration 
involving the O9$^-$-plane). The above tadpole cancellation condition
uniquely fixes the twisted Chan-Paton matrix (up to equivalent 
representations):
\begin{equation}
 \gamma_\theta={\rm diag}(\omega,\omega^{-1})\otimes I_8~.
\end{equation}
Thus, just as in the case of $b=2$ $\Omega J^\prime$ orientifold, this model
with D5-branes has $U(8)$ gauge group and one massless hypermultiplet in 
${\bf 36}$ of $U(8)$. (The closed string sector is the same as in the $\Omega
J^\prime$ orientifold.) Note that if we attempted to place some D5-branes in
the bulk or at other O5-planes, we would not have been able to cancel twisted
tadpoles. Indeed, D5-branes away from the origin must be placed in a
${\bf Z}_3$ symmetric fashion, that is, the number of D5-branes away from the
origin must be a multiple of 3. These branes are then permuted by the action
of the ${\bf Z}_3$ orbifold, and the corresponding part of the twisted 
Chan-Paton matrix $\gamma_\theta$ is traceless\footnote{Note that this
is the case even if we place D5-branes (in a fashion compatible with the
action of the $\Omega R J^\prime$ orientifold) 
at ${\bf Z}_3$ orbifold fixed points 
not fixed under $R$. That is, the corresponding part of the twisted Chan-Paton
matrix $\gamma_\theta$ must be traceless to satisfy tadpole cancellation
conditions.}. 
The part corresponding to the
D5-branes at the origin then cannot have trace equal $-8$, so that the
twisted tadpole cancellation conditions cannot be satisfied. The point we have 
just discussed can be understood in terms of the effective field theory 
language as follows. Note that moving D5-branes off the O5$^+$-plane 
(located at the origin) into the
bulk would correspond to giving the appropriate VEV to a massless 
hypermultiplet. The only massless hypermultiplet in this model (with all 
D5-branes at the origin) is in ${\bf 36}$ of $U(8)$. If we could give a VEV to
this hypermultiplet, it would break $U(8)$ down to $SO(8)$. (Note that under 
the breaking $SU(8)\rightarrow SO(8)$ we have ${\bf 36}={\bf 1}+{\bf 35}$.)
However, to satisfy the D-flatness conditions we must have at least two 
hypermultiplets in ${\bf 36}$ of $U(8)$, so that the aforementioned Higgsing is
not possible\footnote{One way to see this is as follows. Note that the $U(1)$
subgroup of the $U(8)$ gauge group is anomalous. When all D5-branes are
on top of the O5$^+$-plane at the origin of $T^4/{\bf Z}_3$, this $U(1)$ 
factor is broken via the generalized Green-Schwarz mechanism involving the
twisted closed string sector hypermultiplet coming from the ${\bf Z}_3$ 
fixed point at the origin. However, if we move D5-branes away from the
origin, the $U(1)$ breaking can no longer involve this twisted 
hypermultiplet - the latter is localized at the corresponding fixed point. 
In fact, the singlet of $SO(8)$ in the decomposition ${\bf 36}={\bf 1}+
{\bf 35}$ under $SU(8)\rightarrow SO(8)$ carries a non-zero $U(1)$ charge 
(which is equal $+2$ in the aforementioned normalization). This is precisely
the singlet whose VEV would measure the separation between the O5$^+$-plane at
the origin and the D5-branes in the bulk had the Higgsing been possible. 
However, in this case this singlet would also have to be the one eaten in 
Higgsing the $U(1)$ factor, so that we would have only the $SO(8)$ gauge bosons
but no massless hypermultiplets coming from D5-branes in the bulk. This, 
however, would mean that there is no modulus corresponding to the separation
between the O5$^+$-plane and D5-branes. We, therefore, conclude that the
aforementioned Higgsing is indeed impossible.}. 
This is in accord with the fact that no distribution of D5-branes
in the bulk giving rise to the $SO(8)$ gauge symmetry could possibly be 
${\bf Z}_3$ symmetric. Thus, D5-branes are {\em stuck} at the O5$^+$-plane 
located at the origin of $T^4/{\bf Z}_3$ in this model.

{}Finally, let us discuss the $b=4$ $\Omega R J^\prime$ orientifold. In this
case we have 10 O5$^-$-planes and 6 O5$^+$-branes, and the unique distribution
of O5-planes compatible with the ${\bf Z}_3$ symmetry is the following:
those at the fixed points $(0,0),(e_a/2,d_a/2)$ are of the O5$^-$ type, while
those at the fixed points $(0,d_a/2),(e_a/2,0)$ are of the O5$^+$ type.
In this case we have only 8 D5-branes all of which must be placed at the
origin, so that the twisted tadpole cancellation condition
\begin{equation}
 {\rm Tr}(\gamma_\theta)=8
\end{equation} 
can be satisfied. Note that the sign in this case is plus instead of minus as 
in the $b=2$ case as here the O5-plane at which the D5-planes are placed is
of the O5$^-$ type. The unique solution to the above tadpole condition is
$\gamma_\theta=I_8$, so that the gauge group of this model is $SO(8)$ with
no massless hypermultiplets in the open string sector. The closed string 
sector spectrum is the same as in the $b=2$ case. In fact, the massless 
spectrum of this model is the same as that of the $\Omega J^\prime$ 
orientifold with $b=4$. Note that, just as in the $b=2$ case, 
in the $b=4$ model D5-branes are also stuck at the origin of 
$T^4/{\bf Z}_3$ - there are no massless open string hypermultiplets in this 
model.  

{}Before we end this subsection let us note that in the case of $\Omega 
J^\prime$ orientifolds the inability to Higgs the gauge group is interpreted 
as impossibility of turning on Wilson lines compatible with the ${\bf Z}_3$
symmetry in such a way that all tadpoles are canceled.

\subsection{The ${\bf Z}_6$ Models}    

{}In this subsection we will discuss the $\Omega J^\prime$ orientifolds of
Type IIB on $T^4/{\bf Z}_6$ with $b=2,4$. As before, we will assume that 
$T^4=T^2\otimes T^2$. Let $g$ be the generator of ${\bf Z}_6$. Then, since 
${\bf Z}_6\approx{\bf Z}_2\otimes {\bf Z}_3$, we can write $g=R\theta$, where
$R$ and $\theta$ are the generators of the ${\bf Z}_2$ and ${\bf Z}_3$ 
subgroups, respectively, with the following actions on the coordinates
$z_1,z_2$ parametrizing the two 2-tori: $Rz_{1,2}=-z_{1,2}$, $\theta z_1=
\omega z_1$, $\theta z_2=\omega^{-1} z_2$ ($\omega\equiv \exp(2\pi i/3)$). 
The action of $J^\prime$ on the
closed string untwisted as well as $R$ twisted sector fields is trivial, while
$J^\prime$ interchanges the $\theta$ twisted sector with the $\theta^{-1}$
twisted sector, as well as the $g$ twisted sector with the $g^{-1}$ twisted 
sector.

{}To begin with let us discuss the closed sting sector of the above 
orientifold.
The untwisted sector gives rise to the six dimensional ${\cal N}=1$ 
supergravity multiplet plus one tensor multiplet and 2 hypermultiplets.
The $g$ and $g^{-1}$ twisted sectors together give rise to one tensor multiplet
and one hypermultiplet. The $\theta$ and $\theta^{-1}$ twisted sectors together
give rise to 5 tensor multiplets and 5 hypermultiplets. As to the $R$ twisted
sector, we must consider $b=0,2,4$ cases separately as the $\Omega$ projection
acts differently on the corresponding fixed points for different values of $b$.

{}In the $b=0$ case before the ${\bf Z}_3$ projection in the $R$ twisted 
sector we have 16 fixed points. At all of these fixed points we have 
O5$^-$-planes, so that each of them gives rise to a hypermultiplet (but no 
tensor multiplets). Note that the fixed point at the origin is invariant under
the ${\bf Z}_3$ twist $\theta$. The other 15 fixed points fall into 5 distinct
groups, each group containing 3 fixed points which are permuted by the action
of $\theta$. In each of these 5 groups we can form one linear combination of
the corresponding 3 fixed points which is invariant under $\theta$. Thus,
we have total of 6 hypermultiplets coming from the $R$ twisted sector - one
from the origin, and the other 5 from the rest of the fixed points.

{}Next, let us consider the $b=2$ case. Here we have 12 O5$^-$-planes and 
4 O5$^+$-planes. At the fixed point at the origin we have an O5$^+$-plane.
Thus, this fixed point gives rise to a tensor multiplet. The other
3 fixed points
at which we have O5$^+$-planes together give rise to another tensor multiplet.
Finally, the 12 fixed points at which we have O5$^-$-planes together
give rise to 4 hypermultiplets. Thus, the $R$ twisted sector gives rise to 4
hypermultiplets and 2 tensor multiplets in the $b=2$ case.

{}In the $b=4$ case we have 10 O5$^-$-planes and 6 O5$^+$-planes. At the fixed
point at the origin we have an O5$^-$-plane. This fixed point, therefore, 
gives rise to a hypermultiplet. The other 9 fixed points at which we have
O5$^-$-planes together give rise to 3 additional hypermultiplets. Finally,
the 6 fixed points at which we have O5$^+$-planes together give rise to
2 tensor multiplets. Thus, the $R$ twisted sector gives rise to 4 
hypermultiplets and 2 tensor multiplets in the $b=4$ case, which is the same 
as in the $b=2$ case\footnote{This corrects the error in \cite{KST}, where
the $R$ twisted sector in the $b=2,4$ ${\bf Z}_6$ models 
was thought to give rise to 6 hypermultiplets and no
tensor multiplets.}.  

{}Next, let us discuss the open string sector in the $b=2$ model. 
In this model we have 32 D9-branes and 16 D5-branes.
Note that
from our discussion of the corresponding ${\bf Z}_3$ model it follows that
all D5-branes in this models must be placed at the O5$^+$-plane at the
origin of $T^4/{\bf Z}_6$. This then, following our discussion in subsection A,
implies that the twisted Chan-Paton matrix $\gamma_{R,5}$ must have eigenvalues
$\pm 1$, and so must the matrix $\gamma_{R,9}$. From this it follows that the
Wilson lines in the 99 sector must be of the $D_4$ type, which is consistent
with our discussion in subsection B\footnote{These points
were missed in \cite{KST} in the discussion of this model. There it was
erroneously assumed that the O5-plane at the origin, at which all D5-branes
were placed, is of the O5$^-$ type. Consequently, the matrices $\gamma_{R,5}$
and $\gamma_{R,9}$ were assumed to have eigenvalues $\pm i$, and, in the
language we are using here, the Wilson lines in the 99 sector were assumed to
be of the $D_4^\prime$ type. From the above discussions it should be clear that
such a setup would be inconsistent as it is not even ${\bf Z}_3$ symmetric
(so that the ${\bf Z}_3$ orbifolding procedure would be inconsistent).
In fact, it is not difficult to show that with these
assumptions the corresponding massless spectrum would have to be anomalous
(as some of the tadpoles would not be canceled). The corresponding
spectrum given in \cite{KST}, however, is free of, say, the $R^4$ anomaly. One
of the errors made in \cite{KST} that had lead to this seemingly consistent 
spectrum, as we have already mentioned, was the incorrect computation of the
number of twisted tensor multiplets in the closed string spectrum. Another 
error, related to the multiplicity of states,  
was made in the discussion of the 59 sector. In fact, this point is rather 
non-trivial, and we will discuss it in more detail in a moment. Finally, the 
$\theta$ projection was carried out erroneously in \cite{KST}, which was 
already noticed in \cite{BST}. All these 
errors added up to give the erroneous spectrum reported in \cite{KST}.}. 

{}Here we can ask whether the $b=2$ ${\bf Z}_6$ model is consistent once we
make the aforementioned choices. First, recall that the $b=2$ ${\bf Z}_2$ model
with the Wilson lines of the $D_4$ type (and the twisted Chan-Paton matrices 
with $\nu^2=+1$ - see subsection A) suffers from the presence of 
half-hypermultiplets in real representations in the 59 sector. The 
$b=2$ ${\bf Z}_6$ model, therefore, is expected to have a similar problem as 
well. However, as we have already mentioned, the 99 ${\bf Z}_2$ discrete gauge
symmetry is broken by the ${\bf Z}_3$ twist. This might at first seem to imply 
that in the ${\bf Z}_6$ model unlike the ${\bf Z}_2$ model we might be able
to avoid the difficulty with the 59 half-hypermultiplets. In fact, if this were
so, then this model at first might seem to be consistent even for the conformal
field theory orbifold - recall from subsection B that in the $b=2$ 
$\Omega R J^\prime$ ${\bf Z}_3$ model we cannot move D5-branes away from
the O5$^+$-plane at the origin of K3 (and, similarly, in the $b=2$ $\Omega 
J^\prime$ ${\bf Z}_3$ model we cannot Higgs the 99 gauge group by turning on
Wilson lines). This then implies that in the $b=2$ ${\bf Z}_6$ model
{\em a priori} we do not have one of the problems we have encountered in the 
corresponding ${\bf Z}_2$ model, in particular, that related to the 
inconsistencies arising once we move D5-branes from an O5$^+$-plane to an 
O5$^-$-plane, or {\em vice-versa}. However, such a conclusion would immediately
run into a puzzle with our discussion of the role of the twisted 
$B$-flux - recall that we do {\em not} expect to be able to consistently 
have O5$^+$-planes within the conformal field theory orbifold 
if the orbifold group contains the ${\bf Z}_2$ generator 
$R$ (albeit, O5$^+$-planes are perfectly consistent with conformal field
theory orbifolds of odd order). In fact, as we will see in a moment, in the
$b=2$ ${\bf Z}_6$ model there is indeed a subtle inconsistency in the 59 
sector\footnote{It is then 
not difficult to show that similar conclusions hold for 
the $b=4$ ${\bf Z}_6$ model as well.}.

{}Thus, let us understand the 59 sector in this model. In fact, to understand
the point we would like to make here it suffices to consider the 59 sector
before the ${\bf Z}_2$ orbifold projection. We can alternatively view this
as introducing D5-brane probes \cite{Do} in the $b=2$ $\Omega J^\prime$ ${\bf
Z}_3$ model. Note that the 
59 sector states are in bifundamental representations
of the 55 and 99 gauge groups. In fact, for our purposes here the precise 55
quantum numbers are not going to be relevant. This is related to the fact that
in the 55 sector we have just the ${\bf Z}_3$ orbifold projection, which is
straightforward to carry out. However, in the 99 sector we have the additional
projections coming from the ${\bf Z}_2\otimes {\bf Z}_2$ freely acting 
orbifold. Thus, we
would like to understand how ${\bf 32}$ of $SO(32)$ decomposes under the
gauge group left unbroken after the ${\bf Z}_2\otimes {\bf Z}_2$ freely 
acting orbifold as well as the ${\bf Z}_3$ orbifold projections\footnote{Note 
that before any of the orbifold projections the 95 sector states are in the
following half-hypermultiplet of $SO(32)_{99}\otimes Sp(16)_{55}$: 
${1\over 2}({\bf 32};{\bf 16})$.}. 
Before we do this, however, it is instructive
to consider the analogous decomposition for the adjoint of $SO(32)$. Note 
that the Wilson line $\gamma_{S_1,9}$ breaks $SO(32)$ down to $U(16)$. On the 
other hand, the twisted Chan-Paton matrix $\gamma_{\theta,9}$ breaks $SO(32)$
down to $SO(16)\otimes U(8)$. Now, the maximal common subgroup of
$U(16)$ and $SO(16)\otimes U(8)$ is $U(8)\otimes U(8)$. Under the breaking
$SO(32)\supset U(8)\otimes U(8)$ the adjoint of $SO(32)$ decomposes as follows:
\begin{eqnarray}
 {\bf 496}=&&({\bf 64},{\bf 1})_{1,1}\oplus({\bf 1},{\bf 64})_{1,1}\oplus
 ({\bf 8},{\overline {\bf 8}})_{1,\omega^{-1}}\oplus
 ({\overline {\bf 8}},{\bf 8})_{1,\omega}\oplus\nonumber\\
           &&({\bf 28},{\bf 1})_{-1,1}\oplus ({\bf 1},{\bf 28})_{-1,1}
               \oplus ({\bf 8},{\bf 8})_{-1,\omega}\oplus\nonumber\\
           &&({\overline {\bf 28}},{\bf 1})_{-1,1}\oplus
              ({\bf 1},{\overline {\bf 28}})_{-1,1}
               \oplus ({\overline {\bf 8}},
                      {\overline {\bf 8}})_{-1,\omega^{-1}}~, 
\end{eqnarray} 
where the subscript indicates the ${\bf Z}_2$ valued phase due to the 
$\gamma_{S_1,9}$ projection as well as the ${\bf Z}_3$ valued phase due to the
$\gamma_{\theta,9}$ projection. The states with the ${\bf Z}_2$ phase $-1$
are all heavy, so that the massless states all have the ${\bf Z}_2$ phase $+1$.
Such states with the ${\bf Z}_3$ phase 1 are the gauge bosons of $U(8)\otimes
U(8)$, while the states with the ${\bf Z}_3$ phases $\omega$ and $\omega^{-1}$
combine into one massless hypermultiplet in $({\bf 8},{\overline {\bf 8}})$ of 
$U(8)\otimes U(8)$. 

{}Next, the gauge group left unbroken after the full ${\cal T}$ orbifold
projection can be determined as follows. The action of the second 
Wilson line $\gamma_{S_2,9}$ amounts to permuting the two $U(8)$ subgroups in
$U(8)\otimes U(8)$ (left unbroken by $\gamma_{S_1,9}$ and $\gamma_{\theta,9}$)
accompanied by the complex conjugation. Thus, for instance, $({\bf 8},{\bf 1})$
of $U(8)\otimes U(8)$ is mapped to $({\bf 1},{\overline {\bf 8}})$ by the 
action of $\gamma_{S_2,9}$. This implies that the final unbroken gauge group is
$U(8)$, and the massless matter consists of one hypermultiplet in ${\bf 36}$ of
$U(8)$. Note that normally we would expect the appearance of the ${\bf Z}_2$ 
discrete gauge subgroup in the breaking $U(8)\otimes U(8)\rightarrow 
U(8)_{\rm diag}\otimes {\bf Z}_2$. However, as we will show in a moment, this 
${\bf Z}_2$ discrete gauge group is actually broken in the case under 
consideration. Note that it is the 59
sector massless states that are expected to carry non-trivial
99 ${\bf Z}_2$ discrete gauge quantum 
numbers. However, as we will see momentarily, the 59 sector states in this 
model do not carry well defined gauge quantum numbers at all.

{}To see this, let us discuss the decomposition of ${\bf 32}$ of $SO(32)$
under $SO(32)\rightarrow U(8)\otimes U(8)\rightarrow U(8)$. Under the
first breaking we have:
\begin{equation}\label{32}
 {\bf 32}=({\bf 8},{\bf 1})_1\oplus ({\overline {\bf 8}},{\bf 1})_1\oplus
 ({\bf 1},{\bf 8})_\omega\oplus ({\bf 1},{\overline{\bf 8}})_{\omega^{-1}}~.
\end{equation} 
Here we have only shown the ${\bf Z}_3$ valued phases due to the 
$\gamma_{\theta,9}$ projection. In fact, we did not give the 
phases (which are actually ${\bf Z}_4$ valued) 
due to the $\gamma_{S_1,9}$ projection for the reason that the
Wilson lines (more precisely, the freely acting ${\bf Z}_2\otimes {\bf Z}_2$ 
orbifold generators) do {\em not} act in the 59 sector (just as they do not 
act in the 55 sector). One way to see this is to note that the Wilson lines
can only act on states with Kaluza-Klein momenta (but not windings) coming 
from the compact directions. This can also be seen by noting that continuously 
turning on Wilson lines (which is equivalent to Higgsing the 99 gauge group 
by giving VEVs to the 99 hypermultiplets) in, say, the $b=0$ ${\bf Z}_2$ model
does {\em not} change the number of 59 hypermultiplets. In fact, this is 
precisely the key reason for the problem we are going to point out next. Note
that under the action of $\gamma_{S_2,9}$ the state $({\bf 8},{\bf 1})_1$
is mapped to the state $({\bf 1},{\overline{\bf 8}})_{\omega^{-1}}$, and, 
similarly, the state $({\overline {\bf 8}},{\bf 1})_1$ is mapped to the state
$({\bf 1},{\bf 8})_\omega$. Thus, the linear combinations that carry well 
defined gauge quantum numbers under the unbroken $U(8)$ gauge group are 
given by:
\begin{eqnarray}
 &&|{\bf 8}\rangle_\pm\equiv{1\over\sqrt{2}}\left(|({\bf 8},{\bf 1})_1\rangle
 \pm |({\bf 1},{\overline{\bf 8}})_{\omega^{-1}}\rangle\right)~,\\
 &&|{\overline {\bf 8}}
 \rangle_\pm\equiv{1\over\sqrt{2}}\left(|({\overline {\bf 8}},{\bf 1})_1
 \rangle
 \pm |({\bf 1},{\bf 8})_\omega \rangle\right)~,
\end{eqnarray} 
where the subscript on the left hand side indicates the ${\bf Z}_2$ gauge
quantum numbers with the ${\bf Z}_2$ subgroup arising in the breaking
$U(8)\otimes U(8)\rightarrow U(8)_{\rm diag}\otimes {\bf Z}_2$. Note, however,
that the states $|{\bf 8}\rangle_\pm$ and $|{\overline {\bf 8}} \rangle_\pm$
do {\em not} carry well defined ${\bf Z}_3$ quantum numbers. This implies that
if in the 59 sector we perform the ${\bf Z}_3$ orbifold projection by keeping
the ${\bf Z}_3$ invariant states, then the latter will not have consistent 
couplings to the $U(8)$ gauge bosons. If instead we keep the above linear 
combinations with consistent couplings to the $U(8)$ gauge bosons, then these
states will be incompatible with the ${\bf Z}_3$ orbifold projection. Either 
way we have an inconsistency in the 59 sector of this model, which, in turn, is
consistent with our discussions in subsection A\footnote{Here the following 
remark is in order. Consider probe D1-branes in the $\Omega J^\prime$ 
${\bf Z}_3$ models with $B$-flux. Then in the 19 open string sector we would
encounter the same problem as that we have just discussed for D5-brane probes.
This might signal that there could be a non-perturbative inconsistency in
these ${\bf Z}_3$ models, which would have to be visible in the dual heterotic
compactification. In particular, on the heterotic side this inconsistency 
might manifest itself via the world-sheet theory of the fundamental heterotic
string being inconsistent. Note, however, that such a problem is absent in 
the $\Omega R J^\prime$ ${\bf Z}_3$ orientifolds with $B$-flux which could,
therefore, be consistent even non-perturbatively. Here we note that the
$\Omega J^\prime$ and $\Omega R J^\prime$ ${\bf Z}_3$ models with $B$-flux are
different at the massive (which are non-BPS) 
levels, which might be the reason why one set of these
models could have different non-perturbative behavior compared with the 
other.}.

{}Before we end this subsection, we would like to discuss the massless spectra
of the ${\bf Z}_6$ models with $B$-flux. More precisely, here we can discuss 
the 99 and 55 as well as closed string sectors. The closed string sector in
both $b=2$ and $b=4$ models contains (together with the six dimensional ${\cal 
N}=1$ supergravity multiplet) 9 tensor multiplets and 12 hypermultiplets. The
gauge group of the $b=2$ model is $[U(4)\otimes U(4)]_{99}\otimes [U(4)\otimes
U(4)]_{55}$. The massless hypermultiplets in the 99 and 55 sectors are given 
by:
\begin{eqnarray}
 &&({\bf 4},{\bf 4};{\bf 1},{\bf 1})_{99}~,\\
 &&({\bf 1},{\bf 1};{\bf 4},{\bf 4})_{55}~.
\end{eqnarray}  
As to the 95 sector, the irreducible $R^4$ and $F^4$ anomaly cancellation
would require that we have the following massless hypermultiplets:
\begin{eqnarray}
 &&({\bf 4},{\bf 1};{\bf 4},{\bf 1})_{95}~,\\
 &&({\bf 1},{\bf 4};{\bf 1},{\bf 4})_{95}~.
\end{eqnarray} 
However, as we discussed above, the 59 sector states in this model do {\em not}
carry well defined gauge quantum numbers, hence inconsistency in this model.

{}Finally, in the $b=4$ model the gauge group is $U(4)_{99}\otimes U(4)_{55}$,
and there are no massless hypermultiplets in the 99 and 55 
sectors\footnote{This corrects the error in the corresponding spectrum given in
\cite{KST}.}. As to the 95 sector, the irreducible $R^4$ and $F^4$ anomaly 
cancellation would require that we have {\em two} massless hypermultiplets
in $({\bf 4};{\bf 4})_{95}$. Note, however, that as in the $b=2$ model the 95
sector states in this model do not carry well defined gauge charges. This is
due to the fact that the 99 $({\bf Z}_2)^2$ discrete gauge symmetry is 
completely broken by the ${\bf Z}_3$ twist. This, in turn, is consistent with 
the fact that the 95 sector states do not carry well defined gauge quantum 
numbers - it would otherwise be difficult to understand how we can have {\em 
two} copies of the aforementioned 95 hypermultiplets as there is no discrete 
gauge symmetry to distinguish the corresponding vertex 
operators\footnote{Here we note that before the ${\bf Z}_3$ orbifold 
projection, that is, in the corresponding
$b=4$ ${\bf Z}_2$ model we have {\em four}
distinct vertex operators in the 59 sector distinguished by their
charges under the 99 $({\bf Z}_2)^2$ discrete symmetry. To have an anomaly 
free model, however, only {\em two} of these vertex operators would have to
survive the ${\bf Z}_3$ projection, while the other two would have to be
projected out. Thus, the ${\bf Z}_3$ twist would have to act non-trivially 
on the corresponding quantum numbers. In fact, it indeed does, except that
its action is incompatible with the 99 gauge quantum numbers - as we have 
already explained, the ${\bf Z}_3$ invariant states in the 59 sector do not 
possess well defined gauge quantum numbers. The reason for this is that
the ${\bf Z}_3$ twist does not commute with the already non-commuting 
Wilson lines. The situation in the $b=2$ ${\bf Z}_6$ model
is analogous to that we have just 
described for the $b=4$ model, except that there are additional subtleties due
to the fact that before the ${\bf Z}_3$ projection we have half-hypermultiplets
in the corresponding ${\bf Z}_2$ model.}. 
Thus, we 
conclude that the ${\bf Z}_6$ models with $B$-flux cannot be made completely
consistent within this framework\footnote{The above massless
spectra (with the guessed
95 matter content) are free of the irreducible anomalies, so that there might
exist consistent string constructions giving rise to these spectra. However, 
the $\Omega J^\prime$ orientifolds of Type IIB on $T^4/{\bf Z}_6$ with $
B$-flux do not seem to be the corresponding constructions.}.

\subsection{The ${\bf Z}_4$ Models}

{}In this section we will discuss the $\Omega J^\prime$ orientifolds of Type 
IIB on $T^4/{\bf Z}_4$ (for simplicity we will assume that 
$T^4=T^2\otimes T^2$) with $b=2,4$. Let $g$ be the generator of ${\bf Z}_4$.
Its action on the complex coordinates $z_1,z_2$ parametrizing the two 2-tori
is given by: $gz_1=iz_1, gz_2=-iz_2$. Note that $g^2\equiv R$ is the generator 
of the ${\bf Z}_2$ subgroup of ${\bf Z}_4$ ($Rz_{1,2}=-z_{1,2}$). The action 
of $J^\prime$ on the closed string untwisted as well as $R$ twisted sector 
fields is trivial, while $J^\prime$ interchanges the $g$ twisted sector with 
the $g^{-1}$ twisted sector.

{}Lets us first discuss the closed string sector of the above orientifold.
The untwisted sector gives rise to the six dimensional ${\cal N}=1$ 
supergravity multiplet plus one tensor multiplet and 2 hypermultiplets.
The $g$ and $g^{-1}$ twisted sectors together give rise to 4 tensor multiplets
and 4 hypermultiplets. As to the $R$ twisted sector, as in the ${\bf Z}_6$ 
case, we must consider $b=0,2,4$ cases separately.

{}In the $b=0$ case we have 16 fixed points in the $R$ twisted sector
before the ${\bf Z}_4$ projection. At all of these fixed points we have 
O5$^-$-planes, so each of them gives rise to a hypermultiplet (but no tensor
multiplets). Let $e_a$ and $d_a$ be the vielbeins corresponding to the two
2-tori. Then the 4 fixed points at $(0,0),(e_3/2,0),(0,d_3/4),(e_3/2,d_3/2)$
are invariant under the ${\bf Z}_4$ twist $g$. The other 12 fixed points
fall into 6 distinct groups, each group containing 2 fixed points which are
permuted by the action of $g$ (note that $g$ acts as a ${\bf Z}_2$ twist
on these fixed points which follows from the fact that by definition
$g^2=R$ must be 1 on all points fixed under $R$). In each of these 6 groups
we can form one linear combination of the corresponding 2 fixed points which
is invariant under $g$. Thus, we have total of 10 hypermultiplets coming from 
the $R$ twisted sector - 4 from the points fixed under both $R$ and $g$, and 6
from the points fixed under $R$ but not $g$.

{}Next, let us consider the $b=2$ case. Here we have 12 O5$^-$-planes and
4 O5$^+$-planes. For definiteness let us assume that the $B$-flux is turned
on on the first $T^2$. Then the ${\bf Z}_4$ symmetry requires one of the
following two distributions of the O5-planes. We can have 4 O5$^+$-planes at
the fixed points $(0,0),(0,d_a/2)$, or we can have 4 O5$^+$-planes
at the fixed points $(e_3/2,0),(e_3/2,d_a/2)$. Both setups give equivalent
models, so we will focus on the latter setup for definiteness. Note that the
fixed points $(e_3/2,0)$ and $(e_3/2,d_3/2)$ are also fixed under $g$, so that
these fixed points give rise to one tensor multiplet each. On the other hand,
the two fixed points $(e_3/2,d_1/2)$ and $(e_3/2,d_2/2)$ are permuted by the
action of $g$, so that they together give rise to one tensor multiplet. 
Finally, it is not difficult to see that the rest of the fixed points (at which
we have O5$^-$-planes) give rise to 7 hypermultiplets. Thus, in the
$b=2$ case we have 3 tensor multiplets and 7 hypermultiplets coming from
the $R$ twisted sector\footnote{This corrects the error in \cite{KST}, where 
the $R$ twisted sector in the $b=2$ ${\bf Z}_4$ model was thought to give
rise to 2 tensor multiplets and 8 hypermultiplets.}.

{}In the $b=4$ case we have 10 O5$^-$-planes and 6 O5$^+$-planes. One of the
four equivalent distributions of O5-planes consistent with the ${\bf Z}_4$
symmetry is the following. At the fixed points $(0,0)$ and $(e_a/2,d_b/2)$
we have O5$^-$-planes, while at the fixed points $(0,d_a/2)$ and $(e_a/2,0)$
we have O5$^+$-planes. After the $g$ projection, the latter give rise to 4
tensor multiplets, while the former give rise to 6 hypermultiplets. Thus,
in the $b=4$ case
we have 4 tensor multiplets and 6 hypermultiplets in the $R$ twisted 
sector\footnote{This corrects the error in \cite{KST}, where 
the $R$ twisted sector in the $b=4$ ${\bf Z}_4$ model was thought to give
rise to 3 tensor multiplets and 7 hypermultiplets.}.
  
{}Next, let us discuss the open string sector. First let us focus on the
$b=2$ model. Note that we have the $B$-flux in the direction of the first 
$T^2$, so that we have the corresponding non-commuting Wilson lines in the
directions $e_1$ and $e_2$. Note that the action of $g$ on the corresponding
shifts $S_a$ is given by (here $S_3\equiv S_1 S_2$):
\begin{equation}
 gS_1 g^{-1}=S_2~,~~~gS_2g^{-1}=S_1^{-1}=S_1~,~~~gS_3g^{-1}=S_1^{-1}S_2=S_3~,
\end{equation}
where we have used the fact that the shifts $S_i$ are ${\bf Z}_2$ valued, that
is, $S_i^2=1$. This implies that similar relations must also hold for the
corresponding Chan-Paton matrices $\gamma_{S_a,9}$ and $\gamma_{g,9}$. There
are, however, immaterial sign ambiguities in these relations such as
whether to require $\gamma_{g,9}\gamma_{S_3,9}\gamma_{g,9}^{-1}=
+\gamma_{S_3,9}$ or $-\gamma_{S_3,9}$. This ambiguity is related to the fact 
that even though the $S_a$ shifts are ${\bf Z}_2$ valued, the corresponding
$\gamma_{S_a}$ matrices can be ${\bf Z}_4$ valued in the sense that
$\gamma_{S_a}^2$ need not be the identity matrix $I_{32}$ but can also be
equal $-I_{32}$ (in the latter case $-\gamma_{S_a}=\gamma_{S_a}^{-1}$). 
In the following it will be convenient to use the following choices for 
these signs: 
\begin{equation}\label{conjZ_4}
 \gamma_{g,9}\gamma_{S_1,9}\gamma_{g,9}^{-1}=\gamma_{S_2,9}~,~~~
 \gamma_{g,9}\gamma_{S_2,9}\gamma_{g,9}^{-1}=\gamma_{S_1,9}~,~~~
 \gamma_{g,9}\gamma_{S_3,9}\gamma_{g,9}^{-1}=-\gamma_{S_3,9}~.
\end{equation}
A solution to these conditions can be written in terms of 16 copies of the
corresponding 2 dimensional representations:
\begin{eqnarray}
 &&\gamma_{S_1,9}=\kappa\sigma_3\otimes I_{16}~,\\
 &&\gamma_{S_2,9}=\kappa\sigma_1\otimes I_{16}~,\\
 &&\gamma_{S_3,9}=i\sigma_2\otimes I_{16}~,\\
 &&\gamma_{g,9}=\zeta_g\otimes \Gamma_g~,
\end{eqnarray}
where $\kappa^2=1$ corresponds to the $D_4^\prime$ type of Wilson 
lines, while $\kappa^2=-1$ corresponds to the $D_4$ type of Wilson lines. 
Note that both of these choices are {\em a priori} allowed as (\ref{conjZ_4})
implies that
\begin{equation}
 \gamma_{S_1,9}^2=\gamma_{S_2,9}^2~,
\end{equation}
but does not relate $\gamma_{S_3,9}^2$ to $\gamma_{S_i,9}^2$.

{}Note that in the above solution for $\gamma_{S_a,9}$ and $\gamma_{g,9}$
we have assumed (for definiteness) that $\eta_{12}=\kappa^2$, albeit this
can be relaxed. The $2\times 2$ matrix $\zeta_g$, 
whose eigenvalues are $1,-1$, is given by:
\begin{equation}
 \zeta_g\equiv{1\over \sqrt{2}}\left(\sigma_3+\sigma_1\right)~.
\end{equation}
Finally, the matrix $\Gamma_g$ is determined as follows. First, the twisted
tadpole cancellation conditions imply that $\gamma_{g,9}$ must be traceless
\cite{GJ}.
It then follows that $\Gamma_g$ must be traceless as well. Next, we have
\begin{equation}
 \gamma_{R,9}=\gamma_{g,9}^2=I_2\otimes \Gamma_g^2~.
\end{equation}
Note that for $\kappa^2=+1$ (that is, for the $D_4^\prime$ type of Wilson
lines) we must have $\gamma_{R,9}$ (which is also traceless)
with eigenvalues $\pm i$, while for $\kappa^2=-1$ (that is, for the $D_4$ type
of Wilson lines) we must have $\gamma_{R,9}$ with eigenvalues $\pm 1$. This
fixes $\Gamma_g$ (up to equivalent representations) as follows:
\begin{eqnarray}
 &&\kappa^2=+1:~~~\Gamma_g={\rm diag}(\alpha,\alpha^{-1},-\alpha,-\alpha^{-1})
       \otimes I_4~,\\
 &&\kappa^2=-1:~~~\Gamma_g={\rm diag}(1,-1,i,-i)
       \otimes I_4~,
\end{eqnarray} 
where $\alpha\equiv\exp(2\pi i/8)$.

{}The above discussion can be generalized to the $b=4$ case as well. Here we 
have two sets of Wilson lines $\gamma_{S_a,9}$ and $\gamma_{T_a,9}$ 
corresponding to the first and the second $T^2$, respectively.
A solution for the Wilson lines and the twisted Chan-Paton matrix 
$\gamma_{g,9}$ satisfying all the required consistency conditions is given by:
\begin{eqnarray}
 &&\gamma_{S_1,9}=\kappa\sigma_3\otimes I_2\otimes I_{8}~,\\
 &&\gamma_{S_2,9}=\kappa\sigma_1\otimes I_2\otimes I_{8}~,\\
 &&\gamma_{S_3,9}=i\sigma_2\otimes I_2\otimes I_{8}~,\\
 &&\gamma_{T_1,9}=I_2\otimes \sigma_3\otimes I_{8}~,\\
 &&\gamma_{T_2,9}=I_2\otimes \sigma_1\otimes I_{8}~,\\
 &&\gamma_{T_3,9}=I_2\otimes i\sigma_2\otimes I_{8}~,\\
 &&\gamma_{g,9}=\zeta_g\otimes \zeta_g \otimes {\widetilde \Gamma}_g~,
\end{eqnarray}
where the $8\times 8$ matrix ${\widetilde \Gamma}_g$ is given by
\begin{eqnarray}
 &&\kappa^2=+1:~~~{\widetilde \Gamma}_g=
 {\rm diag}(\alpha,\alpha^{-1},-\alpha,-\alpha^{-1})
       \otimes I_2~,\\
 &&\kappa^2=-1:~~~{\widetilde \Gamma}_g={\rm diag}(1,-1,i,-i)
       \otimes I_2~.
\end{eqnarray} 
Here we note that in the $b=4$ case for 
$\kappa^2=+1$ the 99 gauge group before the $\gamma_{g,9}$
projection is $SO(8)\otimes ({\bf Z}_2)^2$, 
while for $\kappa^2=-1$ it is $Sp(8)\otimes 
({\bf Z}_2)^2$. In the $b=2$ case
with the aforementioned choice of Wilson lines the 99 gauge group before
the $\gamma_{g,9}$ projection is $SO(16)\otimes {\bf Z}_2$ for $\kappa^2=+1$, 
while for $\kappa^2=-1$ it is $Sp(16)\otimes {\bf Z}_2$.

{}Note that the twisted Chan-Paton matrix $\gamma_{g,5}$ is fixed (up
to equivalent representations) once
$\gamma_{g,9}$ is fixed. In fact, without loss of generality 
we can choose it to be given by\footnote{Here we are assuming that all 
D5-branes are placed at an O5-plane located at a ${\bf Z}_4$ fixed point.
As we will show in a moment, other {\em a priori} allowed
configurations would be continuously
connected to these ones had the ${\bf Z}_4$ models with $B$-flux
been consistent, so this 
assumption is not restrictive.}
\begin{eqnarray}
 &&b=2:~~~\gamma_{g,5}=\Gamma_g~,\\
 &&b=4:~~~\gamma_{g,5}={\widetilde \Gamma}_g~.
\end{eqnarray}
This implies that before the $\gamma_{g,5}$ projection (assuming that 
we place 
all D5-branes at the same O5-plane)
the 55 gauge group
is $SO(32/2^{b/2})$ for $\kappa^2=+1$, and $Sp(32/2^{b/2})$ for $\kappa^2=-1$
($b=2,4$). That is, in the former case we must place D5-branes at an 
O5$^-$-plane, while in the latter case we must place D5-branes at an 
O5$^+$-plane. This is in complete parallel with our discussion of the
corresponding ${\bf Z}_2$ models.

{}In fact, here we run into the same problem as in the ${\bf Z}_2$ case 
if we attempt to interpret the above orientifold in the context of the
conformal field theory orbifold. More precisely, there are two separate issues
here. As we will point out in a moment, in the ${\bf Z}_4$ models with $B$-flux
we appear to have a problem with the 59 sector states analogous to that in the
${\bf Z}_6$ models. That is, the 59 vertex operators are not well defined. 
However, the issue that we would like to discuss first depends only on the
structure of the 55 (and 99) sector states, whose vertex operators are well
defined. So for a moment we will (erroneously)
assume that the ${\bf Z}_4$ orbifold action yields consistent 59 vertex 
operators corresponding to a discrete gauge group which is 
$({\bf Z}_2)^{\otimes (b/2)}$ or a subgroup thereof. 
With this assumption we can derive the spectra of the corresponding models.

{}First, let us start with the $b=2$ model with $\kappa^2=+1$. The gauge
group is $[U(4)\otimes U(4)\otimes {\bf D}]_{99}
\otimes [U(4)\otimes U(4)]_{55}$, and the massless 
hypermultiplets are
given by:
\begin{eqnarray}
 &&({\bf 6},{\bf 1};{\bf 1},{\bf 1})_{99}~,~~~
   ({\bf 1},{\bf 6};{\bf 1},{\bf 1})_{99}~,~~~
   ({\bf 4},{\bf 4};{\bf 1},{\bf 1})_{99}~,\\
 &&({\bf 1},{\bf 1};{\bf 6},{\bf 1})_{55}~,~~~
   ({\bf 1},{\bf 1};{\bf 1},{\bf 6})_{55}~,~~~
   ({\bf 1},{\bf 1};{\bf 4},{\bf 4})_{55}~,\\
 &&({\bf 4}_D,{\bf 1};{\bf 4},{\bf 1})_{95}~,~~~
 ({\bf 1},{\bf 4}_D;{\bf 1},{\bf 4})_{95}~,
\end{eqnarray} 
where the subscript $D$ in the 95 sector indicates the 99 ${\bf D}$ discrete
gauge quantum numbers. Here we encounter the first signs of trouble. In 
particular, note that for no choice of the discrete gauge symmetry ${\bf D}$ 
can we cancel, say, the $R^4$
irreducible anomaly (recall that the closed string sector contains 
13 hypermultiplets
and 8 tensor multiplets in this model). 

{}Next, consider the $b=4$ model with $\kappa^2=+1$. The gauge
group is $[U(2)\otimes U(2)\otimes {\bf D}]_{99}
\otimes [U(2)\otimes U(2)]_{55}$, and the massless
hypermultiplets are
given by:
\begin{eqnarray}
 &&({\bf 1}_a,{\bf 1};{\bf 1},{\bf 1})_{99}~,~~~
   ({\bf 1},{\bf 1}_a;{\bf 1},{\bf 1})_{99}~,~~~
   ({\bf 2},{\bf 2};{\bf 1},{\bf 1})_{99}~,\\
 &&({\bf 1},{\bf 1};{\bf 1}_a,{\bf 1})_{55}~,~~~
   ({\bf 1},{\bf 1};{\bf 1},{\bf 1}_a)_{55}~,~~~
   ({\bf 1},{\bf 1};{\bf 2},{\bf 2})_{55}~,\\
 &&({\bf 2}_D,{\bf 1};{\bf 2},{\bf 1})_{95}~,~~~
   ({\bf 1},{\bf 2}_D;{\bf 1},{\bf 2})_{95}~,
\end{eqnarray} 
where 
${\bf 1}_a$ is the antisymmetric representation of $U(2)$ (it is a singlet of
$SU(2)$ but is charged under the $U(1)$ factor).
Note that this spectrum, just as in the $b=2$ case, cannot be made 
free of, say, the $R^4$ 
irreducible anomaly (the closed string sector contains 12 hypermultiplets
and 9 tensor multiplets in this model).

{}The gauge group of the $\kappa^2=-1$ model with rank $b$ $B$-flux is
$[U(K)\otimes Sp(K)\otimes Sp(K)\otimes {\bf D}]_{99}
\otimes [U(K)\otimes Sp(K)\otimes Sp(K)]_{55}$, where $K\equiv 8/2^{b/2}$
($b=2,4$).
The massless hypermultiplets are given by:
\begin{eqnarray}
 &&({\bf K},{\bf K},{\bf 1};{\bf 1},{\bf 1},{\bf 1})_{99}~,~~~
   ({\bf K},{\bf 1},{\bf K};{\bf 1},{\bf 1},{\bf 1})_{99}~,\\
 &&({\bf 1},{\bf 1},{\bf 1};{\bf K},{\bf K},{\bf 1})_{55}~,~~~
   ({\bf 1},{\bf 1},{\bf 1};{\bf K},{\bf 1},{\bf K})_{55}~,\\
 &&({\bf K}_D,{\bf 1},{\bf 1};{\bf K},{\bf 1},{\bf 1})_{95}~,~~~
 {1\over 2}({\bf 1},{\bf K}_D,{\bf 1};{\bf 1},{\bf K},{\bf 1})_{95}~,~~~
   {1\over 2}({\bf 1},{\bf 1},{\bf K}_D;{\bf 1},{\bf 1},{\bf K})_{95}~. 
\end{eqnarray} 
Just as for the $\kappa^2=+1$ models, these spectra cannot be made free of the
$R^4$ anomaly
(the closed string matter contents in the $\kappa^2=-1$ models 
are the same as in the 
corresponding $\kappa^2=+1$ models).
Moreover, the 95 sector spectrum is actually anomalous as it contains 
half-hypermultiplets in real representations\footnote{Here we note that
the above spectrum differs in the 95 sector from that given for these models 
in the second reference in \cite{Ka3}. However, neither spectra are completely
consistent due to the 
problem with the 95 vertex operators we are going to point
out in a moment. In fact, this problem is at least partially responsible for
an ambiguity in determining the 95 sector spectrum in these models, which had
led to that reported in the second reference in \cite{Ka3}.}. 
This is similar to what happens in the corresponding ${\bf Z}_2$ models. 

{}In fact, the main reason why we have 
given the above spectra for both $\kappa^2=
\pm 1$ models is that, if we assume that they are realized as orientifolds of 
the conformal field theory $T^4/{\bf Z}_4$ orbifold, then they are 
continuously connected. This is in complete parallel with the corresponding 
discussion in the ${\bf Z}_2$ models. Thus, for instance, in the
conformal field theory orbifold setup there is no 
obstruction to moving D5-branes in, say, the $b=2$ model with $\kappa^2=+1$
from an O5$^-$-plane to an O5$^+$-plane. This then leads to the inconsistency
in the 59 sector spectrum where half-hypermultiplets arise in various
inappropriate representations. To see that the Higgsing corresponding
to such motion of D5-branes is allowed, note that in this model (that is, the
$b=2$ model with $\kappa^2=+1$) we can simultaneously 
turn on VEVs of the hypermultiplets
$({\bf 1},{\bf 1};{\bf 6},{\bf 1})_{55}$, 
$({\bf 1},{\bf 1};{\bf 1},{\bf 6})_{55}$ and 
$({\bf 1},{\bf 1};{\bf 4},{\bf 4})_{55}$ in such a way that the D-flatness 
conditions are satisfied. The maximal unbroken 55 gauge 
subgroup then is $Sp(4)$. This
corresponds to moving together 2 dynamical 5-branes off the O5$^-$-plane. Note
that each dynamical 5-brane is made of 8 D5-branes - one pairing is due to the
orientifold projection, while another
grouping in multiples of 4 is due to the
${\bf Z}_4$ orbifold projection. In this subspace of the moduli space, which 
corresponds to these two dynamical 5-branes sitting on top of each other in
the bulk, we have one 55 hypermultiplet in ${\bf 1}\oplus{\bf 5}$ of 
$Sp(4)_{55}$. In fact, the VEV of the 
singlet measures the separation between the D5-branes and the O5$^-$-plane. We
can split the two dynamical 5-branes from each other by giving a VEV to the
hypermultiplet in ${\bf 5}$ of $Sp(4)$, which breaks the gauge group down to
$Sp(2)\otimes Sp(2)$. Thus, each dynamical 5-brane, as expected, gives rise
to one $Sp(2)$ subgroup, and the leftover {\em two} singlet hypermultiplets
measure the individual locations of the two dynamical 5-branes in the bulk. 
Note that once all 16 D5-branes approach an O5$^+$-plane, the 55 gauge group 
is enhanced to $[U(4)\otimes Sp(4)\otimes Sp(4)]_{55}$. In fact, the matter
content of the 55 sector in the $b=2$ model with $\kappa^2=-1$ is just right
to Higgs the latter gauge group down to $Sp(4)_{55}$ plus the hypermultiplet in
${\bf 1}\oplus {\bf 5}$ (or its $Sp(2)\otimes Sp(2)$ subgroup with two singlet
hypermultiplets). This then implies that we will have inconsistencies arising
in the 59 sector once we, say, move D5-branes from an O5$^-$-plane to an 
O5$^+$-plane in the $b=2$ model with $\kappa^2=+1$. In fact, these 
inconsistencies are of the same type as those we have encountered in the 
corresponding ${\bf Z}_2$ models in subsection A.

{}The above discussion, just as in the corresponding ${\bf Z}_2$ cases, leads
us to the conclusion, which is consistent with our discussions in subsection 
A, that we cannot consistently have O5$^+$-planes in the context of the
conformal field theory $T^4/{\bf Z}_4$ orbifold. This is not surprising as 
${\bf Z}_4$ contains a ${\bf Z}_2$ subgroup, and we have already come to
such a conclusion in the $T^4/{\bf Z}_2$ case. In the latter models we also
pointed out a possibility of deforming the orbifold by partially blowing-up
the collapsed ${\bf P}^1$'s and turning off the twisted $B$-flux at the 
${\bf Z}_2$ fixed points where we expect the O5$^+$-planes. The resulting
K3 compactification could then {\em a priori} be the correct framework
for considering the corresponding orientifolds with $B$-flux, at least, in the
${\bf Z}_2$ case we did not find any obvious inconsistencies with such a
possibility. So here we can ask whether a similar possibility exists in the
${\bf Z}_4$ cases as well. The answer to this question appears to be negative
for at least two reasons. First, the massless spectra of the $\kappa^2=+1$
(as well as $\kappa^2=-1$) models cannot be made anomaly free. We would like to
discuss the second reason next. 

{}The point is that, just as in the ${\bf Z}_6$ models with $B$-flux (and for
essentially the same reason), the 99 $({\bf Z}_2)^{\otimes(b/2)}$ discrete
gauge symmetry is broken by the ${\bf Z}_4$ orbifold action. This, in turn,
implies that the above spectra, in particular, in the 59 sector, are not 
completely correct. More concretely, in the 59 sector the states invariant
under the $\gamma_{g,9}$ (together with $\gamma_{g,5}$) ${\bf Z}_4$ orbifold
projection do {\em not} carry well defined gauge quantum numbers at all. Just 
as in the ${\bf Z}_6$ cases, this is due to the fact that $\gamma_{g,9}$ does 
not commute with already non-commuting Wilson lines $\gamma_{S_i,9}$.

{}Let us illustrate this point for the $b=2$ model with $\kappa^2=+1$ (all 
other cases can be treated similarly). Note that in this case $\gamma_{S_a,9}=
\gamma_a\otimes I_{16}$ (where $\gamma_a$ are the corresponding $2\times 2$ 
matrices), and $\gamma_{g,9}=\zeta_g\otimes \Gamma_g$. The
non-commutativity between $\gamma_{g,9}$ and $\gamma_{S_a,9}$ is in the 
$2\times 2$ block corresponding to $\gamma_a$ and $\zeta_a$. So, we can ignore
the other $16\times 16$ block structure for our purposes here. Moreover, just 
as in the ${\bf Z}_6$ cases, the important point is how ${\bf 32}$ of $SO(32)$
(that is, the 99 gauge group before the ${\bf Z}_4$ orbifold as well as 
${\bf Z}_2\otimes {\bf Z}_2$ freely acting
orbifold projections) is decomposed under the
unbroken 99 gauge group - the 55 part of the corresponding vertex operators
cannot possibly give any trouble here. So, for simplicity we will discuss
the following Chan-Paton matrices: $\gamma_{S_a,9}=\gamma_a\otimes I_{16}$,
$\gamma_{g,9}=\zeta_g\otimes I_{16}$. (This choice does not satisfy the 
twisted tadpole cancellation conditions, but this is not going to be
relevant here.) Moreover, for definiteness we will choose $\kappa=1$, so that
$\gamma_1=\sigma_3$, $\gamma_2=\sigma_1$, $\gamma_3=i\sigma_2$. 

{}The unbroken gauge group with the aforementioned choice of Chan-Paton 
matrices is $SO(16)$. This gauge group arises as follows. First,
$\gamma_{S_1,9}$
breaks $SO(32)$ down to $SO(16)\otimes SO(16)$. Then $\gamma_{S_2,9}$ breaks
$SO(16)\otimes SO(16)$ down to $SO(16)_{\rm diag}\otimes {\bf Z}_2$. Finally,
$\gamma_{g,9}$ breaks the ${\bf Z}_2$ subgroup, but does not affect the 
$SO(16)_{\rm diag}$ subgroup. To understand this, let us see how ${\bf 32}$ of 
$SO(32)$ decomposes under these breakings. First, under $\gamma_{S_1,9}$ we 
have
\begin{equation}\label{32=16+16}
{\bf 32}=({\bf 16},{\bf 1})\oplus({\bf 1},{\bf 16})~.  
\end{equation}
Next, under $\gamma_{S_2,9}$ the two $SO(16)$ subgroups are permuted, so
that we have
\begin{equation}\label{32=2X16}
 {\bf 32}={\bf 16}_+\oplus{\bf 16}_-~,
\end{equation}
where the subscript indicates the 99 ${\bf Z}_2$ discrete gauge charges. The
latter can be quantified as follows. Note that the eigenvectors of the matrix
$\gamma_1=\sigma_3$ are 2-component column-vectors $\psi_\pm$ with the 
components $\psi_+^{\uparrow}=1$, $\psi_+^{\downarrow}=0$, 
$\psi_-^{\uparrow}=0$, $\psi_-^{\downarrow}=1$. Note that the vertex operators
in (\ref{32=16+16}) for $({\bf 16},{\bf 1})$ and $({\bf 1},{\bf 16})$ are 
proportional to $\psi_+$ and $\psi_-$, respectively. However, the eigenvectors
of the matrix $\gamma_2=\sigma_1$ are 
\begin{equation}
 \chi_\pm ={1\over\sqrt{2}}\left(\psi_+\pm \psi_-\right)~.
\end{equation}
In fact, the vertex operators for ${\bf 16}_+$ and ${\bf 16}_-$ in 
(\ref{32=2X16}) are proportional to $\chi_+$ and $\chi_-$, respectively. Thus,
$\chi_\pm$ provide precisely the part of the vertex operators 
corresponding to the 99 ${\bf Z}_2$ discrete gauge 
quantum numbers. Note, however, 
that $\chi_\pm$ are {\em not} eigenvectors of $\zeta_g$ as $\zeta_g$ and
$\gamma_2$ do not commute. In particular, we have
\begin{equation}
 \zeta_g\chi_\pm =\psi_\pm~,
\end{equation}
so that the action of $\gamma_{g,9}$ on the 59 sector states mixes states
with different gauge quantum numbers. That is, the 59 states with 
distinct gauge
quantum numbers under the unbroken $SO(16)$ subgroup are incompatible with the
${\bf Z}_4$ orbifold projection. On the other hand, the 59 states invariant
under the ${\bf Z}_4$ orbifold projection do not possess well defined 
gauge quantum numbers. We therefore conclude that we have an inconsistency at
the level of the 59 vertex operators in this and other ${\bf Z}_4$ models with 
$B$-flux.

\subsection{Summary}

{}Let us briefly summarize the results of this section. In subsection A
we have argued that the $\Omega$ orientifold of Type IIB on the conformal
field theory $T^4/{\bf Z}_2$ orbifold with $B$-flux is inconsistent. This
can be seen by noting that in the toroidal orbifold with $B$-flux we would have
to introduce O5$^+$-planes, which cannot be placed at the conformal field
theory $T^4/{\bf Z}_2$ orbifold fixed points. The obstruction here
is due to the
twisted half-integer $B$-flux 
(which makes the conformal field theory orbifold non-singular)
stuck in the corresponding collapsed 
${\bf P}^1$'s.
We have also pointed out that perhaps by deforming the orbifold from the
conformal field theory point (such a deformation would involve blowing up the
fixed points where we expect the O5$^+$-planes and turning off the 
corresponding twisted 
$B$-flux) we could make the corresponding orientifold background consistent.
Such a K3 compactification, however, would not correspond to an exactly 
solvable conformal field theory, and it is unclear at present how to make 
the corresponding statements quantitatively precise. In particular, the moduli
space corresponding to the motion of D5-branes as well as that corresponding 
to turning on continuous Wilson lines in the 99 sector would have to have some
funny properties which we do not know how to argue for (or, for that sake,
against). Thus, for instance, D5-branes should not be able to come on top
of the O5$^+$-planes - this could in principle 
be a property of compactifications on such a K3 as the corresponding moduli
space is no longer flat, but it is unclear whether it really is. If, however, 
this is indeed the case, then the ${\bf Z}_2$ models with $B$-flux could be
consistent in the context of such K3 compactifications if we place D5-branes
on top of an O5$^-$-plane or in the bulk with the
gauge bundle corresponding to the ${\bf Z}_2$ orbifold twist ``without vector
structure'' in the language of \cite{BLPSSW}. However, attempts to construct
models with ${\bf Z}_2$ gauge bundles ``with vector structure'' run into 
various inconsistencies. In fact, these inconsistencies can be seen in the
language of the effective field theory where they manifest themselves via
anomalies. In particular, the fact that the conformal field theory 
$T^4/{\bf Z}_2$ orbifold cannot be the correct background for the corresponding
orientifold with $B$-flux can also be seen in the effective field theory 
language. Thus, for instance, in the conformal field theory orbifold case 
there is no obstruction to moving D5-branes from an O5$^-$-plane to an 
O5$^+$-plane, and at the latter points in the moduli space the massless 
spectra in the 59 sector are anomalous.

{}In fact, it is always the 59 sector where the trouble shows up. This is,
in turn, not an accident. Thus, there is no reason for any inconsistencies to
arise in the 99 or 55 sectors (at least perturbatively). However, the 59
sector by definition (since it contains hypermultiplets in the bifundamental
representations of the 99 and 55 gauge groups) requires certain vector
structure contrary to the lack of vector structure required by the fact that
once we turn on the $B$-flux the generalized second Stieffel-Whitney class
is non-vanishing. This appears to be the key reason for the aforementioned
inconsistencies arising in these models.

{}Another (perhaps indirect) hint of this is what we have found in the 
${\bf Z}_6$ and ${\bf Z}_4$ models with $B$-flux, namely, that the 59 sector
states, if properly projected by the orbifold action, do not carry well defined
gauge quantum numbers. Even though in the previous subsections we have only
used examples to illustrate this point, the corresponding underlying reason
can be stated quite generally. 

{}Thus, consider a general setup with Type IIB compactified
on $T^d$, where for definiteness we will choose $d$ to be even. 
Let us introduce some number $N$ of D9-branes wrapping $T^d$. Note that
to cancel tadpoles we would have to introduce the O9$^-$-plane and set $N=32$.
However, we will not do so here, that is, we will not introduce any orientifold
planes at all, and we will
let $N$ be arbitrary - the point we would like to make here
is independent of the tadpole cancellation requirements. Next, let us
introduce some number $N^\prime$ of D$p$-branes transverse to $T^d$. 
Even though this is
not particularly important for our discussion here, we will assume that $9-p$
is a multiple of 4 so that supersymmetry is not completely broken in this
background. Note that in the $9p$ open string sector we have no momenta or
windings (but only oscillator excitations) in the directions of $T^d$.

{}Next, we would like to turn on non-zero $B$-flux on $T^d$. We can do this
by considering a freely acting orbifold whose generators $S_i$, $i=1,\dots,d$,
are shifts in the corresponding directions of $T^d$ (and, therefore, they 
commute). Note that here we can have
discrete torsion $\Delta_{ij}$ between different $S_i$ generators. In 
particular, we must have $\Delta_{ii}=+1$, but $\Delta_{ij}$ can be equal
$-1$ for $i\not=j$ as long as both $S_i$ and $S_j$ have even orders. The matrix
$B_{ij}$ corresponding to the $B$-field can then be compactly written as 
\begin{equation}
 B_{ij}={1\over 2\pi i}\ln\left(\Delta_{ij}\right)~.
\end{equation}  
Note that $B_{ij}\sim B_{ij}+1$.

{}The action of the shifts $S_i$ on the 99 sector Chan-Paton charges 
is described by $N\times N$ matrices $\gamma_{S_i,9}$, and
corresponds to turning on Wilson lines. The string
consistency requires that
\begin{equation}
 \gamma_{S_i,9}\gamma_{S_j,9}=\Delta_{ij} \gamma_{S_j,9}\gamma_{S_i,9}~,
\end{equation}
so that if we have non-trivial discrete torsion, then the corresponding
Wilson lines are non-commuting. Here we note that the action of the shifts
on the $pp$ open string sector is trivial. In the 99 sector, however, it
breaks the gauge group $U(N)$ to its subgroup $G$. Suppose now that we have
non-trivial discrete torsion between some generators $S_i$. That is, let
the rank $b$ of the matrix $B_{ij}$ be non-zero. Then the rank of the unbroken 
gauge group $r(G)<N$. In fact, $r(G)=N/2^{b/2}$. The coset $U(N)/G$ is given
by the discrete gauge group $D\equiv ({\bf Z}_2)^{\otimes (b/2)}$. Let 
$d_\alpha$ denote the corresponding discrete charges, $\alpha=1,\dots,|D|$
($|D|=2^{b/2}$). Then it is not difficult to see that the fundamental ${\bf N}$
of $U(N)$ decomposes as follows:
\begin{equation}\label{fundamentalNN}
 {\bf N}=\bigoplus_{\ell,\alpha} ({\bf N}_\ell,d_\alpha)~, 
\end{equation}
where ${\bf N}_\ell$ are the fundamental representations of the corresponding 
subgroups of $G=\bigotimes_\ell U(N_\ell)$. (Note that $r(G)=\sum_\ell 
N_\ell$.)
On the other hand, the adjoint of
$U(N)$ decomposes as follows:
\begin{equation}
 {\bf Adj}={\bf N}\otimes{\overline {\bf N}}=\bigoplus_{\ell, \ell^\prime,
 \alpha,\beta}
 ({\bf N}_\ell\otimes {\overline {\bf N}}_{\ell^\prime},d_\alpha\otimes
 {\overline d}_\beta)~.
\end{equation}
Note that the gauge bosons of the unbroken gauge group $G$ come from the subset
with $\ell=\ell^\prime$ and $\beta=\alpha$. This subset reads:
\begin{equation}\label{subsetdd}
 \bigoplus_{\ell,\alpha} ({\bf Adj}_\ell,d_\alpha\otimes{\overline
 d}_\alpha)~.
\end{equation}
In fact, this subset contains exactly $|D|$ copies of the adjoint of $G$
as $\left| \bigoplus_\alpha d_\alpha\otimes{\overline d}_\alpha 
\right| = |D|$. Note that in 
the 99 sector only 
one 
of these copies survives the full freely acting orbifold projection
with respect to the action of $\gamma_{S_i,9}$. It is then not difficult to
see that the remaining states are given by\footnote{Thus, for instance, 
consider the case where $b=d$, and
$\Delta_{12}=\Delta_{34}=\dots=\Delta_{b-1,b}=-1$ with all other 
$\Delta_{ij}=+1$. Then we can group the  
$b$ independent projections $\gamma_{S_i,9}$ as follows. First, $b/2$ 
projections $\gamma_{S_1,9},\gamma_{S_3,9},\dots,\gamma_{S_{b-1},9}$ break 
the original $U(N)$ gauge group down to its subgroup of rank $N$. Then, 
the rest of the projections 
$\gamma_{S_2,9},\gamma_{S_4,9},\dots,\gamma_{S_b,9}$ break this gauge group 
to its subgroup $G\otimes D$. Note that the number of gauge bosons in 
(\ref{subsetdd}) is $|G| |D|=2^{b/2} |G|$. This corresponds to keeping the 
states with trivial discrete gauge charges (indeed, the representations
$\bigoplus_{\alpha} d_\alpha\otimes{\overline  d}_\alpha$ correspond to 
precisely such discrete gauge charges). These states are invariant under
half of the $\gamma_{S_i,9}$ projections, say, the
$\gamma_{S_2,9},\gamma_{S_4,9},\dots,\gamma_{S_b,9}$ projections. On the 
hand, the number of states in (\ref{subsetdd}) which are also invariant under 
the remaining projections 
$\gamma_{S_1,9},\gamma_{S_3,9},\dots,\gamma_{S_{b-1},9}$ is $|G|$. These 
states are given by (\ref{Dinv}).}
\begin{equation}\label{Dinv}
 \bigoplus_\ell ({\bf Adj}_\ell, D_{\rm inv})~,
\end{equation}
where the vertex operator corresponding to the discrete gauge charge
$D_{\rm inv}$ is given by
\begin{equation}
 |D_{\rm inv}\rangle={1\over \sqrt{D}}\sum_{\alpha} |d_\alpha\otimes
 {\overline d}_\alpha\rangle~.
\end{equation}
In contrast, in the $9p$ sector
no states are projected out by the action of this freely acting orbifold - 
it actually does not act in the $9p$ sector (neither does it act in the $pp$
sector). So instead the fundamental ${\bf N}$ 
of $U(N)$ in the $9p$ sector simply
decomposes under the unbroken gauge group as in (\ref{fundamentalNN}).

{}Now suppose we orbifold $T^d$ with the $B$-field turned on by the action
of the orbifold point group $\Gamma$ such that some of the elements of $\Gamma$
do not commute with the shifts $S_i$. It is necessary for the consistency of 
the theory that the shifts $S_i$ and the twists in $\Gamma$ form a larger
orbifold group ${\widetilde \Gamma}$, which is non-Abelian. Let us consider
one non-trivial element $g$ (such that $g^2\not=1$) that does not commute
with  some of the shifts $S_i$. Then this implies that the corresponding 
Chan-Paton matrices $\gamma_{S_i,9}$ and $\gamma_{g,9}$ also do not commute.
It is, however, important to note that the action of the twisted 
Chan-Paton matrix $\gamma_{g,9}$ does not reduce the rank of the unbroken
group $G_g\subset G$, that is, $r(G_g)=r(G)$. This follows from the fact that
$g$ simply permutes elements of ${\widetilde \Gamma}$ that are pure shifts
among each other. In fact, this implies that
we can always find $\gamma_{g,9}$ such that the
unbroken gauge group $G_g=G$. In this case the action of $\gamma_{g,9}$ is
non-trivial only on the discrete gauge quantum 
numbers\footnote{Our discussion here straightforwardly generalizes to 
the most general case. However, to illustrate the point we would like to make 
here, it suffices to consider $\gamma_{g,9}$ of the aforementioned type.}. 
In fact, this action
breaks the discrete gauge symmetry $D$ either completely or to its smaller
subgroup. Thus, the action of $\gamma_{g,9}$ on the discrete quantum numbers
$d_\alpha$ is given by:
\begin{equation}
 \gamma_{g,9}:~d_\alpha\rightarrow\sum_{\alpha^\prime} c_{\alpha\alpha^\prime}
 d_{\alpha^\prime}~,
\end{equation} 
where the matrix $c_{\alpha\alpha^\prime}$ is {\em not} diagonal as 
$\gamma_{g,9}$ does not commute with the Wilson lines. This then implies that  
the $9p$ sector states invariant under the $g$ projection cannot have well
defined gauge quantum numbers under the unbroken subgroup $G$. This can be
seen by noting that the states in (\ref{Dinv}) (that correspond to 
the gauge
bosons of the unbroken gauge group $G$) are invariant under the action of $g$
as\footnote{Here we are using the fact that $\sum_\alpha c_{\alpha\beta} 
{\overline c}_{\gamma\alpha}=\delta_{\beta\gamma}$, which is the statement
that $\gamma_{g,9}\gamma_{g,9}^{-1}=1$, in particular, when acting on the
discrete gauge charges $d_\alpha$ and ${\overline d}_\alpha$. Also, note that
Chan-Paton matrices are unitary, so that the action of $g$ on ${\overline
{\bf N}}$ of $U(N)$ is given by the matrix $\gamma_{g,9}^{-1}$ - see below.}
\begin{equation}
 \gamma_{g,9}:|D_{\rm inv}\rangle \rightarrow {1\over\sqrt{D}}
 \sum_\alpha \sum_{\beta,\gamma} c_{\alpha\beta} 
 {\overline c}_{\gamma\alpha} |d_\beta\otimes{\overline
 d}_\gamma \rangle={1\over \sqrt{D}} \sum_\alpha |d_\alpha\otimes{\overline
 d}_\alpha\rangle=|D_{\rm inv}\rangle~,
\end{equation}
that is, the aforementioned states are actually invariant under the 
$\gamma_{g,9}$ action, so that they provide the correct basis for the gauge 
bosons. On the other hand, the states in the $9p$
sector that are invariant under the $g$
projection are in a basis different from that of the $G$ gauge bosons. This can
be seen by noting that in the basis where $\gamma_{g,9}$ acts diagonally on
${\bf N}$ of $U(N)$ we have:
\begin{equation}
 {\bf N}=\bigoplus_{\ell,k} ({\bf N}_\ell,k)~,
\end{equation} 
where the vertex operators corresponding to the quantum numbers $k$ are given
by:
\begin{equation}
 |k\rangle\equiv \sum_\alpha f_{k\alpha}|d_\alpha\rangle~.
\end{equation}
Here $f_{k\alpha}$ are the eigenvectors of the matrix $c_{\alpha
\alpha^\prime}$:
\begin{equation}
 \sum_\alpha f_{k\alpha}c_{\alpha\alpha^\prime}=\lambda_k f_{k\alpha^\prime}~,
\end{equation}
where $\lambda_k$ are the eigenvalues of $c_{\alpha\alpha^\prime}$. It is 
important to note that for each $k$ $f_{k\alpha}\not=0$ for at least
two different values of $\alpha$, which follows from the fact that the twist
$g$ and the Wilson lines do not commute. Next, consider a massless $9p$ sector
state containing $({\bf N}_\ell,k)$. Its scattering with its own
conjugate state then will contain ${\bf Adj}_\ell$ together with the
following vertex operator corresponding to the discrete gauge quantum numbers:
\begin{equation}
 |k\rangle\otimes |{\overline k}\rangle=\sum_{\alpha\alpha^\prime}
 f_{k\alpha}{\overline f}_{k\alpha^\prime}|d_\alpha\otimes {\overline
 d}_{\alpha^\prime}\rangle~.
\end{equation}
Thus, the aforementioned $9p$ sector states 
would scatter into states, in particular, gauge bosons, 
containing non-diagonal
terms with $|d_\alpha\otimes {\overline d}_{\alpha^\prime}\rangle$ with
$\alpha\not=\alpha^\prime$ absent in (\ref{Dinv}). Thus, we see that the 
$g$ invariant states in the $9p$ sector indeed do not carry well defined gauge
quantum numbers. 

{}The reason why $g$ acts so differently in the 99 and $9p$ sectors is actually
very simple. The Chan-Paton part of the $9p$ vertex operators is proportional
to 
\begin{equation}
 V_{9p}\sim\lambda_9 {\overline \lambda}_p~,
\end{equation}
while in the 99 sector we have 
\begin{equation}
 V_{99}\sim \lambda_9{\overline \lambda}_9~,
\end{equation}
where $\lambda_9$ and ${\overline \lambda}_9$ correspond to the fundamental and
antifundamental of the 99 gauge group $U(N)$, while
$\lambda_p$ and ${\overline \lambda}_p$ correspond to the fundamental and
antifundamental of the $pp$ gauge group $U(N^\prime)$. 
Thus, the action of $g$ in the
$9p$ sector is given by
\begin{equation}
 g:~\lambda_9{\overline \lambda}_p
 \rightarrow \gamma_{g,9}\lambda_9{\overline \lambda}_p
 \gamma_{g,p}^{-1}~,
\end{equation} 
while in the 99 sector we have 
\begin{equation}
 g:~\lambda_9{\overline \lambda}_9 
 \rightarrow \gamma_{g,9}\lambda_9{\overline \lambda}_9
 \gamma_{g,9}^{-1}~.
\end{equation}
Thus, the action of $\gamma_{g,9}$ in the 99 sector is bilinear, while in the
$9p$ sector it is linear, and this is precisely the reason why the two actions
are incompatible in the presence of non-commuting Wilson lines 
(which affect the 99 quantum numbers only) as explained
above.

{}The above discussion implies that we indeed have a conflict between
the facts that gauge bundles of D9-branes wrapped on tori
with $B$-flux lack vector structure, while the presence of $9p$ sectors
with D$p$-branes transverse to such tori implies the presence of certain
vector structure. In fact, the inconsistencies we have encountered in 
orientifolds of Type IIB on $T^4/{\bf Z}_M$ orbifolds with $B$-flux simply
imply that the corresponding choices of the gauge bundles are not consistent
within this framework (albeit consistent choices could be found for the cases
without the $B$-flux). The obstruction to having consistent gauge bundles, 
once again, is related to the lack of vector structure.

\section{Four Dimensional Orientifolds with $B$-flux}

{}In this section we will discuss four dimensional orientifolds with $B$-flux.
In particular, we will consider compactifications with ${\cal N}=1$ as well as
${\cal N}=2$ supersymmetry. More concretely, we will discuss orientifolds of 
Type IIB on $T^6/\Gamma$ orbifolds. If $\Gamma$ is a (non-trivial)
subgroup of $SU(2)$, then we have ${\cal N}=2$ supersymmetry, and if $\Gamma$ 
is a subgroup of $SU(3)$ but not of $SU(2)$, then we have ${\cal N}=1$ 
supersymmetry. In subsection A we will discuss ${\cal N}=2$ examples. In 
subsections B,C,D,E we will discuss ${\cal N}=1$ examples
with the orbifold groups
$\Gamma={\bf Z}_3,{\bf Z}_7,{\bf Z}_3\otimes {\bf Z}_3,
{\bf Z}_6$, respectively.  

\subsection{The ${\cal N}=2$ Models}

{}In this subsection we will discuss orientifolds of Type IIB on $T^2
\otimes {\rm K3}$, 
where ${\rm K3}=T^4/{\bf Z}_M$ ($M=2,3,4,6$). {\em A priori}
the $B$-flux can be 
turned on either on $T^2$ or K3 or both. However, 
we will focus on the cases with $B$-flux such that we will {\em not} 
encounter the difficulties analogous to those we have found in six dimensional
orientifolds with $B$-flux. In particular, the latter will always be either
transverse to or inside of the world-volumes of
{\em all} D-branes present in a given model\footnote{If only one type of 
D-branes is present, which is the case in the ${\bf Z}_3$ models, then we can
have $B$-flux simultaneously turned on 
in the directions inside of as well as transverse to the D-brane
world-volumes.}.

{}Let us discuss this point in a bit more detail. What we have found in the
previous section is that if we simultaneously have D-branes wrapping tori
(or, more precisely, orbifolds thereof) and D-branes transverse to such tori,
then we run into various subtle inconsistencies. The latter would not, for 
instance, occur if all D-branes where transverse to $B$-flux. Similarly, such 
inconsistencies would also be absent if all D-branes are wrapping such tori.
This is precisely the strategy we are going to employ here to construct
consistent four dimensional ${\cal N}=2$ supersymmetric orientifolds with
$B$-flux. Note that in the case of K3 orientifolds with both D9- and D5-branes
we cannot have such a setup. But with three compact complex dimensions
this now becomes possible to achieve.

{}To begin with, we can consider the following setup. Consider the
$\Omega J^\prime$ orientifold\footnote{The action of $J^\prime$ was defined 
in the previous section. In particular, note that $J^\prime=1$ in the ${\bf 
Z}_2$ models.} 
of Type IIB on $T^2
\otimes {\rm K3}$, where ${\rm K3}=
T^4/{\bf Z}_M$ ($M=2,3,4,6$). For $M=2,4,6$, where
we have both D9- and D5-branes, we will
assume that the $B$-field on K3 is trivial, while we have non-zero $B$-flux
on $T^2$. In the $M=3$ case {\em a priori} we can have $B$-flux on both $T^2$
as well as K3.

{}Let us first consider the models with $M=2,4,6$. The $B$-flux on $T^2$ can
be described in terms of the freely acting ${\bf Z}_2\otimes {\bf Z}_2$ 
orbifold with discrete torsion. Note that the latter now commutes with the
${\bf Z}_M$ orbifold action. In fact, the spectra of these models can be
obtained by compactifying the corresponding K3 models of 
\cite{GJ} with trivial $B$-flux on $T^2$ with $B$-flux\footnote{Note that
in the open string sector this is equivalent to compactifying a six
dimensional gauge theory on a non-commutative $T^2$.}. The closed string 
spectra are given by a straightforward dimensional reduction of the 
corresponding six dimensional spectra given in \cite{GJ}\footnote{Note, in 
particular, that the $B$-flux on $T^2$ does {\em not} affect the number of
six dimensional tensor multiplets, which upon the dimensional reduction 
actually give rise to four dimensional vector multiplets.}. As to the open 
string sector, its massless spectrum is obtained by performing the freely
acting ${\bf Z}_2\otimes {\bf Z}_2$ orbifold projections. 
Note that now the Wilson lines act in
both 99 as well as 55 sectors. The corresponding Chan-Paton matrices 
$\gamma_{S_i,9}$ and $\gamma_{S_i,5}$ must be the same (up to equivalent 
representations). This is necessary for the action of the ${\bf Z}_2\otimes
{\bf Z}_2$ freely acting orbifold on the 59 sector states, which
is given by the matrices $\gamma_{S_i,9}
\gamma_{S_i,5}^{-1}$, to be consistent. Moreover, the twisted Chan-Paton 
matrices $\gamma_{R,9}$ and $\gamma_{R,5}$ (which up to equivalent 
representations must be the same so that the ${\bf Z}_2\subset {\bf Z}_M$ 
orbifold twist $R$ acts
consistently on the 59 sector states) have eigenvalues $\pm i$ (and not
$\pm 1$). This follows from the corresponding statement for the six dimensional
${\bf Z}_2$ model with trivial $B$-flux. Note that depending upon whether the
Wilson lines on $T^2$ are of the $D_4$ or $D_4^\prime$ type we have different
models, which, however, are connected as they belong to different points
of the Coulomb branch of the same ${\cal N}=2$ gauge theory.

{}Note that the ranks
of both the 99 and 55 gauge groups are equal\footnote{Here we assume that
the Wilson lines in both the 99 and 55 sectors correspond to the points
of the respective moduli spaces which can be described by the freely
acting ${\bf Z}_2\otimes {\bf Z}_2$ orbifold.} 8. 
In fact, both gauge groups contain ${\bf Z}_2$ discrete
gauge subgroups, but there are no massless states carrying non-trivial
$[{\bf Z}_2]_{99}$ or $[{\bf Z}_2]_{55}$ charges. This, in particular, applies
to the $59$ sector states as well where the multiplicity of states now is
$\xi_{59}=1$. This is because now the Wilson lines {\em do} act in the 59 
sector. In particular, the 59 states carrying non-trivial
$[{\bf Z}_2]_{99}$ or $[{\bf Z}_2]_{55}$ charges are now massive - they are
at heavy Kaluza-Klein levels corresponding to the compactification on $T^2$.
Here we note that in the limit of the large volume $T^2$ the effect of the 
$B$-flux is ameliorated. Thus, as was pointed out in \cite{Ka4}, in this limit
in all three 99, 55 and 59 sectors the Kaluza-Klein states with non-trivial
$[{\bf Z}_2]_{99}$ and/or $[{\bf Z}_2]_{55}$ charges become massless (along
with the Kaluza-Klein states with trivial discrete gauge charges) so that
the freely acting orbifold action is ameliorated, and the massless six 
dimensional states arising in this limit are in the representations of
the full rank $16+16$ 99 plus 55 gauge symmetry. That is, in this limit
we recover the six dimensional K3 orientifolds of \cite{GJ}. 

{}For illustrative purposes let us consider the ${\bf Z}_2$ example. The gauge
group (at the ${\bf Z}_2\otimes {\bf Z}_2$ freely acting orbifold points) is
$U(8)_{99}\otimes U(8)_{55}$. (Here we drop the $[{\bf Z}_2]_{99}\otimes
[{\bf Z}_2]_{55}$ discrete gauge subgroup
as no massless states carry non-trivial 
charges with respect to the latter.) If the Wilson lines on $T^2$ are of the 
$D_4$ type, then the massless hypermultiplets are given by:
\begin{equation}
 2\times ({\bf 36};{\bf 1})~,~~~2\times ({\bf 1};{\bf 36})~,~~~
 ({\bf 8};{\bf 8})~.
\end{equation}
On the other hand, if the Wilson lines on $T^2$ are of the 
$D_4^\prime$ type, then the massless hypermultiplets are given by:
\begin{equation}
 2\times ({\bf 28};{\bf 1})~,~~~2\times ({\bf 1};{\bf 28})~,~~~
 ({\bf 8};{\bf 8})~.
\end{equation} 
As we have already mentioned, these points belong to the same branch of the
moduli space corresponding to the Coulomb branch. 

{}In fact, most of the above discussion also applies to the
$M=3$ cases, except that here we can also have $B$-flux on K3, so that
the corresponding four dimensional models are obtained by compactifying
the six dimensional $\Omega J^\prime$ ${\bf Z}_3$ models with $b=0,2,4$ 
on $T^2$ with $B$-flux, and the former are recovered in the large $T^2$ 
limit\footnote{Here we can also consider $\Omega R J^\prime$ 
${\bf Z}_3$ orientifolds with $B$-flux turned on both on $T^2$ and K3.
In the limit of large volume $T^2$ we recover the six dimensional $\Omega R 
J^\prime$ ${\bf Z}_3$ models (with D5-branes only)
discussed in subsection B of section III. Here we note that there is 
a possibility of a 
non-perturbative inconsistency in the four dimensional 
$\Omega J^\prime$ ${\bf Z}_3$ models with $B$-flux turned on on K3
(regardless of the $B$-flux on $T^2$). This is
in complete parallel with our discussion for the corresponding six dimensional
$\Omega J^\prime$ ${\bf Z}_3$ models. Note, however, that such a 
non-perturbative inconsistency is not expected in the four dimensional
$\Omega J^\prime$ ${\bf Z}_3$ models with $B$-flux turned on on $T^2$ only.
Neither should it occur in the four dimensional $\Omega R J^\prime$ ${\bf Z}_3$
models with $B$-flux on K3 (regardless of the $B$-flux on $T^2$).}.

{}Another class of orientifolds we can consider here is the following.
Let $z_1$ parametrize $T^2$, and $z_2,z_3$ parametrize $T^4$ in ${\rm K3}=
T^4/{\bf Z}_M$. For simplicity let us assume that $T^4=T^2\otimes T^2$, where
$z_2,z_3$ parametrize these two 2-tori. (To avoid confusion, from now on
we will refer to the 2-torus parametrized by $z_1$ as ${\widetilde T}^2$.)
Then we can consider the $\Omega 
R^\prime J^\prime$ orientifold of Type IIB on $T^2\otimes (T^4/{\bf Z}_M)$,
where the action of $R^\prime$ is given by $R^\prime z_1=-z_1$, $R^\prime
z_2=-z_2$, $R^\prime z_3=z_3$.
For $M=3$ we then have only D5-branes wrapping the 2-torus in $T^4$ (or, more 
precisely, an orbifold thereof) parametrized by $z_3$. That is, the locations
of these D5-branes are given by points in the directions $z_1,z_2$. These 
${\bf Z}_3$ models with $B$-field turned on in various directions are 
straightforward to analyze along the lines of our previous discussions. We
will therefore focus on the models with $M=2,4,6$, where we have two types of
D-branes (intersecting at right angles). 
Thus, we have D5-branes wrapping the $T^2$ parametrized by $z_3$.
We also have D5$^\prime$-branes wrapping the $T^2$ parametrized by $z_2$.
This follows from the fact that we have O5-planes whose world-volumes coincide
with the set of points fixed under $R^\prime$, and we also have 
O5$^\prime$-planes whose world-volumes coincide with the set of points fixed 
under $RR^\prime$, $R$ being the generator of the ${\bf Z}_2$ subgroup of
${\bf Z}_M$. Let ${\widetilde R} z_1=-z_1$. There are four points on 
${\widetilde T}^2$ fixed under the action of ${\widetilde R}$: $\xi_0=0$,
$\xi_a=e_a/2$, $a=1,2,3$, 
where $e_i$, $i=1,2$, are the vielbeins on ${\widetilde T}^2$,
and $e_3\equiv -e_1-e_2$. Note that we have total of 12 O5$^-$-planes and
4 O5$^+$-planes. Similarly, we have total of 12 O5$^{\prime -}$-planes and 
4 O5$^{\prime +}$-planes. (This follows from the fact that we have 
half-integer $B$-flux on ${\widetilde T}^2$ but trivial $B$-flux on K3.) 
For definiteness let us assume that 4 O5-planes
corresponding to the $\xi_0$ fixed point are of the O5$^+$ type. Then the other
12 O5-planes corresponding to the fixed points $\xi_a$ are of the O5$^-$ type.
What about the O5$^\prime$-planes? It is not difficult to see that with the
above assumption the 4 O5$^\prime$-planes corresponding to the fixed point
$\xi_0$ must be of the O5$^{\prime +}$ type. Similarly, the other 12 
O5$^\prime$-planes corresponding to the fixed points $\xi_a$ must be of the 
O5$^{\prime -}$ type. This can be seen as follows. First note that the 
consistency of the $R$ projection in the $55^\prime$ sector requires that 
$\gamma_{R,5}$ and $\gamma_{R,5^\prime}$ both have either $\pm i$ or $\pm 1$
eigenvalues. Second, arguments similar to those in subsection A of section III
imply that if $\gamma_{R,5}$ has eigenvalues $\pm 1$, then the 
O5$^\prime$-planes corresponding to a given fixed point on ${\widetilde T}^2$
are of the type opposite to that of the corresponding O5-planes. That is, if
the latter are, say, of the O5$^-$ type, then the former are of the 
O5$^{\prime +}$ type. On the other hand, if 
$\gamma_{R,5}$ has eigenvalues $\pm i$, then the 
O5$^\prime$-planes corresponding to a given fixed point on ${\widetilde T}^2$
are of the same type as the corresponding O5-planes\footnote{In fact, one can 
check the above statements in the following simple way. Note that if we
T-dualize, say, on $T^2$ parametrized by $z_2$ (note that
there are no subtleties with this T-duality procedure as the $B$-flux on this
$T^2$ is trivial), then D5-branes turn into D7-branes, while D5$^\prime$-branes
turn in to D3-branes. Analogous statements also apply to the corresponding
O-planes. Thus, after T-duality transformation we obtain a background
with O3- and O7-planes as well as the corresponding D-branes. For this system
we can straightforwardly repeat the argument in the beginning of subsection A 
of section III (which was carried out for the 59 system but is identical
for the 37 system), which (after T-dualizing back to the $55^\prime$ system)
leads precisely to the aforementioned conclusions.
Equivalently, we can carry out these arguments for the $55^\prime$ system
by noting that in the $55^\prime$ sector we have 4 Neumann-Dirichlet
boundary conditions just as in the 37 sector or 59 sector.}. 
Since we must have 4 
O5$^+$-planes and 12 O5$^-$-planes as well as 4 O5$^{\prime +}$-planes and
12 O5$^{\prime -}$-planes (which follows from the tadpole cancellation 
conditions), it is then clear that $\gamma_{R,5}$ must have eigenvalues 
$\pm i$, and O5- and O5$^\prime$-planes corresponding to a given fixed point
on ${\widetilde T}^2$ must be of the same type\footnote{Note that this 
statement for the $\Omega R^\prime J^\prime$ orientifolds
is the analogue of the statement for the corresponding $\Omega J^\prime$
orientifolds that the Wilson lines in the 99 and 55 sectors must be of the
same type - see above for details.}. 

{}The massless spectra of the $\Omega R^\prime J^\prime$ orientifolds
are the same as those of the corresponding $\Omega J^\prime$ orientifolds.
However, the massive spectra differ. Note, for instance, that in the $\Omega
J^\prime$ orientifold we have 32 D9-branes as well as 32 D5-branes. The rank of
the 99 and 55 gauge groups, however, is 8, and we have the $[{\bf Z}_2]_{99}
\otimes [{\bf Z}_2]_{55}$ discrete gauge symmetry under which some massive
Kaluza-Klein states are charged non-trivially. Such a discrete gauge symmetry
is absent in the corresponding $\Omega R^\prime J^\prime$ orientifolds - we
have only 16 D5-branes and 16 D5$^\prime$-branes. Furthermore, if we take the
size of ${\widetilde T}^2$ in the $\Omega R^\prime J^\prime$ orientifolds
to infinity, we will {\em not} obtain six dimensional theories - Lorentz
invariance in these backgrounds is always broken to that of a four dimensional
theory. In particular, as was pointed out in \cite{Ka4}, the rank of the gauge
group in such models in {\em not} enhanced in the large ${\widetilde T}^2$ 
limit - this is,
in fact, a direct consequence of the fact that the number of each type of 
branes is only 16. What happens in the large ${\widetilde T}^2$ limit is that
the O-planes corresponding to the fixed points $\xi_a$ on ${\widetilde T}^2$
are removed to infinity and decouple from the remaining O-planes
corresponding to the fixed point $\xi_0$.

{}Before we end this subsection, we would like to make a remark on the 
motion of branes in the $\Omega R^\prime J^\prime$ models. In particular, note
that the motion of branes in the direction of ${\widetilde T}^2$ corresponds 
to different points on the Coulomb branch of the ${\cal N}=2$ gauge theory.
Suppose that D5-branes and D5$^\prime$-branes are on top of the respective
O-planes corresponding to the {\em same} fixed point on ${\widetilde T}^2$.
Then we have a non-trivial $55^\prime$ massless matter content. If we move,
say, D5-branes off the corresponding O5-plane while leaving D5$^\prime$-branes
untouched, then this corresponds to Higgsing the 55 sector gauge group, while
the $5^\prime 5^\prime$ gauge group is untouched. Note that the $55^\prime$ 
states in this process become heavy (due to Higgsing). In the brane language 
this is simply the statement that $55^\prime$ strings now cannot have zero
length, so that the corresponding states are always heavy. On the other hand,
if we move D5- and D5$^\prime$-branes off the corresponding fixed points 
together, then the $55^\prime$-sector states remain massless. In the
gauge theory language this corresponds to a special subspace of the
Coulomb branch where the $55^\prime$ hypermultiplets remain massless as the
mass term coming from the coupling to the 55 Higgs field is precisely cancelled
by the mass term coming from the coupling to the $5^\prime 5^\prime$ Higgs
field.

\subsection{The ${\cal N}=1$ ${\bf Z}_3$ Models}

{}In this subsection we would like to discuss four dimensional ${\cal N}=1$ 
orientifolds of Type IIB on $T^6/{\bf Z}_3$, where the generator $\theta$ of
${\bf Z}_3$ acts on the complex coordinates $z_I$, $I=1,2,3$,
parametrizing $T^6$ as follows: $\theta z_I=\omega z_I$, where $\omega\equiv
\exp(2\pi i/3)$. {\em A priori} we can consider various orientifolds
with O9-, O7-, O5- or O3-planes with $B$-flux turned on inside of and/or
transverse to their world-volumes. Since other cases are straightforward to
consider along the lines of our previous discussions, here we will focus on 
the models with O3-planes\footnote{Here we note that if we have $B$-flux 
inside of the world-volumes of O$p$-planes and the corresponding D$p$-branes
in this background, then, just as in the corresponding
six dimensional ${\bf Z}_3$ models, there is a possibility of a 
non-perturbative inconsistency. If we, however, consider the models with 
O3-planes, such a non-perturbative inconsistency is not expected to arise.}. 
Thus, we will discuss $\Omega R^\prime J^\prime (-1)^{F_L}$ 
orientifolds of Type IIB on $T^6/{\bf Z}_3$, where $R^\prime z_I=-z_I$, and
the action of $J^\prime$ is analogous\footnote{Note, however, that, unlike the
six dimensional cases, in the four dimensional orientifolds where some of the
orbifold group elements twist all three complex coordinates $z_I$,
the action of 
$J^\prime$ (which is trivial in the untwisted sector, while it maps a 
$g$ twisted sector to its inverse $g^{-1}$ twisted sector) does not seem to
have a well defined geometric interpretation \cite{KST1}. Here we will ignore 
potential difficulties with such an interpretation (which could be seen 
\cite{KST1}, for instance, by considering a map \cite{sen1} of these
orientifolds to F-theory 
\cite{vafa}), and assume that such an action is well defined in the
conformal field theory context.} 
to that in the six dimensional 
orientifolds discussed in subsections B,C,D of section III.
In such an orientifold we have $n_{f-}=32+32/2^{b/2}$ O3$^-$-planes, and
$n_{f+}=32-32/2^{b/2}$ O3$^+$-planes, where $b$ is the rank of the $B$-flux. 
Moreover, it is not difficult to show \cite{Ka2}
that the twisted tadpole cancellation
conditions read: 
\begin{equation}
 {\rm Tr}(\gamma_\theta)=-(-1)^{b/2}\times 4~.
\end{equation}
Note that for the untwisted Chan-Paton matrix we have ${\rm Tr}(\gamma_I)=
32/2^{b/2}$.

{}Let us consider the $b=2,4,6$ cases separately\footnote{Here we should 
point out that these models were originally discussed in \cite{ABPSS1}. More
precisely, the spectra of the $b=2,6$ models given in \cite{ABPSS1} 
were erroneous as it was not realized there that the consistent orientifold
projection in these cases is of the $Sp$ type. This was originally corrected
in \cite{Ka2}, and more recently in \cite{An}.}.\\
$\bullet$ For $b=2$ we have $n_{f-}=48$ O3$^-$-planes and $n_{f+}=16$ 
O3$^+$-planes. The ${\bf Z}_3$ symmetry requires that the O3-plane at the
origin of $T^6$ be of the O3$^+$ type. If we place all 16 D3-branes on top
of this O3-plane, then the gauge group is (note that ${\rm Tr}(\gamma_\theta)=
+4$ in this case) $U(4)\otimes Sp(8)$, and the massless open string sector 
contains chiral supermultiplets in
\begin{equation}
 \Phi_s=3\times
 ({\bf 10},{\bf 1})~,~~~
 Q_s=3\times ({\overline {\bf 4}},{\bf 8})~,~~~s=1,2,3~. 
\end{equation}
There is a non-trivial superpotential in this model given by:
\begin{equation}
 {\cal W}=\epsilon_{ss^\prime s^{\prime\prime}}\Phi_s Q_{s^\prime}
 Q_{s^{\prime\prime}}~.
\end{equation}
$\bullet$ For $b=4$ we have $n_{f-}=40$ O3$^-$-planes and $n_{f+}=24$ 
O3$^+$-planes. The ${\bf Z}_3$ symmetry requires that the O3-plane at the
origin of $T^6$ be of the O3$^-$ type. If we place all 8 D3-branes on top
of this O3-plane, then the gauge group is (note that ${\rm Tr}(\gamma_\theta)=
-4$ in this case) $U(4)$, and the massless open string sector 
contains chiral supermultiplets in $\Phi_s=3\times {\bf 6}$. There are no
renormalizable couplings in this model.\\
$\bullet$ For $b=6$ we have $n_{f-}=36$ O3$^-$-planes and $n_{f+}=28$ 
O3$^+$-planes. The ${\bf Z}_3$ symmetry requires that the O3-plane at the
origin of $T^6$ be of the O3$^+$ type. If we place all 4 D3-branes on top
of this O3-plane, then the gauge group is (note that ${\rm Tr}(\gamma_\theta)=
+4$ in this case) $Sp(4)$, and there are no massless open string sector 
states in this model.

\subsection{The ${\cal N}=1$ ${\bf Z}_7$ Models}

{}In this subsection we discuss four dimensional ${\cal N}=1$ orientifolds
of Type IIB on $T^6/{\bf Z}_7$, where the generator $g$ of ${\bf Z}_7$ acts
on the complex coordinates $z_I$, $I=1,2,3$, parametrizing $T^6$ as follows:
$g z_1=\alpha$, $g z_2=\alpha^2 z_2$, $g z_3=\alpha^4 z_3$, where
$\alpha\equiv \exp(2\pi i/7)$. As in the previous section, let us focus on the
cases where we have O3-planes. Thus, consider the $\Omega R^\prime J^\prime
(-1)^{F_L}$ orientifold of Type IIB on $T^6/{\bf Z}_7$ ($R^\prime
z_I=-z_I$). 
Here we note that the
rank of the $B$-flux can take only two values in this case: $b=0,6$. The
reason why is that only for these values of $b$ does the ${\bf Z}_7$ orbifold 
act crystallographically on $T^6$. Another way of seeing this is as follows.
Note that we have $n_{f\mp}=32\pm 32/2^{b/2}$ O3$^\mp$-planes for the rank $b$
$B$-flux. The O3-plane at the origin is invariant under the ${\bf Z}_7$
twist $g$. However, O3-planes at other fixed points of $R^\prime$ must come in
groups of 7 such that they are permuted by the ${\bf Z}_7$ orbifold within each
group. However, for $b=2,4$ neither $n_{f-}-1$ nor $n_{f+}-1$ are divisible
by 7. For $b=0,6$ both $n_{f-}-1$ and $n_{f+}$ are divisible by 7, so we 
conclude that in these cases the O3-plane at the origin is of the
O3$^-$ type. In the $b=6$ case
we must place all 4 D3-branes at this O3-plane, which
is consistent with the twisted tadpole cancellation condition ${\rm Tr}
(\gamma_g)=+4$ \cite{Ka2}. The gauge group in this case is $SO(4)$ with no
massless open string matter. 

\subsection{The ${\cal N}=1$ ${\bf Z}_3\otimes {\bf Z}_3$ Models}

{}In this subsection we would like to discuss four dimensional ${\cal N}=1$
orientifolds of Type IIB on $(T^2\otimes T^2\otimes T^2)/({\bf Z}_3
\otimes {\bf Z}_3)$, where the action of the generators $\theta$ and 
$\theta^\prime$ of the two ${\bf Z}_3$ subgroups on the complex coordinates
$z_I$, $I=1,2,3$, parametrizing the three 2-tori is as follows
($\omega\equiv \exp(2\pi i/3)$): 
$\theta z_1=\omega z_1$, $\theta z_2=\omega^{-1} z_2$, $\theta z_3=z_3$,
$\theta^\prime z_1=z_1$, $\theta^\prime z_2=\omega z_2$, $\theta^\prime
 z_3=\omega^{-1} z_3$. Here we will focus on the $\Omega R^\prime J^\prime 
(-1)^{F_L}$ orientifolds ($R^\prime z_I=-z_I$). Using our previous results it
is then no difficult to show that we have the following 
models\footnote{Solutions to the tadpole cancellation conditions for these
models were found in \cite{Ka2}.}:\\
$\bullet$ $b=2$. The gauge group is $U(4)\otimes U(4)$ with the following
massless open string sector chiral multiplets
\begin{equation}
 Q=({\overline {\bf 4}},{\overline{\bf 4}})~,~~~
 R=({\bf 4},{\overline{\bf 4}})~,~~~\\
 \Phi=({\bf 1},{\bf 10})~,
\end{equation} 
and the superpotential
\begin{equation}
 {\cal W}=QR\Phi~.
\end{equation}
$\bullet$ $b=4$. The gauge group is $U(4)$, and in the open string sector 
we have a massless chiral multiplet in ${\bf 6}$ of $U(4)$. There are no 
renormalizable couplings in this case.\\ 
$\bullet$ $b=6$. The gauge group is $Sp(4)$, and there is no massless matter
in the open string sector.

\subsection{The ${\cal N}=1$ ${\bf Z}_6$ Models}

{}In this subsection we would like to discuss four dimensional ${\cal N}=1$
orientifolds of Type IIB on $T^6/{\bf Z}_6$, where the generators $\theta$ and
$R$ of the ${\bf Z}_3$ and ${\bf Z}_2$ subgroups of the ${\bf Z}_6$ orbifold 
group act on the complex coordinates $z_I$, $I=1,2,3$, parametrizing the 
three 2-tori (we assume that $T^6=T^2\otimes T^2\otimes
T^2$) as follows ($\omega\equiv\exp(2\pi i/3)$): 
$\theta z_I=\omega z_I$, $Rz_1=z_1$, $R z_{2,3}=-z_{2,3}$.
Note that the $\Omega J^\prime$ 
${\bf Z}_6$ model with trivial $B$-flux was originally
constructed in the second reference in \cite{KS}. In \cite{ST}\footnote{This
model, among other four dimensional ${\cal N}=1$ models with and without 
$B$-flux, was also discussed in \cite{Ka2}.} 
the following 
model with $B$-flux was discussed\footnote{This model was discussed in
\cite{ST} in the phenomenological context of TeV-scale brane world.
For a partial list of other recent developments in these directions, 
see \cite{TeV}.}. 
Consider the $\Omega J^\prime$ 
orientifold of Type IIB on $(T^2\otimes T^2\otimes T^2)/{\bf Z}_6$ with
$b=2$ $B$-flux turned on on the second or third $T^2$ only. 
It should be clear that
in this model we are going to have inconsistencies similar to those we
have encountered in the six dimensional ${\bf Z}_6$ models with 
$B$-flux\footnote{Note that the O5-plane at the $R$ fixed point $z_2=z_3=0$
must be of the O5$^+$ type for the corresponding background to be ${\bf
Z}_3$ symmetric. This would then require that the twisted Chan-Paton matrices
$\gamma_{R,5}$ as well as $\gamma_{R,9}$ have eigenvalues $\pm 1$.
In \cite{ST}, however, these Chan-Paton matrices were assumed to have 
eigenvalues $\pm i$.}. To avoid these difficulties we can assume that $B$-flux
is turned on {\em inside} of both D9- and D5-branes present in this case.
That is, consider the $\Omega J^\prime$ ${\bf Z}_6$ model with $b=2$ 
$B$-flux turned on on the first $T^2$. The corresponding Wilson lines must be
of the same type in both 99 and 55 sectors. In fact, to be compatible with
the ${\bf Z}_3$ orbifold action, they must be of the $D_4$ (and not 
$D_4^\prime$) type. The important point here is that the multiplicity of
states in the 59 sector (before the ${\bf Z}_3$ orbifold projection) is 
$\xi_{59}=1$ (and not 2, which would be the case in the model of \cite{ST} had
it been consistent). 

{}Instead of describing the massless spectrum of the above model, we will
discuss that of a different model\footnote{The reason for this is that in the 
former model there is a possibility of a non-perturbative inconsistency
arising along the lines of our previous discussions, while in the model we are
going to discuss next such an inconsistency is not expected to arise.} 
(these two models actually have identical
massless spectra albeit their massive spectra are different). Thus, consider
the $\Omega R^\prime J^\prime$ orientifold of Type IIB on $(T^2\otimes T^2
\otimes T^2)/{\bf Z}_6$, where $R^\prime z_{1,2}=-z_{1,2}$, $Rz_3=z_3$.
Let us turn on $b=2$ $B$-flux on the first $T^2$ parametrized by $z_1$.
Note that in this model we have two types of D-branes (intersecting at right
angles). Thus, we have D5-branes wrapping the $T^2$ parametrized by $z_3$.
We also have D5$^\prime$-branes wrapping the $T^2$ parametrized by $z_2$. Just
as in the previous subsection, the O5- and O5$^\prime$-planes corresponding to
the same fixed point on $T^2$ parametrized by $z_1$ are of the same type. 
Moreover, to be compatible with the ${\bf Z}_3$ orbifold action, the
4 O5-planes corresponding to the origin of this $T^2$ must be of the 
O5$^+$ type, and, similarly, the 4 O5$^\prime$-planes corresponding   
to the origin of this $T^2$ must be of the 
O5$^{\prime +}$ type. The other 12 O5-planes are of the O5$^-$ type, and, 
similarly, the other 12 O5$^\prime$-planes are of the O5$^{\prime -}$ type.
Let us place all 16 D5-branes and 16 D5$^\prime$-branes at 
an O5$^+$-plane and an O5$^{\prime +}$-plane, respectively. The solution to the
twisted tadpole cancellation conditions reads (up to equivalent 
representations) \cite{Ka2}:
\begin{eqnarray}
 && 
 \gamma_{\theta,5}=\gamma_{\theta,5^\prime}={\rm diag}(1,1,\omega,\omega^{-1})
 \otimes I_4~,\\
 && 
 \gamma_{R,5}=\gamma_{R,5^\prime}=I_4
 \otimes i\sigma_3 \otimes I_2~.
\end{eqnarray} 
The gauge group of this model is $[U(2)\otimes U(2)\otimes U(4)]_{55}\otimes
[U(2)\otimes U(2)\otimes U(4)]_{99}$, and the massless open string 
chiral matter reads:
\begin{eqnarray}
 &&\Phi_{1,2}=
 2\times ({\bf 3},{\bf 1},{\bf 1};{\bf 1},{\bf 1},{\bf 1})_{55}~,~~~
 {\widetilde \Phi}_{1,2}=
 2\times ({\bf 1},{\overline {\bf 3}},{\bf 1};
 {\bf 1},{\bf 1},{\bf 1})_{55}~,~~~\\
 && P_{1,2}=
 2\times ({\overline {\bf 2}},{\bf 1},{\overline {\bf 4}};
     {\bf 1},{\bf 1},{\bf 1})_{55}~,~~~
 {\widetilde P}_{1,2}=
 2\times ({\bf 1},{\bf 2},{\bf 4};
     {\bf 1},{\bf 1},{\bf 1})_{55}~,\\
 && P_3= ({\overline {\bf 2}},{\bf 1},{\bf 4};
     {\bf 1},{\bf 1},{\bf 1})_{55}~,~~~
 {\widetilde P}_3=
 ({\bf 1},{\bf 2},{\overline {\bf 4}};
     {\bf 1},{\bf 1},{\bf 1})_{55}~,~~~R=({\bf 2},{\overline {\bf 2}},{\bf 1};
     {\bf 1},{\bf 1},{\bf 1})_{55}~,\\
 &&\Phi^\prime_{1,2}=
 2\times ({\bf 1},{\bf 1},{\bf 1};{\bf 3},{\bf 1},{\bf 1})_{5^\prime 
 5^\prime}~,~~~ 
 {\widetilde \Phi}^\prime_{1,2}=
 2\times ({\bf 1},{\bf 1},{\bf 1};{\bf 1},{\overline {\bf 3}},
 {\bf 1})_{5^\prime 5^\prime}~,\\
 && P^\prime_{1,2}=
 2\times ({\bf 1},{\bf 1},{\bf 1};{\overline {\bf 2}},{\bf 1},
 {\overline {\bf 4}})_{5^\prime 5^\prime}~,~~~
 {\widetilde P}^\prime_{1,2}=
 2\times (
     {\bf 1},{\bf 1},{\bf 1};{\bf 1},{\bf 2},{\bf 4})_{5^\prime 
 5^\prime}~,\\
 && P^\prime_3=(
     {\bf 1},{\bf 1},{\bf 1};{\overline {\bf 2}},{\bf 1},
 {\bf 4})_{5^\prime 5^\prime}~,~~~
 {\widetilde P}^\prime_3=(
     {\bf 1},{\bf 1},{\bf 1};{\bf 1},{\bf 2},{\overline 
 {\bf 4}})_{5^\prime 5^\prime}~,~~~
 R^\prime=(
     {\bf 1},{\bf 1},{\bf 1};{\bf 2},{\overline {\bf 2}},
 {\bf 1})_{5^\prime 5^\prime}~,\\
 && S=({\bf 2},{\bf 1},{\bf 1};{\bf 2},{\bf 1},{\bf 1})_{55^\prime}~,~~~
    T=({\bf 1},{\bf 2},{\bf 1};{\bf 1},{\bf 1},{\bf 4})_{55^\prime}~,~~~
    U=({\bf 1},{\bf 1},{\bf 4};{\bf 1},{\bf 2},{\bf 1})_{55^\prime}~,\\
 &&{\widetilde S}=({\bf 1},{\overline {\bf 2}},{\bf 1};{\bf 1},
 {\overline {\bf 2}}, {\bf 1})_{55^\prime}~,~~~ 
 {\widetilde T}=({\overline {\bf 2}},{\bf 1},{\bf 1};{\bf 1},{\bf 1},
 {\overline{\bf 4}})_{55^\prime}~,~~~ 
 {\widetilde U}=({\bf 1},{\bf 1},{\overline {\bf 4}};{\overline {\bf 2}},
 {\bf 1},{\bf 1})_{55^\prime}~,
\end{eqnarray}
where ${\bf 2}$ and ${\overline {\bf 2}}$ of $U(2)$ carry the $U(1)$ charges
$+1$ and $-1$, respectively, while ${\bf 3}$ and ${\overline {\bf 3}}$ of 
$U(2)$ carry the $U(1)$ charges $+2$ and $-2$, respectively. Similarly,
${\bf 4}$ and ${\overline {\bf 4}}$ of $U(4)$ carry the $U(1)$ charges
$+1$ and $-1$, respectively. Note that this spectrum is the same as that
in \cite{ST} except for the $55^\prime$ sector multiplicity of states, which 
in the above model is 1, while in \cite{ST} it was assumed to be 2 (in the
corresponding 59 sector)\footnote{Here we should point out that the
aforementioned multiplicity being 1 is consistent with the $U(1)$ anomaly
cancellation in the above model via the generalized Green-Schwarz mechanism,
while attempts to implement the latter with the aforementioned multiplicity
being 2 (as in \cite{ST}) appear to run into various difficulties 
\cite{cvetic1}.}. 

{}The above model is expected to be consistent. The superpotential in this 
model is given by:
\begin{eqnarray}
 {\cal W}=&&P_1 {\widetilde P}_2 R+ P_2 {\widetilde P}_1 R +
 \Phi_1 P_2 P_3 +\Phi_2 P_1 P_3 + {\widetilde \Phi}_1 {\widetilde P}_2
 {\widetilde P}_3 +
 {\widetilde \Phi}_2 {\widetilde P}_1 {\widetilde P}_3+\nonumber\\
 &&
 P_1^\prime {\widetilde P}_2^\prime R^\prime + P_2^\prime 
 {\widetilde P}_1^\prime R^\prime +
 \Phi_1^\prime P_2^\prime P_3^\prime +\Phi_2^\prime P_1^\prime P_3^\prime + 
 {\widetilde \Phi}_1^\prime {\widetilde P}_2^\prime {\widetilde P}_3^\prime +
 {\widetilde \Phi}_2^\prime {\widetilde P}_1^\prime {\widetilde P}_3^\prime+
 \nonumber\\
 &&S {\widetilde U} P_3+ U {\widetilde S} {\widetilde P}_3 +
 T {\widetilde T} R+
 S {\widetilde T} P_3^\prime+
 T {\widetilde S} {\widetilde P}_3^\prime +
 U {\widetilde U}  R^\prime~.
\end{eqnarray}
Using this superpotential it is not difficult to see that if we Higgs the gauge
group along the lines of \cite{ST}\footnote{In particular, 
assume that the $S$ and ${\widetilde S}$ fields acquire
non-zero VEVs, which break the
gauge group down to $U(2)_{\rm diag}\otimes U(2)_{\rm diag} \otimes U(4)_{55}
\otimes U(4)_{5^\prime 5^\prime}$. Then the observation of \cite{ST} is that
one can treat, say, the 
$U(2)_{\rm diag}\otimes U(2)_{\rm diag} \otimes U(4)_{55}$ part of this
gauge group as the Pati-Salam gauge symmetry. Here we should point out that
there are certain subtleties related to the $U(1)$ factors in the
aforementioned Higgsing, which we will not discuss in this paper as the key 
point here is that even if such Higgsing is possible, the number of
chiral Pati-Salam generations still cannot be 3.}, 
then
the number of remaining chiral generations for the Pati-Salam gauge group is
2 (and not 3 as it was originally intended in \cite{ST}), which is due to the
fact that the multiplicity of the $55^\prime$
states in the above model is only 1 (and not
2 as was assumed in \cite{ST} for the 59 sector states).

{}Before we end this subsection, let us note that in orientifolds of 
the above ${\bf Z}_6$ orbifold compactification we cannot turn on $b=4,6$
$B$-flux without running into the aforementioned inconsistencies - for $b=4,6$
we cannot avoid having two sets of D-branes such that $B$-flux is inside of
the world-volumes of one set of D-branes while the other set is transverse to 
it\footnote{Note that, for essentially the same reasons, 
the ${\cal N}=1$ ${\bf Z}_2\otimes {\bf Z}_2$ as well
as ${\bf Z}_2\otimes {\bf Z}_2\otimes {\bf Z}_3$ models with $B$-flux discussed
in \cite{Ka2} also suffer from various subtle inconsistencies for all three
values of $b=2,4,6$. Note, however, that the ${\bf Z}_2\otimes {\bf Z}_2\otimes
{\bf Z}_3$ model with trivial $B$-flux originally constructed in the first 
reference in \cite{Ka1} is consistent. This model when interpreted in the
phenomenological context has three chiral generations, and its phenomenological
implications were studied in \cite{KTKT}.}.

\section{Comments}

{}In this section we would like to comment on various issues related
to discussions in the previous sections. To begin with, let us note that
non-trivial multiplicity of states, which is due to non-trivial $B$-flux, 
in the 59 sectors of some of the models discussed
in the previous sections appears to be in conflict with ${\bf Z}_M$ 
orbifold projections (with $M=4,6$) acting on the corresponding coordinates
(that is, those transverse to D5-branes). This observation might be relevant 
for other compactifications with such non-trivial multiplicity of states
in $pp^\prime$ sectors. In particular, such sectors arise in some
of the models recently discussed in \cite{BLUMEN}, where the orientifold
action involves complex conjugation of the compact coordinates. The analogue
of the quantized $B$-flux in such backgrounds is given by the components of the
complex structure corresponding to the $B$-flux in the K{\"a}hler structure
under the interchange of the complex and K{\"a}hler structures. In fact, the
origin of non-trivial multiplicity of states in such orientifolds is
also analogous to that in orientifolds with non-zero $B$-flux, so 
{\em a priori} there might be a possibility of subtle inconsistencies, similar
to those we have found in the latter backgrounds, also arising in the former
ones. It would be interesting to understand this issue in a bit more detail,
but this would be outside of the scope of this paper.

{}The second comment concerns possible generalizations to non-supersymmetric
backgrounds with $B$-flux. Recently compact non-supersymmetric orientifolds 
have been constructed in, for instance, \cite{uranga,quevedo} by introducing 
both D-branes as well as anti-D-branes in orientifolds of Type IIB on orbifolds
that preserve some number of supersymmetries. However, regardless of the
$B$-flux such models have one peculiar feature that some of the NS-NS tadpoles
must be non-vanishing. The reason for this is that in the aforementioned 
construction one introduces only O$^{--}$- and/or O$^{++}$-planes 
(see section II for
notations) which are supersymmetric in the sense that the NS-NS and R-R
tadpoles for each of these O-planes are identical. However, if we now introduce
both D-branes and anti-D-branes, then it is clear that we cannot cancel both
NS-NS and R-R tadpoles simultaneously. Thus, a D-brane has R-R charge $+1$, 
while
an anti-D-brane has R-R charge $-1$. 
Both of these objects, however, give rise to
identical NS-NS tadpoles. Thus, if we require R-R tadpole cancellation, then
we have some uncanceled NS-NS tadpoles. In \cite{uranga} the standard
argument was employed that such tadpoles might not be dangerous as they could
possibly be dealt with via the Fischler-Susskind mechanism. However, 
{\em a priori} it is unclear what the corresponding consistent backgrounds
would be if any\footnote{This is analogous to the situation in the
non-supersymmetric $Sp(32)$ theory discussed in section II. In fact, some
of the aforementioned backgrounds could be thought of as compactifications
of this theory.}.

{}Actually, we can understand this point in a bit more detail. If some
of the twisted NS-NS tadpoles are non-zero, then we would have to employ
the Fischler-Susskind mechanism for twisted (as well as untwisted) scalar
fields that couple to D-branes and O-planes. Shifting the VEVs of the
twisted scalars, however, implies that we blow up the orbifold, and the
consistent background is no longer an exactly solvable conformal field theory.
Orientifolds of such backgrounds are difficult to study, so it is unclear
what the resulting theory would look like if there at all exists the 
corresponding consistent background. One things, however, is quite clear -
{\em a priori}
there is no reason to believe that, if such a non-orbifold background indeed 
exists, the open string spectrum of the theory would be the same.

{}There is, however, a way to avoid the aforementioned difficulty by 
considering backgrounds where all twisted NS-NS tadpoles cancel. Nonetheless,
we have some uncanceled untwisted NS-NS tadpoles, which imply that VEVs 
of some of the untwisted closed string scalars must be shifted. This, in turn,
gives us a hint of what the corresponding consistent backgrounds could
look like. Thus, imagine that we have D$p$-branes as well as 
D${\overline p}$-branes transverse to some compact coordinates. To avoid
appearance of open string tachyons, we must place D$p$-branes and 
D${\overline p}$-branes far enough apart from each other. In fact, to avoid
a possibility of brane-anti-brane annihilation, we can assume that branes
and anti-branes are stuck at, say, the corresponding orbifold
fixed points (for definiteness we will assume that at the orbifold fixed
point located at the origin we have branes and not anti-branes). 
Then tachyons are absent if the separation between the fixed
points (related to the compactification radii)
is large enough. However, supersymmetry is broken, and the
cosmological constant is non-zero. In fact, it depends upon the 
compactification radii. For large enough values of the latter the vacuum
energy monotonically decreases with radii. In fact, in the decompactification 
limit the branes and anti-branes decouple\footnote{This is the case if the 
number of decompactified dimensions is larger than 2.} 
from each other, and the resulting
theory with branes located at the origin is now supersymmetric. In fact,
all tadpoles in this theory (that is, both R-R and NS-NS tadpoles) cancel.
Thus, here we see that there exists a consistent background for such 
theories which can be reached via the Fischler-Susskind mechanism - it is
a (partially) decompactified background with branes only, which is 
supersymmetric. The latter has lower vacuum energy compared with the
original compactified theory with {\em runaway} scalar potential (as a function
of compactification radii). In this respect such compactifications are 
similar to the standard Scherk-Schwarz type of compactifications
(or generalizations thereof) which are unstable to decompactification and 
eventually end up in a supersymmetric vacuum.

{}Note that we can construct non-supersymmetric theories with both
O$^{--}$/O$^{++}$ (that is, supersymmetric) as well as O$^{-+}$/O$^{+-}$
(that is, non-supersymmetric) orientifold planes, where all R-R as well as 
NS-NS tadpoles cancel. Non-compact versions
of such theories (which contain tachyons) were originally constructed
in \cite{non-susy}\footnote{There such models were discussed in the
context of large $N$ gauge theories, so that the presence of
closed string tachyons did not pose a problem \cite{BKV}.}. One can
easily generalize this construction to compact
cases where one can avoid tachyons by considering (partially) freely acting
orbifolds. However, as usual, the vacuum energy in such models has a runaway
behavior with the stable supersymmetric vacuum reached in a (partial) 
decompactification limit.

{}Finally, we note that some of the discussions of non-perturbative K3
orientifolds with $B$-flux in the third reference in \cite{Ka3}
appear to be affected by the results of this paper, 
in particular, we expect various aforementioned subtleties arising in these
models as well,
so that they
should also be revisited. This, however, is outside of the scope of this paper,
and a more detailed discussion of such and related compactifications
will be given elsewhere\footnote{In certain (limited) cases non-perturbative
orientifolds of \cite{Ka3} can be described in terms of perturbative 
orientifolds of non-geometric backgrounds. The latter approach makes it
possible to better understand the former, including the issues related to 
non-zero $B$-flux.}. Also, some other ${\cal N}=1$ models (such as the 
${\bf Z}_6^\prime$ models discussed in the first reference in \cite{Ka3}) 
should also be considered in this context.

\acknowledgments

{}I would like to thank Ofer Aharony, 
Alex Buchel, Amihay Hanany, Martin Ro{\v c}ek, Gary Shiu, Tom Taylor 
and Henry Tye
for valuable
discussions. This work was supported in part by the National Science 
foundation. Parts of this work were completed while I was visiting ITP at UCSB
as part of the 1999 ITP program ``Supersymmetric Gauge Dynamics and String 
Theory''. I would like to thank the organizers and participants of the program
for creating a stimulating atmosphere, and ITP for their kind hospitality. 
Parts of this manuscript were typed up during my stays 
at Rutgers University and Harvard University. 
I would also like to thank Albert and Ribena Yu for financial support.


\begin{references}

\bibitem{PS} G. Pradisi and A. Sagnotti, Phys. Lett. {\bf B216} (1989) 59;\\
M. Bianchi and A. Sagnotti, Phys. Lett. {\bf B247} (1990) 517; Nucl. Phys. 
{\bf B361} (1991) 539. 

\bibitem{GP} E.G. Gimon and J. Polchinski, Phys. Rev. {\bf D54} (1996) 1667.

\bibitem{GJ} E.G. Gimon and C.V. Johnson, Nucl. Phys. {\bf B477} (1996) 715;\\
A. Dabholkar and J. Park, Nucl. Phys. {\bf B477} (1996) 701.

\bibitem{ABPSS} C. Angelantonj, M. Bianchi, G. Pradisi, A. Sagnotti and Ya.S.
Stanev, Phys. Lett. {\bf B387} (1996) 743. 

\bibitem{KST} Z. Kakushadze, G. Shiu and S.-H.H. Tye, Phys. Rev. {\bf D58}
(1998) 086001.

\bibitem{BL} M. Berkooz and R.G. Leigh, Nucl. Phys. {\bf B483} (1997) 187.

\bibitem{ABPSS1} C. Angelantonj, M. Bianchi, G. Pradisi, A. Sagnotti and Ya.S.
Stanev, Phys. Lett. {\bf B385} (1996) 96.

\bibitem{Ka} Z. Kakushadze, Nucl. Phys. {\bf B512} (1998) 221.

\bibitem{KS} Z. Kakushadze and G. Shiu, Phys. Rev. {\bf D56} (1997) 3686;
Nucl. Phys. {\bf B520} (1998) 75.

\bibitem{Zw} G. Zwart, Nucl. Phys. {\bf B526} (1998) 378.

\bibitem{AFIV} G. Aldazabal, A. Font, L.E. Ib{\'a}{\~n}ez and G. Violero,
Nucl. Phys. {\bf B536} (1998) 29.

\bibitem{KST1} Z. Kakushadze, G. Shiu and S.-H.H. Tye, Nucl. Phys. {\bf
B533} (1998) 25.

\bibitem{Ka1} Z. Kakushadze, Phys. Lett. {\bf B434} (1998) 269; 
Phys. Rev. {\bf D58} (1998) 101901.

\bibitem{ST} G. Shiu and S.-H.H. Tye, Phys. Rev. {\bf D58} (1998) 106007.

\bibitem{Ka2} Z. Kakushadze, Nucl. Phys. {\bf B535} (1998) 311.

\bibitem{LPT} J. Lykken, E. Poppitz and S.P. Trivedi, Nucl. Phys. {\bf B543}
(1999) 105. 

\bibitem{BW} R. Blumenhagen and A. Wisskirchen, Phys. Lett. {\bf B438} (1998)
52.

\bibitem{IRU} L.E. Ib{\'a}{\~n}ez, R. Rabadan and A.M. Uranga, Nucl. Phys.
{\bf B542} (1999) 112.

\bibitem{Ka3} Z. Kakushadze, Phys. Lett. {\bf B455} (1999) 120; hep-th/9904211;
Phys. Lett. {\bf B459} (1999) 497.

\bibitem{cvetic} M. Cveti{\v c}, M. Pl{\" u}macher, J. Wang, hep-th/9911021.

\bibitem{BPS} M. Bianchi, G. Pradisi and A. Sagnotti, Nucl. Phys. {\bf B376}
(1992) 365.

\bibitem{Bi} M. Bianchi, Nucl. Phys. {\bf B528} (1998) 73.

\bibitem{Ka4} Z. Kakushadze, Nucl. Phys. {\bf B544} (1999) 265.

\bibitem{Wi} E. Witten, JHEP {\bf 9802} (1998) 006.

\bibitem{SS} S. Sen and S. Sethi, Nucl. Phys. {\bf B499} (1997) 45.

\bibitem{Po} J. Polchinski, Phys. Rev. {\bf D55} (1997) 6423.

\bibitem{An} C. Angelantonj, hep-th/9908064.

\bibitem{CHL} S. Chaudhuri, G. Hockney and J.D. Lykken, Phys. Rev. Lett.
{\bf 75} (1995) 2264;\\
S. Chaudhuri and J. Polchinski, Phys. Rev. {\bf D52} (1995) 7168.

\bibitem{Su} S. Sugimoto, Prog. Theor. Phys. {\bf 102} (1999) 685.

\bibitem{BeGa} M. Bianchi and A. Sagnotti, Phys. Lett. {\bf B247} (1990) 517;\\
A. Sagnotti, hep-th/9509080;\\
O. Bergman and M.R. Gaberdiel, Nucl. Phys. {\bf B499} (1997) 183.

\bibitem{GR} A. Giveon and M. Ro{\v c}ek, Nucl. Phys. {\bf B380} (1992) 128.
 
\bibitem{NC} Z. Kakushadze, hep-th/9910087.

\bibitem{HKMS} J.A. Harvey, S. Kachru, G. Moore and E. Silverstein, 
hep-th/9909072.

\bibitem{SeWi} N. Seiberg and E. Witten, JHEP {\bf 9909} (1999) 032.

\bibitem{Wi1} E. Witten, Nucl. Phys. {\bf B460} (1996) 506.

\bibitem{BLPSSW} M. Berkooz, R.G. Leigh, J. Polchinski, J.H. Schwarz, 
N. Seiberg and E. Witten, Nucl. Phys. {\bf B475} (1996) 115.

\bibitem{Serone} C.A. Scrucca and M. Serone, hep-th/9907112.

\bibitem{Ha} E. Witten, JHEP {\bf 9807} (1998) 006;\\
K. Hori, Nucl. Phys. {\bf B539} (1999) 35;\\
A. Hanany, B. Kol and A. Rajaraman, JHEP {\bf 9910} (1999) 027;\\
A. Hanany, talk presented at the 1999 ITP program ``Supersymmetric
Gauge Dynamics and String Theory'', 
{\tt http://www.itp.ucsb.edu/online/susy99/hanany}.

\bibitem{Asp} P.S. Aspinwall, Phys. Lett. {\bf B357} (1995) 329.

\bibitem{Sen} A. Sen, Nucl. Phys. {\bf B498} (1997) 135.

\bibitem{BST} A. Buchel, G. Shiu and S.-H.H. Tye, hep-th/9907203.

\bibitem{Do} M.R. Douglas, J. Geom. Phys. {\bf 28} (1998) 255.

\bibitem{sen1} A. Sen, Phys. Rev. {\bf D55} (1997) 7345. 

\bibitem{vafa} C. Vafa, Nucl. Phys. {\bf B469} (1996) 403.

\bibitem{TeV} N. Arkani-Hamed, S. Dimopoulos and G. Dvali, Phys. Lett. 
{\bf B429} (1998) 263; Phys. Rev. {\bf D59} (1999) 086004;\\
K.R. Dienes, E. Dudas and T. Gherghetta, Phys. Lett. {\bf B436} (1998) 55;
Nucl. Phys. {\bf B537} (1999) 47; hep-ph/9807522;\\
I. Antoniadis, N. Arkani-Hamed, S. Dimopoulos and G. Dvali, Phys. Lett. {\bf
B436} (1998) 257;\\
Z. Kakushadze and S-H.H. Tye, Nucl. Phys. {\bf B548} (1999) 180;\\
N. Arkani-Hamed and S. Dimopoulos, hep-ph/9811353;\\
Z. Berezhiani and G. Dvali, Phys. Lett. {\bf B450} (1999) 24;\\
Z. Kakushadze, Nucl. Phys. {\bf B548} (1999) 205; Nucl. Phys. {\bf B552}
(1999) 3; Nucl. Phys. {\bf B551} (1999) 549; Phys. Lett. {\bf B466} (1999)
251, hep-th/9908016;\\
N. Arkani-Hamed, S. Dimopoulos, G. Dvali and J. March-Russell, 
hep-ph/9811448;\\
G. Dvali and S.-H.H. Tye, Phys. Lett. {\bf B450} (1999) 72;\\
Z. Kakushadze and T.R. Taylor, Nucl. Phys. {\bf B562} (1999) 78;\\ 
A.K. Grant and Z. Kakushadze, Phys. Lett. {\bf B465} (1999) 108;\\
N. Arkani-Hamed, S. Dimopoulos, G. Dvali and N. Kaloper, hep-ph/9911386.

\bibitem{cvetic1} M. Cveti{\v c}, private communication.

\bibitem{KTKT} Z. Kakushadze and S.-H.H. Tye, Phys. Rev. {\bf D58} (1998)
126001.

\bibitem{BLUMEN} R. Blumenhagen, L. G{\"o}rlich and B. K{\"o}rs, 
hep-th/9908130; hep-th/9912204.

\bibitem{uranga} G. Aldazabal and A.M. Uranga, JHEP {\bf 9910} (1999) 024. 

\bibitem{quevedo} G. Aldazabal, L.E. Ib{\'a}{\~n}ez and F. Quevedo, 
hep-th/9909172.

\bibitem{non-susy} Z. Kakushadze, Phys. Rev. {\bf D59} (1999) 045007.

\bibitem{BKV} M. Bershadsky, Z. Kakushadze and C. Vafa, Nucl. Phys.
{\bf B523} (1998) 59;\\
Z. Kakushadze, Nucl. Phys. {\bf B529} (1998) 157.

\end{references}
\end{document}